\documentclass[12pt]{article}
\usepackage{amssymb}
\usepackage{amsmath}
\usepackage{graphicx,color}

\numberwithin{equation}{section}

\begin{document}
\begin{titlepage}
%\begin{flushright}
%AAA-BB-111 \\
%\end{flushright}
%
%\vspace*{10mm}
\begin{center}
\baselineskip 25pt 
{\Large\bf
%%%%%%%%%%%%%%%%%%%%%%%%%%%%%%%%%%%%%%%%%%%%%%%%%%%%%%%%%%%%%%%
Classically conformal U(1)$^\prime$ extended standard model, \\
electroweak vacuum stability, and LHC Run-2 bounds 
%%%%%%%%%%%%%%%%%%%%%%%%%%%%%%%%%%%%%%%%%%%%%%%%%%%%%%%%%%%%%%%
}
\end{center}
\vspace{2mm}
\begin{center}
{\large
Arindam Das$^{~a}$ \footnote{adas8@ua.edu},
Satsuki Oda$^{~b,c}$ \footnote{satsuki.oda@oist.jp},
Nobuchika Okada$^{~a}$ \footnote{okadan@ua.edu}, 

\vspace{3mm}
and 
Dai-suke Takahashi$^{~b,c}$ \footnote{daisuke.takahashi@oist.jp}
}
\end{center}
\vspace{2mm}

\begin{center}
{\it
$^{a}$Department of Physics and Astronomy, University of Alabama, \\
Tuscaloosa, Alabama 35487, USA
\vspace{5mm}

$^{b}$Okinawa Institute of Science and Technology Graduate University (OIST), \\ 
Onna, Okinawa 904-0495, Japan
\vspace{5mm}

$^{c}$Research Institute, Meio University, \\
Nago, Okinawa 905-8585, Japan \\
}
\end{center}
\vspace{3mm}
%%%%%%%%%%%%%%%%%%%%%%
\begin{abstract}
%%%%%%%%%%%%%%%%%%%%%%

We consider the minimal U(1)$^\prime$ extension of the standard model (SM) with the classically conformal invariance, 
   where an anomaly-free U(1)$^\prime$ gauge symmetry is introduced along with three generations of right-handed neutrinos 
   and a U(1)$^\prime$ Higgs field. 
Since the classically conformal symmetry forbids all dimensional parameters in the model,  
   the U(1)$^\prime$  gauge symmetry is broken by the Coleman-Weinberg mechanism, 
   generating the mass terms of the U(1)$^\prime$ gauge boson ($Z^\prime$ boson) and the right-handed neutrinos. 
Through a mixing quartic coupling between the U(1)$^\prime$ Higgs field and the SM Higgs doublet field, 
  the radiative U(1)$^\prime$  gauge symmetry breaking also triggers the breaking of the electroweak symmetry. 
In this model context, we first investigate the electroweak vacuum instability problem in the SM. 
Employing the renormalization group equations at the two-loop level 
   and the central values for the world average masses of the top quark ($m_t=173.34$ GeV) 
   and the Higgs boson ($m_h=125.09$ GeV), we perform parameter scans to identify 
   the parameter region for resolving the electroweak vacuum instability problem. 
Next we interpret the recent ATLAS and CMS search limits at the LHC Run-2 
   for the sequential $Z^\prime$ boson to constrain the parameter region in our model. 
Combining the constraints from the electroweak vacuum stability and the LHC Run-2 results, 
  we find a bound on the $Z^\prime$ boson mass as $m_{Z^\prime} \gtrsim 3.5$ TeV.   
We also calculate self-energy corrections to the SM Higgs doublet field through the heavy states, 
   the right-handed neutrinos and the $Z^\prime$ boson, and find the naturalness bound 
   as $m_{Z^\prime} \lesssim 7$ TeV, in order to reproduce the right electroweak scale 
   for the fine-tuning level better than 10\%.  
The resultant mass range of $3.5$ TeV $\lesssim m_{Z^\prime} \lesssim 7$ TeV will be explored 
  at the LHC Run-2 in the near future. 
 
\end{abstract}
\end{titlepage}

%%%%%%%%%%%%%%%%%%%%%%%%%%%%%%%%%%%%
\section{Introduction}
%%%%%%%%%%%%%%%%%%%%%%%%%%%%%%%%%%%%
One of the most serious problems in the standard model (SM) is the so-called gauge hierarchy problem, 
   which has been motivating us to seek new physics beyond the SM for decades.   
The problem originates from the fact that quantum corrections to the self-energy 
  of the SM Higgs doublet field quadratically diverge, and this divergence, once cut off 
  by a physical new physics scale being much higher than the electroweak scale, 
  must be canceled by a fine-tuning of the Higgs mass parameter at the tree level. 
Because of the chiral nature of the SM, the SM Lagrangian possesses the conformal (scale) invariance 
  at the classical level, except for the Higgs mass term.  
It has been argued in \cite{Bardeen} that once the classically conformal invariance and its minimal violation 
  by quantum anomalies are imposed on the SM, it could be free from the quadratic divergences;
  hence, the classically conformal invariance might provide us with a solution to the gauge hierarchy problem.  
This picture nicely fits a setup first investigated by Coleman and Weinberg~\cite{CW}, 
  namely, a U(1) gauge theory with a massless Higgs field. 
In this setup, it has been shown that  the U(1) gauge symmetry is radiatively broken 
  in the Coleman-Weinberg effective potential (Coleman-Weinberg mechanism).

Although it is tempting to apply this Coleman-Weinberg mechanism to the SM Higgs sector, 
  this cannot work with the observed values of top quark and weak boson masses, 
  since the Coleman-Weinberg potential for the SM Higgs field is found to be unbounded 
  from below \cite{Fujikawa}. 
Therefore, in order to pursue this scheme, it is necessary to extend the SM.  
There have been a lot of new physics model proposals (see, for example, \cite{CFmodels, Khoze, IOO1}). 
In this paper, we consider the classically conformal U(1)$^\prime$ extension of the SM proposed in \cite{OOT}, 
   where in addition to the SM particle content, three generations of right-handed neutrinos and a U(1)$^\prime$ 
   Higgs field are introduced. 
By assigning generation-independent U(1)$^\prime$ charges for fermions, 
   making the model free from all gauge and gravitational anomalies,
   and reproducing the Yukawa structure in the SM, 
   it turns out that the U(1)$^\prime$ gauge symmetry is identified as a linear combination 
   of the SM U(1)$_Y$ and the U(1)$_{B-L}$ gauge groups~\cite{ADH}.   
Hence, our model is a generalization on the classically conformal U(1)$_{B-L}$ extension of the SM 
   proposed in \cite{ IOO1}.  
The U(1)$^\prime$ gauge symmetry is radiatively broken by the Coleman-Weinberg mechanism, 
   and the U(1)$^\prime$ gauge field ($Z^\prime$ boson) and the right-handed (Majorana) neutrinos acquire their masses.  
A mixing quartic coupling between the U(1)$^\prime$ Higgs and the SM Higgs doublet fields  
   generates a negative mass squared for the SM Higgs doublet field, and the electroweak symmetry breaking is driven. 
Therefore, the radiative U(1)$^\prime$ gauge symmetry is the sole origin of the mass scale in this model.
With the Majorana heavy neutrinos, the seesaw mechanism \cite{seesaw} is automatically implemented, 
   and tiny active neutrino masses and their flavor mixing are generated after the electroweak symmetry breaking.

The SM Higgs boson has been discovered at the LHC, and 
  this marks the beginning of the experimental confirmation of the SM Higgs sector. 
The observed Higgs boson mass of $m_h=125.09 \pm 0.21 (\rm{stat.}) \pm 0.11 (\rm{syst.})$ GeV 
  from a combined analysis by the ATLAS and the CMS \cite{MhCombine}
  indicates that the electroweak vacuum is unstable \cite{RGErun} 
  since the SM Higgs quartic coupling becomes negative far below the Planck mass, 
  for the top quark pole mass $m_t=173.34 \pm 0.76$ from the combined measurements 
  by the Tevatron and the LHC experiments \cite{MtCombine}.  
Practically, this instability may not be a problem in the SM, since the lifetime of our electroweak 
  vacuum is estimated to be much longer than the age of the universe \cite{meta-stability}. 
However, in the presence of the U(1)$^\prime$ Higgs field, our Higgs potential is a function of 
  two Higgs fields, and there might be a flat path around high mountains of the potential
  toward the true vacuum and make the lifetime of our electroweak vacuum very short.  
Because of a lack of field theoretical technology for analyzing the effective scalar potential 
  with multiscalars in a wide range of field values, it would be the best way to solve the electroweak 
  vacuum instability problem in the context of our model.

In this paper, we investigate the electroweak vacuum stability in the classically conformal U(1)$^\prime$ extended SM.  
In the same model context, the electroweak vacuum stability problem and the current experimental bounds on 
   the model have been investigated in \cite{OOT}. 
The purpose of the present paper is to improve the analysis in \cite{OOT} for the electroweak vacuum stability 
   by the renormalization group equations (RGEs) at the two-loop level, and present a complete result for the parameter scan. 
We also update the constraints on the model parameters by taking into account 
   the recent LHC Run-2 results for the search for $Z^\prime$ boson resonances \cite{ATLAS13_Zp, CMS13_Zp}.  
We find that the LHC Run-2 results dramatically improve those obtained from the Run-1 results. 
In addition, we calculate the SM Higgs self-energy corrections from the effective potential 
   involving the heavy states, the right-handed neutrinos and the $Z^\prime$ boson,
   after the U(1)$^\prime$ symmetry breaking, and derive the naturalness bounds 
   to reproduce the right electroweak scale for a fine-tuning level better than 10\%.

This paper is organized as follows.
Our U(1)$^\prime$ model is defined in the next section. 
In Sec.~\ref{Sec_U1pSB}, we discuss the radiative U(1)$^\prime$ symmetry breaking
  through the Coleman-Weinberg mechanism.  
The electroweak symmetry breaking triggered by the radiative U(1)$^\prime$ gauge symmetry breaking  
  is discussed in Sec.~\ref{Sec_EWSB}.  
In Sec.~\ref{Sec_stability}, we analyze the renormalization group (RG) evolutions of the couplings at the two-loop level 
  and find a region in three dimensional parameter space which can resolve the electroweak vacuum instability 
  and keep all parameters in the perturbative regime up to the Planck mass.   
In Sec.~\ref{Sec_collider}, we analyze the current collider bounds of the model parameters; 
  in particular, the recent ATLAS and CMS results of the search for the $Z^\prime$ boson resonance 
  at the LHC Run-2 are interpreted in the $Z^\prime$ boson case of our model.  
In Sec.~\ref{Sec_naturalness},
  we  evaluate self-energy corrections to the SM Higgs doublet from the effective potential
  and derive the naturalness bounds to reproduce the electroweak scale
  for a fine-tuning level better than 10\%.
We summarize our results in Sec.~\ref{Sec_conclusion}. 
Formulas we used in our analysis are listed in the appendices.

Although Secs.~\ref{Sec_U1p}-\ref{Sec_EWSB} 
  substantially overlap with our previous work \cite{OOT}, 
  we have repeated the similar discussions for convenience. 
Those familiar with the basic properties 
  of the classically conformal U(1)$^\prime$ model discussed in Secs.~\ref{Sec_U1p}-\ref{Sec_EWSB}
  can skip over these sections.

%%%%%%%%%%%%%%%%%%%%%%%%%%%%%%%%%%%%%%%%%%%%%%%%
\section{Classically conformal U(1)$^{\prime}$ extended SM}
%%%%%%%%%%%%%%%%%%%%%%%%%%%%%%%%%%%%%%%%%%%%%%%%
\label{Sec_U1p}

%%%%%%%%%%%%%%%%%%%%%%%%%%%%%%%%%%%%%%%%%%%%%%%
\begin{table}[t]
\begin{center}
\begin{tabular}{c|ccc|rcr}
            & SU(3)$_c$ & SU(2)$_L$ & U(1)$_Y$ & \multicolumn{3}{c}{U(1)$^\prime$} \\
\hline
&&&&&&\\[-12pt]
$q_L^i$    & {\bf 3}   & {\bf 2}& $+1/6$ & $x_q$ 		& = & $\frac{1}{3}x_H + \frac{1}{6}x_\Phi$  \\[2pt] 
$u_R^i$    & {\bf 3} & {\bf 1}& $+2/3$ & $x_u$ 		& = & $\frac{4}{3}x_H + \frac{1}{6}x_\Phi$  \\[2pt] 
$d_R^i$    & {\bf 3} & {\bf 1}& $-1/3$ & $x_d$ 		& = & $-\frac{2}{3}x_H + \frac{1}{6}x_\Phi$  \\[2pt] 
\hline
&&&&&&\\[-12pt]
$\ell_L^i$    & {\bf 1} & {\bf 2}& $-1/2$ & $x_\ell$ 	& = & $- x_H - \frac{1}{2}x_\Phi$   \\[2pt] 
$\nu_R^i$   & {\bf 1} & {\bf 1}& $0$   & $x_\nu$ 	& = & $- \frac{1}{2}x_\Phi$ \\[2pt] 
$e_R^i$   & {\bf 1} & {\bf 1}& $-1$   & $x_e$ 		& = & $- 2x_H - \frac{1}{2}x_\Phi$  \\[2pt] 
\hline
&&&&&&\\[-12pt]
$H$         & {\bf 1} & {\bf 2}& $+1/2$  &  $x_H$ 	& = & $x_H$\hspace*{12.5mm}  \\ 
$\Phi$      & {\bf 1} & {\bf 1}& $0$  &  $x_\Phi$ 	& = & $x_\Phi$  \\ 
\end{tabular}
\end{center}
\caption{
Particle contents. 
In addition to the SM particle contents, the right-handed neutrino $\nu_R^i$ ($i=1,2,3$ denotes the generation index) and a complex scalar $\Phi$ are introduced. 
}
\label{Tab:particle_contents}
\end{table}
%%%%%%%%%%%%%%%%%%%%%%%%%%%%%%%%%%%%%%%%%%%%%%%

The model we investigate is the anomaly-free U(1)$^\prime$ extension of the SM 
  with the classically conformal invariance, which is based on the gauge group 
  SU(3)$_C \times$SU(2)$_L \times$U(1)$_Y \times$U(1)$^\prime$. 
The particle contents of the model are listed in Table~\ref{Tab:particle_contents}. 
In addition to the SM particle content, three generations of right-hand neutrinos $\nu_R^i$ 
  and a U(1)$^\prime$  Higgs field $\Phi$ are introduced. 
The covariant derivatives relevant to U(1)$_Y \times$ U(1)$^\prime$ are defined as 
\begin{equation}
D_\mu \equiv \partial_\mu  %- i g_3 T^a G^{a\mu} - i g_2 \tau^i W^{i\mu}  
			- i\left(\begin{array}{cc} Y_{1} & Y_{X} \end{array}\right )
			\left(\begin{array}{cc} g_{1} & g_{1X} \\g_{X1} & g_{X} \end{array}\right)
			\left(\begin{array}{c} B_{\mu} \\B_{\mu}^\prime \end{array}\right), 
 \label{Eq:covariant_derivative}
\end{equation}
where $Y_{1}$ ($Y_{X}$) are U(1)$_Y$ (U(1)$^\prime$) charge of a particle, 
and the gauge couplings $g_{X1}$ and $g_{1X}$ are introduced associated with a kinetic mixing 
between the two U(1) gauge bosons.

For generation-independent charge assignments,  the U(1)$^\prime$ charges of the fermions 
  are defined to satisfy the gauge and gravitational anomaly-free conditions:
\begin{align}
{\rm U}(1)^\prime \times \left[ {\rm SU}(3)_C \right]^2&\ :&
			2x_q - x_u - x_d &\ =\  0, \nonumber \\
{\rm U}(1)^\prime \times \left[ {\rm SU}(2)_L \right]^2&\ :&
			3x_q + x_\ell &\ =\  0, \nonumber \\
{\rm U}(1)^\prime \times \left[ {\rm U}(1)_Y \right]^2&\ :&
			x_q - 8x_u - 2x_d + 3x_\ell - 6x_e &\ =\  0, \nonumber \\
\left[ {\rm U}(1)^\prime \right]^2 \times {\rm U}(1)_Y&\ :&
			x_q^2 - 2x_u^2 + x_d^2 - x_\ell^2 + x_e^2 &\ =\  0, \nonumber \\
\left[ {\rm U}(1)^\prime \right]^3&\ :&
			6x_q^3 - 3x_u^3 - 3x_d^3 + 2x_\ell^3 - x_\nu^3 - x_e^3 &\ =\  0, \nonumber \\
{\rm U}(1)^\prime \times \left[ {\rm grav.} \right]^2&\ :&
			6x_q - 3x_u - 3x_d + 2x_\ell - x_\nu - x_e &\ =\  0. 
\label{Eq:anomaly_free_cond}
\end{align}
In order to reproduce observed fermion masses  and flavor mixings, 
  we introduce the following Yukawa interactions: 
\begin{equation}
{\cal L}_{\rm Yukawa} = - Y_u^{ij} \overline{q_L^i} \tilde{H} u_R^j
                                - Y_d^{ij} \overline{q_L^i} H d_R^j 
				- Y_\nu^{ij} \overline{\ell_L^i} \tilde{H} \nu_R^j - Y_e^{ij} \overline{\ell_L^i} H e_R^j
				- Y_M^i \Phi \overline{\nu_R^{ic}} \nu_R^i + {\rm h.c.},
\label{Eq:L_Yukawa}
\end{equation}
where $\tilde{H} \equiv i  \tau^2 H^*$, and the third and fifth terms on the right-hand side 
  are for the seesaw mechanism to generate neutrino masses. 
These Yukawa interaction terms impose 
\begin{eqnarray}
x_H 		&=& - x_q + x_u \ =\  x_q - x_d \ =\  - x_\ell + x_\nu \ =\  x_\ell - x_e, \nonumber \\
x_\Phi	&=& - 2x_\nu. 
\end{eqnarray} 
Solutions to these conditions are listed in Table~\ref{Tab:particle_contents} 
  and are controlled by only two parameters, $x_H$ and $x_\Phi$.  
The two parameters reflect the fact that the U(1)$^\prime$  gauge group can be defined as a linear combination 
  of the SM U(1)$_Y$ and the U(1)$_{B-L}$ gauge groups. 
Since the U(1)$^\prime$ gauge coupling $g_{X}$ is a free parameter of the model and 
  it always appears as a product $x_\Phi g_{X}$ or  $x_H g_{X}$,
  we fix $x_\Phi=2$ without loss of generality throughout this paper. 
This convention excludes the case in which the U(1)$^\prime$ gauge group is identical to the SM U(1)$_Y$. 
The choice of $(x_H, x_\Phi)=(0, 2)$ corresponds to the U(1)$_{B-L}$ model. 
Another example is $(x_H, x_\Phi)=(-1, 2)$, which corresponds to the SM with the so-called U(1)$_R$ symmetry. 
When we choose $(x_H, x_\Phi)=(-16/41, 2)$, the beta function of $g_{X1}$ ($g_{1X}$) at the 1-loop level 
   only has terms proportional to $g_{X1}$ ($g_{1X}$) (see Appendix A). 
This is the orthogonal condition between the U(1)$_Y$ and U(1)$^\prime$ at the 1-loop level, 
   under which  $g_{X1}$ and $g_{1X}$ do not evolve once we have set  $g_{X1}=g_{1X}=0$ at an energy scale.
Although it is slightly modified ($x_H$ becomes slightly larger than $-16/41$), 
    we find that the choice of $(x_H, x_\Phi)=(-16/41, 2)$ is a good approximation 
    even at the 2-loop level.

Imposing the classically conformal invariance, the scalar potential is given by
\begin{equation}
V = \lambda_H \! \left( H^\dagger H \right)^2 
	+ \lambda_\Phi \! \left( \Phi^\dagger \Phi \right)^2 
	+ \lambda_{\rm mix} \! \left( H^\dagger H \right) \! \left( \Phi^\dagger \Phi \right) , 
\label{Eq:classical_potential}
\end{equation}
where the mass terms are forbidden by the conformal invariance. 
If $\lambda_{\rm mix}$ is negligibly small, 
  we can analyze the Higgs potential separately for $\Phi$ and $H$ as a good approximation. 
This will be justified in the following sections. 
When the Majorana Yukawa couplings $Y_M^i$ are negligible compared to the U(1)$^\prime$ gauge coupling, 
 the $\Phi$ sector is identical to the original Coleman-Weinberg model \cite{CW}, 
   so the radiative U(1)$^\prime$ symmetry breaking will be achieved. 
Once $\Phi$ develops a vacuum expectation value (VEV) through the Coleman-Weinberg mechanism, 
   the tree-level mass term for the SM Higgs is effectively generated through $\lambda_{\rm mix}$ 
   in Eq.~(\ref{Eq:classical_potential}).
Taking $\lambda_{\rm mix}$ to be negative, the induced mass squared for the Higgs doublet 
  is negative and, as a result, the electroweak symmetry breaking is driven in the same way as in the SM.

%%%%%%%%%%%%%%%%%%%%%%%%%%%%%%%%%%%%%%%%%%%%%%
\section{Radiative U(1)$^\prime$ gauge symmetry breaking}
%%%%%%%%%%%%%%%%%%%%%%%%%%%%%%%%%%%%%%%%%%%%%%
\label{Sec_U1pSB}
Assuming $\lambda_{\rm mix}$ is negligibly small, we first analyze the U(1)$^\prime$ Higgs sector. 
Without mass terms, the Coleman-Weinberg potential \cite{CW} at the 1-loop level is found to be 
 % (in Landau gauge)
\begin{eqnarray}
  V(\phi) =  \frac{\lambda_\Phi}{4} \phi^4 
 + \frac{\beta_\Phi}{8} \phi^4 \left(  \ln \left[ \frac{\phi^2}{v_\phi^2} \right] - \frac{25}{6} \right), 
\label{Eq:CW_potential} 
\end{eqnarray}
where $\phi / \sqrt{2} = \Re[\Phi]$, and 
  we have chosen the renormalization scale to be the VEV  
  of $\Phi$ ($\langle \phi \rangle =v_\phi$).  
Here, the coefficient of the 1-loop quantum corrections is given by 
\begin{eqnarray}
\beta_\Phi	&=& \frac{1}{16 \pi^2} 
		\left[ 20\lambda_\Phi^2 
			+ 6 x_\Phi^4 \left ( g_{X1}^2 + g_{X}^2 \right)^2 - 16\sum_i(Y_M^i)^4 \right] \\
	& \simeq &  \frac{1}{16 \pi^2} 
		\left[ 6 \left(x_\Phi g_{X} \right)^4 - 16\sum_i(Y_M^i)^4 \right] , 
\end{eqnarray}
  where in the last expression, we have used 
  $\lambda_\Phi^2 \ll (x_\Phi g_{X})^4$  as usual in the Coleman-Weinberg mechanism 
   and set $g_{X1}=g_{1X}= 0$ at $ \langle \phi \rangle = v_\phi$,  for simplicity. 
The stationary condition $\left. dV/d\phi\right|_{\phi=v_\phi} = 0$ leads to 
\begin{eqnarray}
   \lambda_\Phi = \frac{11}{6} \beta_\Phi, 
\label{eq:stationary}
\end{eqnarray} 
 and this $\lambda_\Phi$ is nothing but a renormalized self-coupling at $v_\phi$ defined as 
\begin{eqnarray}
 \lambda_\Phi = \frac{1}{3 !}\left. \frac{d^4V(\phi)}{d \phi^4} \right|_{\phi=v_\phi}. 
\end{eqnarray}  
For more detailed discussion, see \cite{Khoze}.

Associated with this radiative U(1)$^\prime$ symmetry breaking 
 (as well as the electroweak symmetry breaking),  
  the U(1)$^\prime$ gauge boson ($Z^\prime$ boson) 
  and the right-handed Majorana neutrinos acquire their masses as 
\begin{eqnarray}
  m_{Z^\prime}  = \sqrt{(x_\Phi g_{X} v_\phi)^2  +  (x_H g_{X} v_h)^2} \simeq   x_\Phi g_{X} v_\phi, 
   \;  \;  m_{N^i} = \sqrt{2} Y_M^i v_\phi, 
\end{eqnarray} 
where $v_h=246$ GeV is the SM Higgs VEV, and we have used $x_\Phi v_\phi  \gg x_H v_h$, 
   which will be verified below. 
In this paper, we assume degenerate masses for the three Majorana neutrinos,  
   $Y_M^i = y_M$ (equivalently, $m_{N^i}=m_N$) for all $i=1,2,3$, for simplicity. 
The U(1)$^\prime$ Higgs boson mass is given by 
\begin{eqnarray}
  m_\phi^2 = \left. \frac{d^2 V}{d\phi^2}\right|_{\phi=v_\phi}  
                    =\beta_\Phi v_\phi^2  \simeq 
  \frac{3}{8 \pi^2} \left( (x_\Phi g_{X})^4 - 8 y_M^4  \right) v_\phi^2 
  \simeq  \frac{3}{8 \pi^2}  \frac{ m_{Z^\prime}^4 - 2 m_N^4}{v_\phi^2}. 
\label{Eq:mass_phi}
\end{eqnarray} 
When the Yukawa coupling is negligibly small, this equation reduces to the well-known relation 
  derived in the original paper by Coleman-Weinberg \cite{CW}. 
For a sizable Majorana mass, this formula indicates that 
  the potential minimum disappears for $ m_N >  m_{Z^\prime}/2^{1/4}$, 
  so there is an upper bound on the right-handed neutrino mass 
  for the U(1)$^\prime$ symmetry to be broken radiatively.  
This is in fact the same reason why the Coleman-Weinberg mechanism
   in the SM Higgs sector fails to break the electroweak symmetry 
   when the top Yukawa coupling is large, as observed. 
In order to avoid the destabilization of the U(1)$^\prime$ Higgs potential, 
    we simply set $m_{Z^\prime}^4 \gg m_N^4$ in the following analysis. 
Note that this condition does not mean that the Majorana neutrinos must be very light, 
  even though a factor difference between $m_{Z^\prime}$ and $m_N$ is enough to satisfy the condition. 
For simplicity, we set $y_M=0$ at $v_\phi$ in the following RG analysis.

%%%%%%%%%%%%%%%%%%%%%%%%%%%%%%%%%%%%%%%
\section{Electroweak symmetry breaking}
%%%%%%%%%%%%%%%%%%%%%%%%%%%%%%%%%%%%%%%
\label{Sec_EWSB}
Let us now consider the SM Higgs sector. 
In our model, the electroweak symmetry breaking is achieved in a very simple way. 
Once the U(1)$^\prime$ symmetry is radiatively broken, 
  the SM Higgs doublet mass is generated through the mixing quartic term between $H$ and $\Phi$ 
  in the scalar potential in Eq.~(\ref{Eq:classical_potential}), 
\begin{equation}
  V(h) = \frac{\lambda_H}{4}h^4 + \frac{\lambda_{\rm mix}}{4} v_\phi^2 h^2,  
\end{equation}
where we have replaced $H$ by $H = 1/\sqrt{2}\, (0,\,h)$ in the unitary gauge. 
Choosing $\lambda_{\rm mix} < 0$, the electroweak symmetry is broken in the same way as in the SM \cite{IOO1}. 
However, we should note that a crucial difference from the SM is that in our model 
  the electroweak symmetry breaking originates from the radiative breaking of the U(1)$^\prime$ gauge symmetry. 
At the tree level, the stationary condition $V^\prime |_{h=v_h} = 0$ 
   leads to the relation 
   $|\lambda_{\rm mix}|= 2 \lambda_H (v_h/v_\phi)^2$, 
   and the Higgs boson mass $m_h$ is given by 
\begin{equation}
  m_h^2 = \left. \frac{d^2 V}{dh^2} \right|_{h=v_h} = |\lambda_{\rm mix}|v_\phi^2 = 2 \lambda_H v_h^2. 
\label{Eq:mass_h}
\end{equation}
In the following RG analysis, 
  this is used as the boundary condition for $\lambda_{\rm mix}$ at the renormalization scale $\mu=v_\phi$. 
Note that since $\lambda_H \sim 0.1$ and $v_\phi \gtrsim 10$ TeV by the large electron-positron collider (LEP) constraint \cite{LEP2A, LEP2B,LEP2Aupdate},  
  $|\lambda_{\rm mix}| \lesssim 10^{-5}$, which is very small.

In our discussion about the U(1)$^\prime$ symmetry breaking, we neglected $\lambda_{\rm mix}$ 
  by assuming it to be negligibly small. 
Here we justify this treatment. 
In the presence of $\lambda_{\rm mix}$ and the Higgs VEV, Eq.~(\ref{eq:stationary}) is modified as 
\begin{eqnarray}
 \lambda_\Phi = \frac{11}{6} \beta_\Phi + \frac{|\lambda_{\rm mix}|}{2} \left( \frac{v_h}{v_\phi} \right)^2 
 \simeq  
  \frac{1}{2 v_\phi^4}  \left(  \frac{11}{8 \pi^2} m_{Z^\prime}^4 + m_h^2 v_h^2    \right).   
\label{eq:consistency} 
\end{eqnarray}
Considering the current LHC Run-2 bound from the search for $Z^\prime$ boson resonances
  \cite{ATLAS13_Zp, CMS13_Zp},  
  $m_{Z^\prime} \gtrsim 3$ TeV,  we find that the first term in the parentheses 
  in the last equality is 5 orders of magnitude greater than the second term, 
  and therefore we can analyze the two Higgs sectors separately.

%%%%%%%%%%%%%%%%%%%%%%%%%%%%%%%
\section{Solving the SM Higgs vacuum instability}
%%%%%%%%%%%%%%%%%%%%%%%%%%%%%%%
\label{Sec_stability}

%%%%%%%%%%%   SM lambda_H and U(1)' lambda_H  %%%%%%%%%%%%%%%%%%%%%%%%%
\begin{figure}[t]
\begin{minipage}{0.5\linewidth}
\begin{center}
\includegraphics[width=0.95\linewidth]{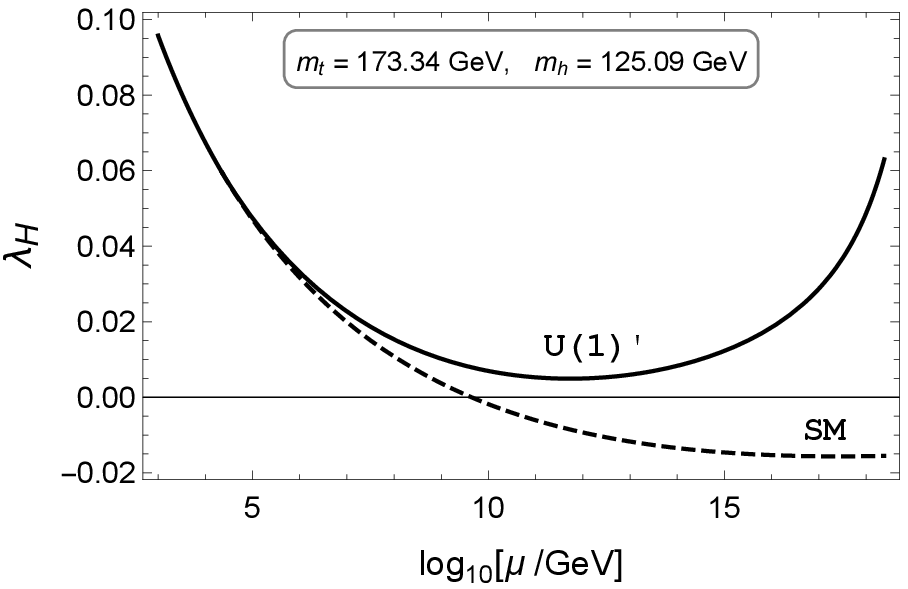}
(a)
\end{center}
\end{minipage}
%%%%%%%%%%%%%   SM lambda_H and U(1)' lambda_H  %%%%%%%%%%%%%%%%%%%%%%%%
\begin{minipage}{0.5\linewidth}
\begin{center}
\includegraphics[width=0.95\linewidth]{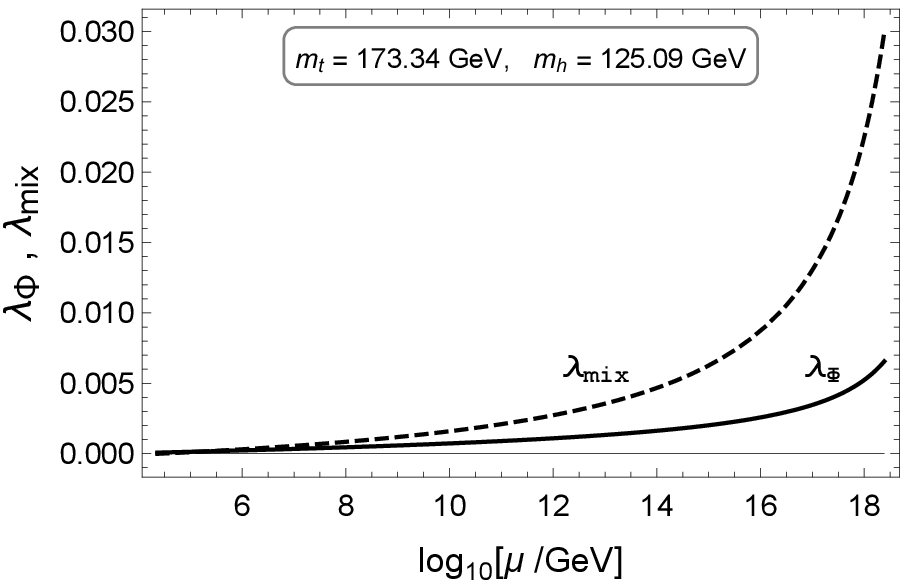}
(b)
\end{center}
\end{minipage}
\caption
{
(a) The evolutions of the Higgs quartic coupling $\lambda_H$ (solid line) 
    for the inputs $m_t=173.34$ GeV and $m_h=125.09$ GeV, 
    along with the SM case (dashed line). 
(b) The RG evolutions of $\lambda_\Phi$ (solid line) and $\lambda_{\rm mix}$ (dashed line). 
Here, we have taken $x_H = 2$, $v_\phi = 23$ TeV and $g_{X}(v_\phi) = 0.09$. 
}
\label{Fig:Higgs_quartics}
\end{figure}
%%%%%%%%%%%%%%%%%%%%%%%%%%%%%%%%%%%%%%%%%%%%%%%%%%%%
%%%%%%%%%%%%%%%%%%%   3D Plotscan I %%%%%%%%%%%%%%%%%%%%
\begin{figure}[t]
\begin{minipage}{0.5\linewidth}
\begin{center}
\includegraphics[width=0.95\linewidth]{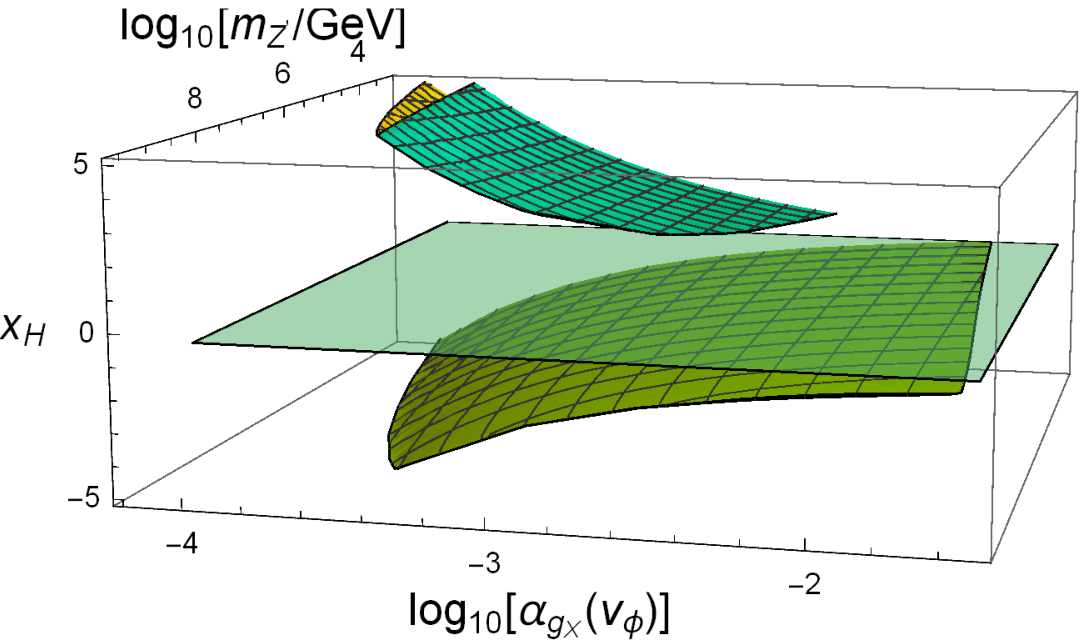}
(a)
\end{center}
\end{minipage}
%%%%%%%%%%%%%   3D Plotscan II  %%%%%%%%%%%%%%%%%%%%%%%%%%
\begin{minipage}{0.5\linewidth}
\begin{center}
\includegraphics[width=0.95\linewidth]{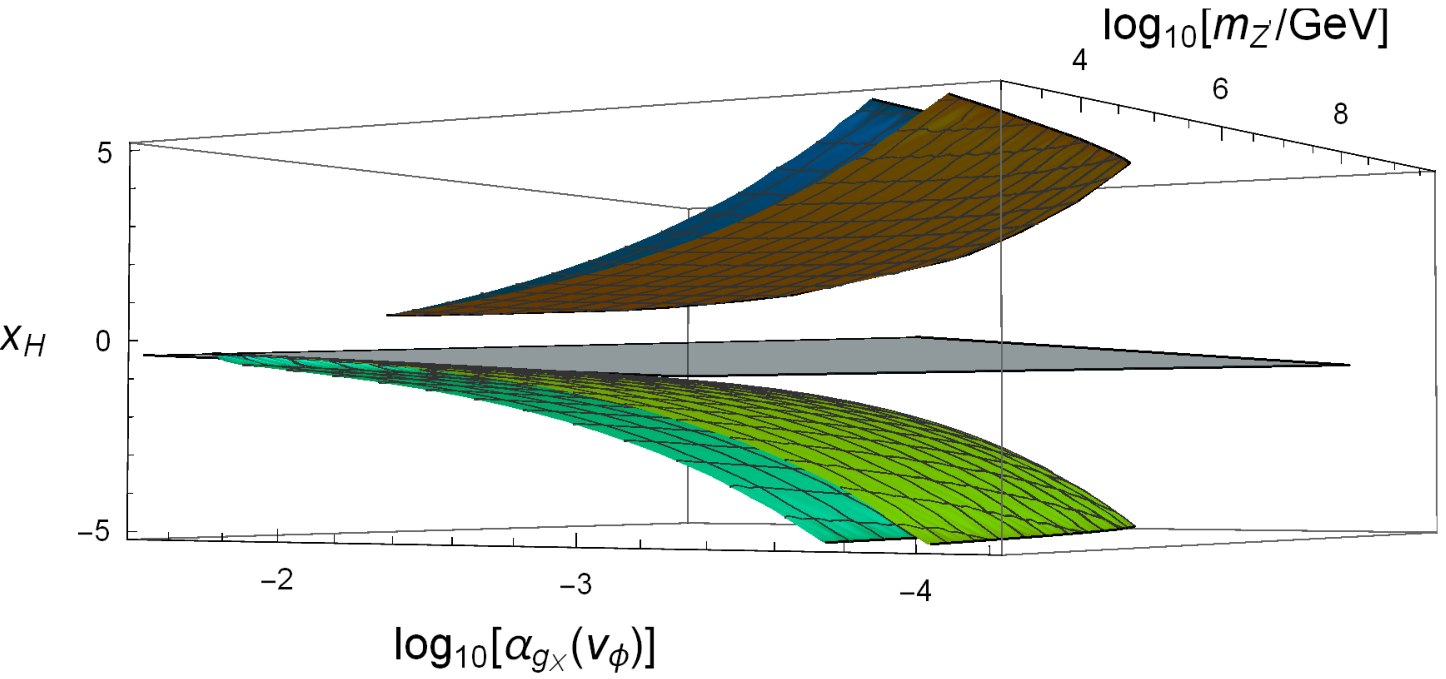}
(b)
\end{center}
\end{minipage}
\caption
{
(a) The result of the three-dimensional parameter scans for $v_\phi$, $g_{X}$ and $x_H$, 
   shown in ($m_{Z^\prime}({\rm GeV}), \alpha_{g_X}, x_H$) parameter space with $m_{Z^\prime} \simeq x_\Phi g_{X} v_\phi$,
   by using the inputs $m_t=173.34$ GeV and $m_h=125.09$ GeV.
As a reference, a horizontal plane for $x_H=-16/41$ is shown, which corresponds to the orthogonal case. 
(b) Same three-dimensional parameter scans as (a), but at a deferent angle.
}
\label{Fig:3DPlot}
\end{figure}
%%%%%%%%%%%%%%%%%%%%%%%%%%%%%%%%%%%%%%%%%%%%%%%%%%%%
%%%%%%%%%%%%%%%%%%%   x_H and m_Z'(g')  %%%%%%%%%%%%%%%%%%%%
\begin{figure}[htbp]
\begin{minipage}{0.5\linewidth}
\begin{center}
\includegraphics[width=0.95\linewidth]{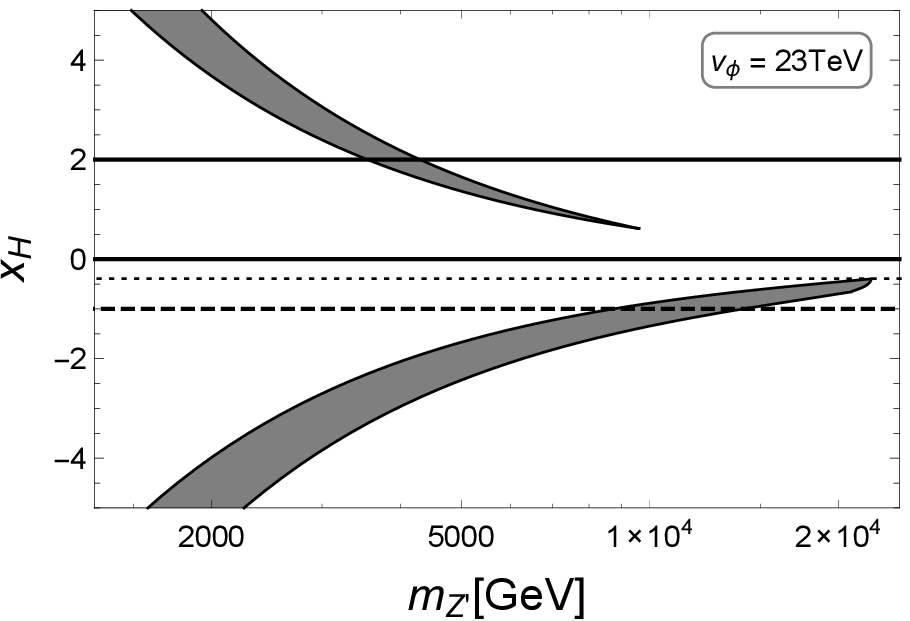}
(a)
\end{center}
\end{minipage}
%%%%%%%%%%%%%   x_H and m_Z'(v_Phi)  %%%%%%%%%%%%%%%%%%%%%%%%%%
\begin{minipage}{0.5\linewidth}
\begin{center}
\includegraphics[width=0.95\linewidth]{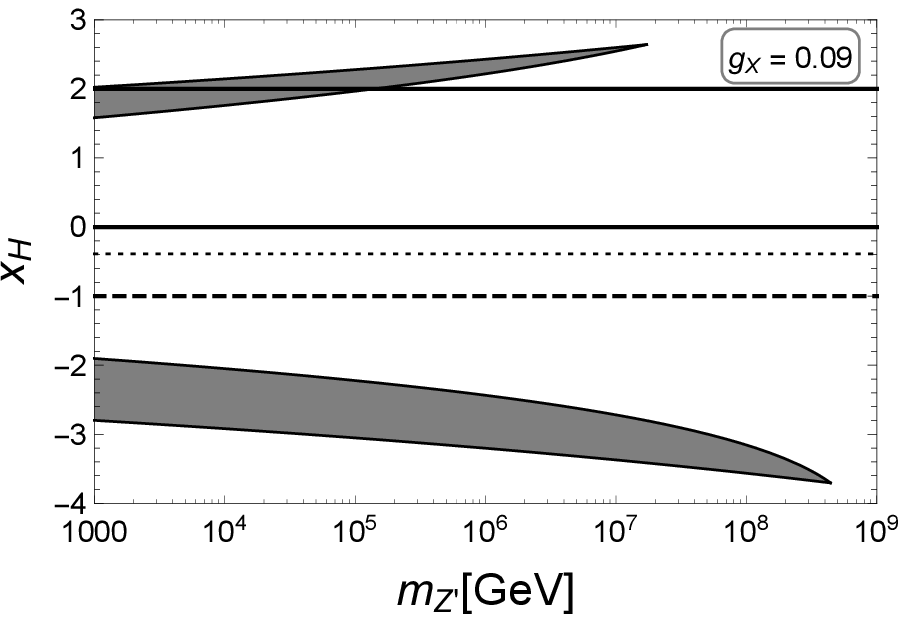}
(b)
\end{center}
\end{minipage}
\caption
{
(a) The result of parameter scan for $x_H$ and $g_{X}$ with a fixed $v_\phi=23$ TeV, 
    shown in the ($m_{Z^\prime}, x_H$) plane with 
    $m_{Z^\prime} \simeq x_\Phi g_{X} v_\phi$. 
As a reference, horizontal lines are depicted for $x_H=2$, $0$ [U(1)$_{B-L}$ case], $-16/41$ [orthogonal case],
 and $-1$ [U(1)$_R$ case]. 
(b) Same as (a), but for a parameter scan for $x_H$ and $v_\phi$ with a fixed $g_{X}=0.09$. 
}
\label{Fig:scan1}
\end{figure}
%%%%%%%%%%%%%%%%%%%%%%%%%%%%%%%%%%%%%%%%%%%%%%%%%%%%
%%%%%%%%%%%%%%%   x_H(+) and m_Z'(g')  enlarge  %%%%%%%%%%%%%%%%%%%%%%%%
\begin{figure}[htbp]
\begin{minipage}{0.5\linewidth}
\begin{center}
\includegraphics[width=0.95\linewidth]{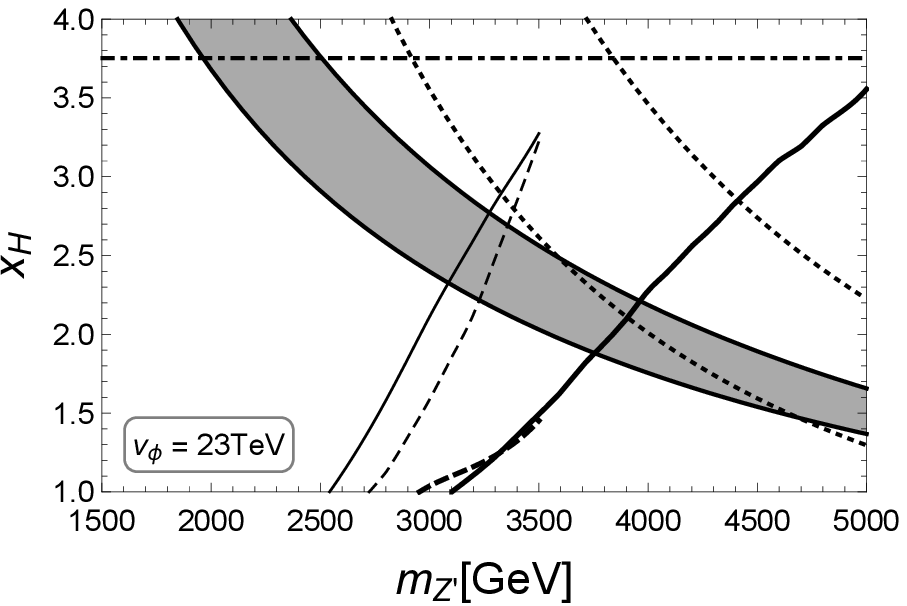}
(a)
\end{center}
\end{minipage}
%%%%%%%%%%%%%%  x_H(-) and m_Z'(g') enlarge %%%%%%%%%%%%%%%%%%%%
\begin{minipage}{0.5\linewidth}
\begin{center}
\includegraphics[width=0.95\linewidth]{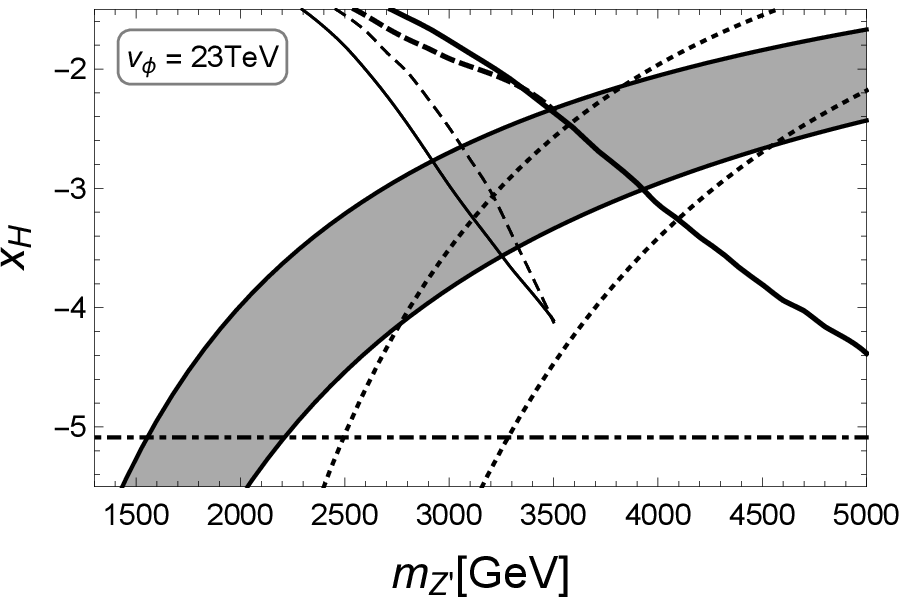}
(b)
\end{center}
\end{minipage}
\caption
{
(a) The allowed positive $x_H$ region at the TeV scale in  Fig.~\ref{Fig:scan1}(a) is magnified, 
    along with the LEP bound (dashed-dotted line), the LHC Run-1 CMS bound (thin dashed line), 
    the LHC Run-1 ATLAS bound (thin solid line), the LHC Run-2 CMS bound (thick dashed line) 
    and the LHC Run-2 ATLAS bound (thick solid line)
    from a direct search for $Z^\prime$ boson resonance. 
    The region on the left side of the lines is excluded.
     Here, the naturalness bounds for 10\% (right dotted line) and 30\% (left dotted line) fine-tuning levels
    are also depicted. 
(b) Same as (a), but for the negative $x_H$ region. 
}
\label{Fig:scan1a_enlarge}
\end{figure}
%%%%%%%%%%%%%%%%%%%%%%%%%%%%%%%%%%%%%%%%%%%%%%%%%%%%

%%%%%%%%%%%%%%%   x_H(+) and m_Z'(v_Phi)  enlarge  %%%%%%%%%%%%%%%%%%%%%%%%
\begin{figure}[htbp]
\begin{minipage}{0.5\linewidth}
\begin{center}
\includegraphics[width=0.95\linewidth]{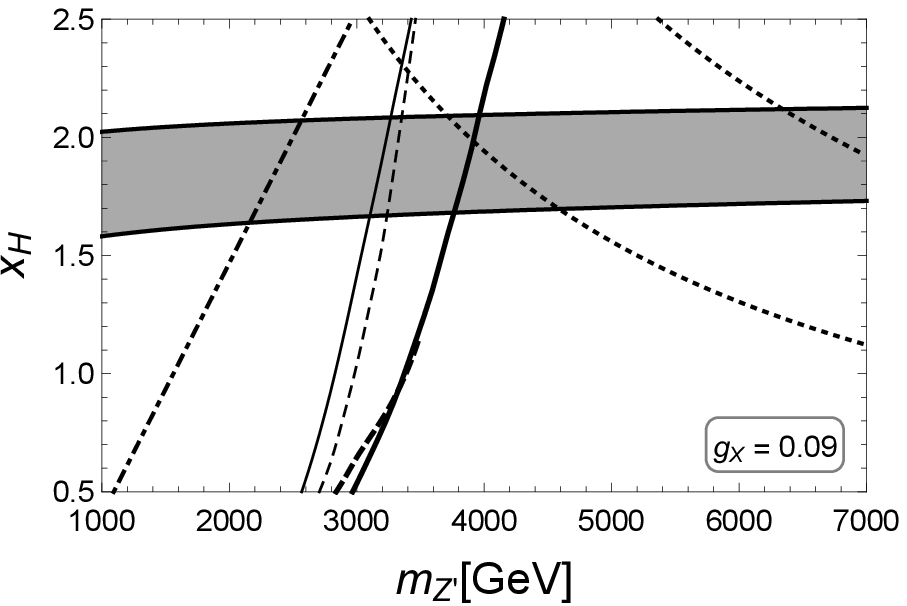}
(a)
\end{center}
\end{minipage}
%%%%%%%%%%%%%%  x_H(-) and m_Z'(v_Phi) enlarge %%%%%%%%%%%%%%%%%%%%
\begin{minipage}{0.5\linewidth}
\begin{center}
\includegraphics[width=0.95\linewidth]{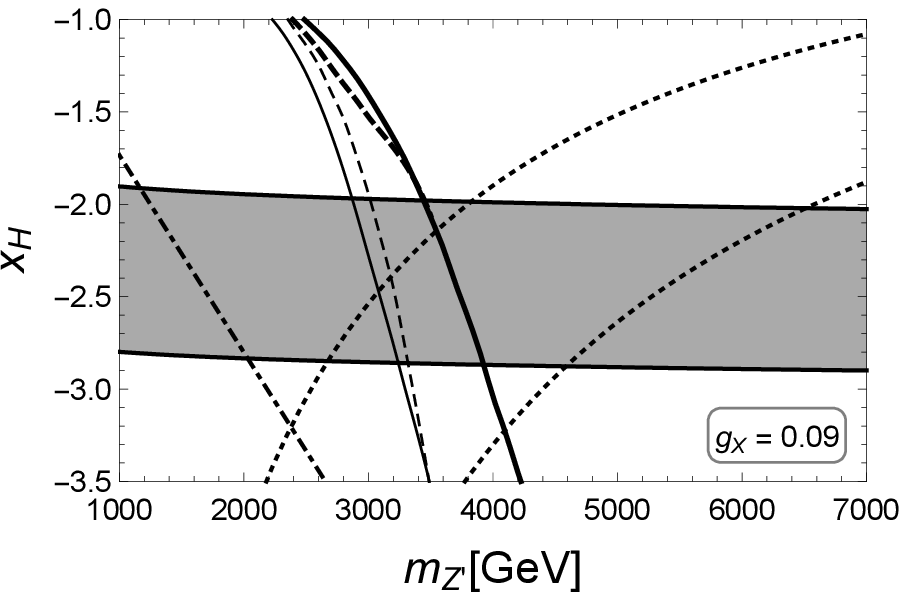}
(b)
\end{center}
\end{minipage}
\caption
{
(a) Same as Fig.~\ref{Fig:scan1a_enlarge}(a), but magnifying Fig.~\ref{Fig:scan1}(b). 
(b) Same as Fig.~\ref{Fig:scan1a_enlarge}(b), but magnifying Fig.~\ref{Fig:scan1}(b). 
}
\label{Fig:scan1b_enlarge}
\end{figure}
%%%%%%%%%%%%%%%%%%%%%%%%%%%%%%%%%%%%%%%%%%%%%%%%%%%%
%%%%%%%%%%%%%%%   ( x_H = 2) g' and m_Z'  %%%%%%%%%%%%%%%%%%%%%%%%
\begin{figure}[htbp]
\begin{minipage}{0.5\linewidth}
\begin{center}
\includegraphics[width=0.95\linewidth]{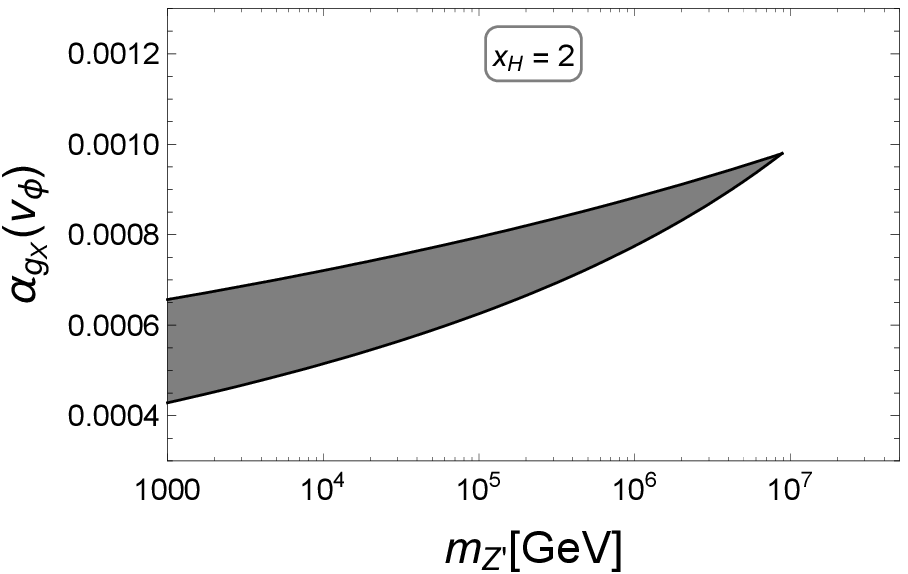}
(a)
\end{center}
\end{minipage}
%%%%%%%%%%%%%%   ( x_H = 2) g' and m_Z' enlarge %%%%%%%%%%%%%%%%%%%%
\begin{minipage}{0.5\linewidth}
\begin{center}
\includegraphics[width=0.95\linewidth]{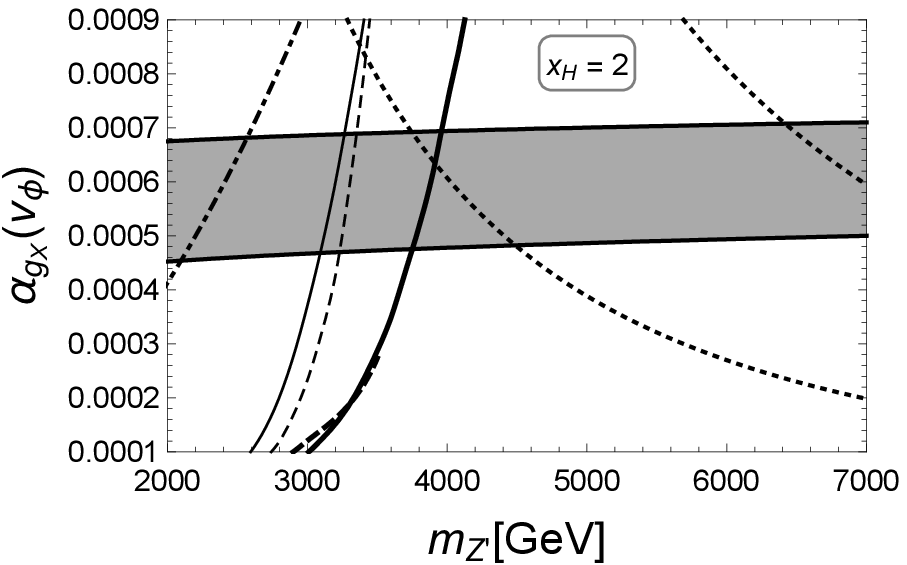}
(b)
\end{center}
\end{minipage}
\caption
{
(a) The result of parameter scan for $v_\phi$ and $g_{X}$ with a fixed $x_H=2$    
     in the ($m_{Z^\prime}, \alpha_{g_{X}}$) plane.
(b) The allowed region at the TeV scale in (a) is magnified, 
    along with the LEP bound (dashed-dotted line), the LHC Run-1 CMS bound (thin dashed line), 
    the LHC Run-1 ATLAS bound (thin solid line), the LHC Run-2 CMS bound (thick dashed line) 
    and the LHC Run-2 ATLAS bound (thick solid line) from direct search for $Z^\prime$ boson resonance. 
    The region on the left side of the lines is excluded.    
    Here, the naturalness bounds for 10\% (right dotted line) and 30\% (left dotted line) fine-tuning levels
    are also depicted. 
}
\label{Fig:scan2_xH=2}
\end{figure}
%%%%%%%%%%%%%%%%%%%%%%%%%%%%%%%%%%%%%%%%%%%%%%%%%%%%

%%%%%%%%%%%%%%%%%   ( x_H = -2.5) g' and m_Z'  %%%%%%%%%%%%%%%%%%%%%
\begin{figure}[htbp]
\begin{minipage}{0.5\linewidth}
\begin{center}
\includegraphics[width=0.95\linewidth]{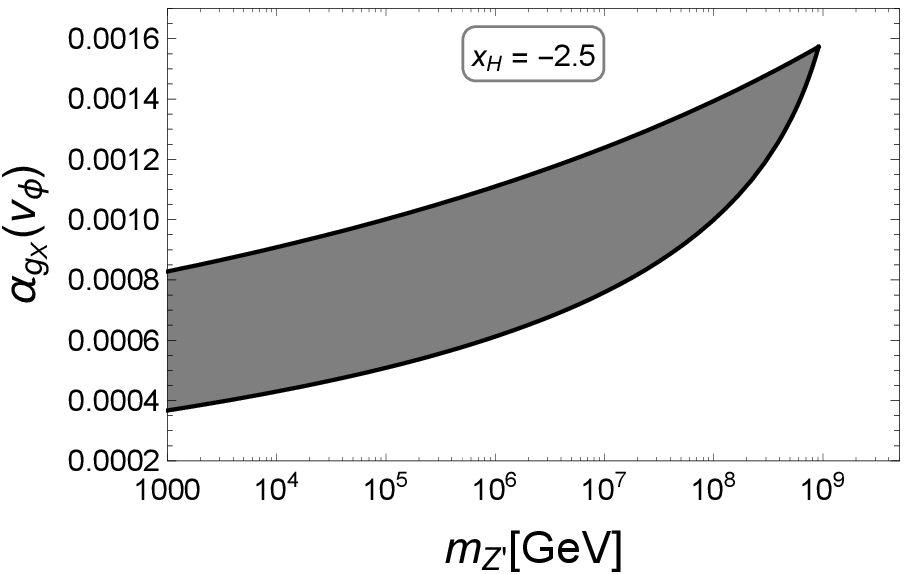}
(a)
\end{center}
\end{minipage}
%%%%%%%%%%%%%%%%%   ( x_H = -2.5) g' and m_Z' enlarge %%%%%%%%%%%%%%%%%
\begin{minipage}{0.5\linewidth}
\begin{center}
\includegraphics[width=0.95\linewidth]{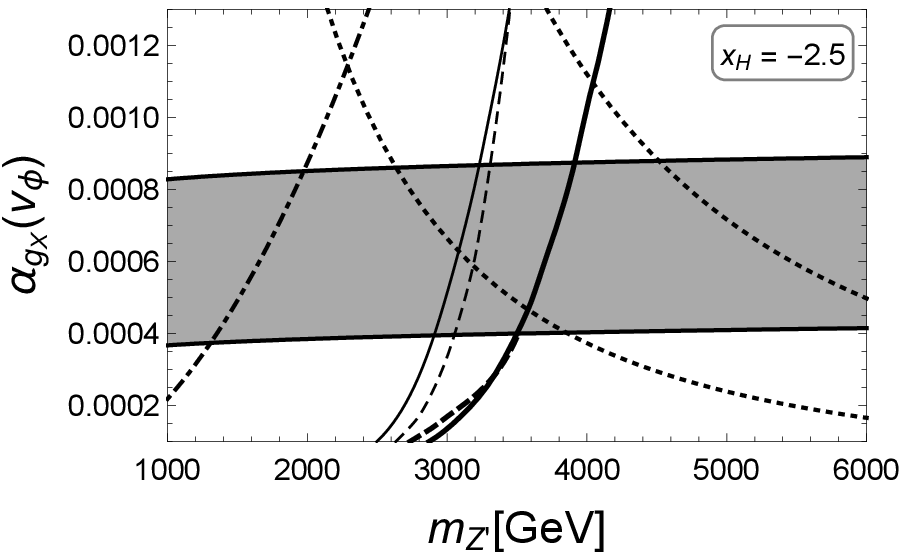}
(b)
\end{center}
\end{minipage}
\caption
{
(a) Same as Fig.~\ref{Fig:scan2_xH=2}(a), but for $x_H=-2.5$. 
(b) Same as Fig.~\ref{Fig:scan2_xH=2}(b), but for $x_H=-2.5$. 
}
\label{Fig:scan2_xH=-2.5}
\end{figure}
%%%%%%%%%%%%%%%%%%%%%%%%%%%%%%%%%%%%%%%%%%%%%%%%%%%%%%%

In the SM with the observed Higgs boson mass of $m_h=125.09$ GeV,  
   the RG evolution of the SM Higgs quartic coupling shows that the running coupling becomes 
   negative at the intermediate scale $\mu \simeq 10^{10}$ GeV \cite{RGErun} 
   for $m_t=173.34$ GeV, and hence the electroweak vacuum is unstable.   
In this section, we investigate RG evolution of the Higgs quartic coupling and a possibility  
  to solve the Higgs vacuum instability problem in our U(1)$^\prime$ extended SM.  
Without the classical conformal invariance, Ref.~\cite{Coriano} (see also \cite{CKL}) 
  has considered the same problem, and identified parameter regions which can resolve the Higgs vacuum instability. 
A crucial difference in our model is that because of the classical conformal invariance 
  and the symmetry breaking by the Coleman-Weinberg mechanism, 
  the initial values of $\lambda_\Phi$ and $\lambda_{\rm mix}$ at $v_\phi$ are not free parameters. 
Therefore, it is nontrivial to resolve the Higgs vacuum instability in the present model.   
The Higgs vacuum stability has been investigated in \cite{Khoze}
  for the classically conformal extension of the SM 
  with an extend gauge group and particle content including a dark matter candidate.

In our RGE analysis, we employ the SM RGEs at the 2-loop level \cite{RGErun}  
  from the top pole mass to the U(1)$^\prime$ Higgs VEV, and connect the RGEs 
  to those of the U(1)$^\prime$ extended SM at the 2-loop level, 
  which are generated by using SARAH \cite{SARAH}. 
RGEs used in our analysis are listed in the appendices. 
For inputs of the Higgs boson mass and top quark pole mass, 
  we employ a central value of the ATLAS and CMS combined measurement
  $m_h=125.09$ GeV \cite{MhCombine},
  while $m_t =173.34$ GeV is the central value of combined results of the Tevatron and the LHC 
  measurements of top quark mass \cite{MtCombine}.  
There are only three free parameters in our model, by which inputs at $v_\phi$ are determined: 
  $x_H$, $v_\phi$, and $g_{X}$.

In Fig.~\ref{Fig:Higgs_quartics}(a), we show the RG evolution of the SM Higgs quartic coupling 
   in our model (solid line), along with the SM result (dashed line).  
Here, we have taken $x_H= 2$, $v_\phi = 23$ TeV, and $g_{X}(v_\phi) = 0.09$ as an example.  
Recall that we have fixed $x_\Phi=2$ without loss of generality. 
The Higgs quartic coupling remains positive all the way up to the Planck mass, 
  so the Higgs vacuum instability problem is solved. 
There are complex, synergetic effects in the coupled RGEs to resolve the Higgs vacuum instability 
  (see the appendices for RGEs). 
For example, the U(1)$_Y$ gauge coupling grows faster than the SM case 
  in the presence of the mixing gauge couplings $g_{X1}$ and $g_{1X}$, 
  which makes the evolution of top Yukawa coupling decrease faster than in the SM case. 
The evolution of the mixing gauge coupling is controlled by the U(1)$^\prime$ gauge coupling. 
Both of them are asymptotic nonfree. 
The gauge couplings positively contribute to the beta function of 
  the SM Higgs quartic coupling, while the top Yukawa coupling gives a negative contribution.  
As a result, the RG evolutions of the gauge and top Yukawa couplings work 
  to change the sign of the beta function of the SM Higgs quartic coupling 
  at $\mu \simeq 10^{12}$ GeV in Fig.~\ref{Fig:Higgs_quartics}(a). 
Figure~\ref{Fig:Higgs_quartics}(b) shows the RG evolutions of the other Higgs quartic couplings. 
Note that the input of $\lambda_\Phi$ and $\lambda_{\rm mix}$ is very small 
  because of the radiative gauge symmetry breaking, and the two couplings remain very small 
  even at the Planck mass.   
Thus, the positive contribution of $\lambda_{\rm mix}$ to the beta function 
  of the SM Higgs quartic coupling is negligible. 
This is in sharp contrast to U(1) extended models without the conformal invariance, 
   where $\lambda_{\rm mix}$ is a free parameter and we can take its input to give a large, 
   positive contribution to the beta function; thus, the Higgs vacuum instability problem 
   is relatively easier to solve.

In order to identify a parameter region to resolve the Higgs vacuum instability, 
  we perform parameter scans for the free parameters $x_H$, $v_\phi$ and $g_{X}$. 
In this analysis, we impose several conditions on the running couplings at $v_\phi \leq \mu \leq M_P$ 
  ($M_P =2.4 \times 10^{18}$ GeV is the reduced Planck mass): 
  stability conditions of the Higgs potential  ($\lambda_H,  \lambda_\Phi > 0$), 
  and the perturbative conditions that all the running couplings remain in the perturbative regime, namely, 
  $g_i^2$ ($i=1,2,3$), $g_{X}^2$, $g_{X1}^2$, $g_{1X}^2<4\pi$
  and $\lambda_H$, $\lambda_\Phi$, $\lambda_{\rm mix}<4 \pi$. 
For theoretical consistency, we also impose a condition that the 2-loop beta functions 
  are smaller than the 1-loop beta functions. 
In Fig.~\ref{Fig:3DPlot}, we show the result of our parameter scans 
  in the three-dimensional parameter space of ($m_{Z^\prime}, \alpha_{g_X}, x_H$), 
  where $\alpha_{g_X}=g_X^2/(4 \pi)$. 
As a reference, we show a horizontal plane corresponding to the orthogonal case $x_H=-16/41$.  
There is no overlapping of the plane with the resultant parameter regions 
  to resolve the electroweak vacuum instability.

In order to discuss our results in detail, we show in Figs.~\ref{Fig:scan1}-\ref{Fig:scan2_xH=-2.5} 
  the parameter scan results on several two-dimensional hypersurfaces in the 3D plot of Fig.~\ref{Fig:3DPlot}. 
In Fig.~\ref{Fig:scan1}, our results are shown for $x_H$ and $g_{X}$ 
  with a fixed $v_\phi=23$ TeV (a) and for $x_H$ and $v_\phi$ with a fixed $g_{X}=0.09$ (b) 
  in the ($m_{Z^\prime}, x_H$) plane, along with the horizontal lines corresponding to 
  $x_H=2$, $0$ [U(1)$_{B-L}$ case], $-16/41$ [orthogonal case], and $-1$ [U(1)$_R$ case]. 
We can see that the resultant parameter space is very restricted. 
For example, the Higgs vacuum instability cannot be resolved
  in the classically conformal U(1)$_{B-L}$ extended SM 
  or the classically conformal orthogonal U(1) extended SM, 
  for the inputs $m_t=173.34$ GeV and $m_h=125.09$ GeV.
The allowed regions at the TeV scale in Figs.~\ref{Fig:scan1}(a) and \ref{Fig:scan1}(b) are magnified
  in Figs.~\ref{Fig:scan1a_enlarge} and \ref{Fig:scan1b_enlarge}, respectively.  
Here we also show the collider bounds, namely,
   the LEP bounds (dashed-dotted lines) \cite{LEP2A, LEP2B,LEP2Aupdate},
   the CMS bounds at the LHC Run-1 (thin dashed lines) \cite{CMS8_Zp}, 
   the ATLAS bounds at the LHC Run-1 (thin solid lines) \cite{ATLAS8_Zp},
   the CMS bounds at the LHC Run-2 (thick dashed lines) \cite{CMS13_Zp}, 
   and the ATLAS bounds at the LHC Run-2 (thick solid lines) \cite{ATLAS13_Zp},
   from the search for $Z^\prime$ boson mediated processes,
   which will be obtained in the next section. 
The region on the left side of the lines is excluded.   
Naturalness bounds (dotted lines), which will be obtained in Sec.~\ref{Sec_naturalness}, are also shown. 
These naturalness bounds for the 10\% fine-tuning level are found to be compatible 
  with the bounds obtained by the LHC Run-2 results. 
The result of parameter scan for $v_\phi$ and $\alpha_{g_{X}}$ 
  with a fixed $x_H=2$ is depicted in Fig.~\ref{Fig:scan2_xH=2}(a),
  and the allowed region at the TeV scale is magnified in Fig.~\ref{Fig:scan2_xH=2}(b),
  along with the collider and naturalness bounds. 
Same plots as Fig.~\ref{Fig:scan2_xH=2} but for $x_H=-2.5$ are shown in Fig.~\ref{Fig:scan2_xH=-2.5}.

%%%%%%%%%%%%%%%%%%%%%%%%%%%%%%%%%%%%%%%%%%%%%%%
\section{LHC Run-2 bounds on the U(1)$^\prime$ $Z^\prime$ boson}
%%%%%%%%%%%%%%%%%%%%%%%%%%%%%%%%%%%%%%%%%%%%%%%
\label{Sec_collider}
The ATLAS and the CMS Collaborations have searched for $Z^\prime$ boson resonance 
  at the LHC Run-1 with $\sqrt{s}=8$ TeV, 
  and continued the search at the LHC Run-2 with $\sqrt{s}=13$ TeV. 
The most stringent bounds on the $Z^\prime$ boson production cross section times branching ratio 
  have been obtained by using the dilepton final state. 
For the so-called sequential SM $Z^\prime$  ($Z^\prime_{SSM}$) model \cite{ZpSSM},
  where the $Z^\prime_{SSM}$ boson has exactly 
  the same couplings with the SM fermions as those of the SM $Z$ boson, 
  the cross section bounds from the LHC Run-1 results lead to lower bounds
  on the $Z^\prime_{SSM}$ boson mass
  as $m_{Z^\prime_{SSM}} \geq 2.90$ TeV from the ATLAS analysis \cite{ATLAS8_Zp} and 
  $m_{Z^\prime_{SSM}} \geq 2.96$ TeV from the CMS analysis \cite{CMS8_Zp}, respectively.  
Very recently, these bounds have been updated by the ATLAS \cite{ATLAS13_Zp}
   and CMS \cite{CMS13_Zp} analysis with the LHC Run-2 at $\sqrt{s}=13$ TeV as 
  $m_{Z^\prime_{SSM}} \geq 3.4$ TeV (ATLAS) and $m_{Z^\prime_{SSM}} \geq 3.15$ TeV (CMS), respectively.     
We interpret these ATLAS and CMS results in the U(1)$^\prime$ $Z^\prime$ boson case  
  and derive an upper bound on $x_H$ or $\alpha_{g_X}$ as a function of $m_{Z^\prime}$.

We calculate the dilepton production cross section
   for the process $pp \to Z^\prime +X \to \ell^{+} \ell^{-} +X$. 
The differential cross section with respect to the invariant mass $M_{\ell \ell}$ of the final state dilepton 
   is described as
\begin{eqnarray}
\frac{d \sigma}{d M_{\ell \ell}}
	= \sum_{a,b}
		\int^{1}_{\frac{M^2_{\ell \ell}}{E^2_{\rm CM}}} d x_1 \frac{2M_{\ell \ell}}{x_1 E^2_{\rm CM}} 
		f_{a}(x_{1}, M^2_{\ell \ell}) f_{b}\left(\frac{M^2_{\ell \ell}}{x_{1} E^2_{\rm CM}}, M^2_{\ell \ell} \right)  
		\hat{\sigma} (\bar{q} q \to Z^\prime \to  \ell^+ \ell^-),
\label{CrossLHC}
\end{eqnarray}
where $f_a$ is the parton distribution function for a parton $a$, 
  and $E_{\rm CM} =13$ TeV ($8$ TeV) is the center-of-mass energy of the LHC Run-2 (Run-1).
In our numerical analysis, we employ CTEQ5M~\cite{CTEQ} for the parton distribution functions. 
In the case of the U(1)$^\prime$ model,
  the cross section for the colliding partons with a fixed $x_\Phi=2$ is given by 
\begin{eqnarray}
\hat{\sigma} (\bar{u} u \rightarrow Z^\prime \to \ell^+ \ell^-)  
	&=& \frac{\pi \alpha_{g_X}^2}{81} 
		\frac{M_{\ell \ell}^2}{(M_{\ell \ell}^2-m_{Z^\prime}^2)^2 + m_{Z^\prime}^2 \Gamma_{Z^\prime}^2}
		(85x_H^4 + 152x_H^3 + 104x_H^2 + 32x_H + 4), 
\nonumber\\
\hat{\sigma}  (\bar{d} d \rightarrow Z^\prime \to \ell^+ \ell^-) 
	&=& \frac{\pi \alpha_{g_X}^2}{81} 
		\frac{M_{\ell \ell}^2}{(M_{\ell \ell}^2-m_{Z^\prime}^2)^2 + m_{Z^\prime}^2 \Gamma_{Z^\prime}^2}
		(25x_H^4 + 20x_H^3 + 8x_H^2 + 8x_H + 4), 
\label{CrossLHC2}
\end{eqnarray}
where the total decay width of the $Z^\prime$ boson is given by
\begin{eqnarray}
\Gamma_{Z^\prime} &=& 
	\frac{\alpha_{g_X} m_{Z^\prime}}{6} \!\!
	\left[ \frac{103x_H^2 + 86x_H + 37}{3} 
		+ \frac{17x_H^2 + 10x_H + 2 + (7x_H^2 + 20x_H + 4)\frac{m_t^2}{m_{Z^\prime}^2}}{3} 
	 		\sqrt{1-\frac{4m_t^2}{m_{Z^\prime}^2}} \right]. 
\nonumber \\
\label{DecayWidthZp}
\end{eqnarray}
Here, we have neglected all SM fermion masses except for $m_t$,
  and we have assumed $m_N^i  > m_{Z^\prime}/2$ for simplicity.
By integrating the differential cross section over a range of $M_{\ell \ell}$ set by the ATLAS
  and CMS analyses, respectively, we obtain the cross section
  as a function of  $x_H$, $\alpha_{g_X}$ and $m_{Z^\prime}$,
  which are compared with the lower bounds obtained by the ATLAS and CMS Collaborations.

%%%%%%%%%%%%%%%%%%%   Sequential SM Z' boson mass by ATLAS  %%%%%%%%%%%%%%%%%%%%
\begin{figure}[t]
%%%%%%%%%%%%%   8 TeV result  %%%%%%%%%%%%%%%%%%%%%%%%%%
\begin{minipage}{0.5\linewidth}
\begin{center}
\includegraphics[width=0.95\linewidth]{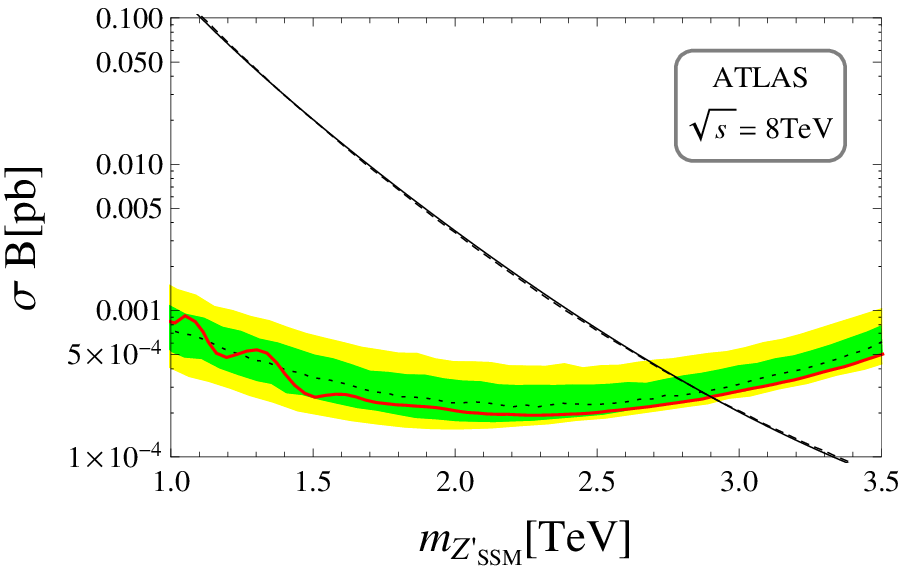}
(a)
\end{center}
\end{minipage}
%%%%%%%%%%%%%   13 TeV result  %%%%%%%%%%%%%%%%%%%%%%%%%%
\begin{minipage}{0.5\linewidth}
\begin{center}
\includegraphics[width=0.95\linewidth]{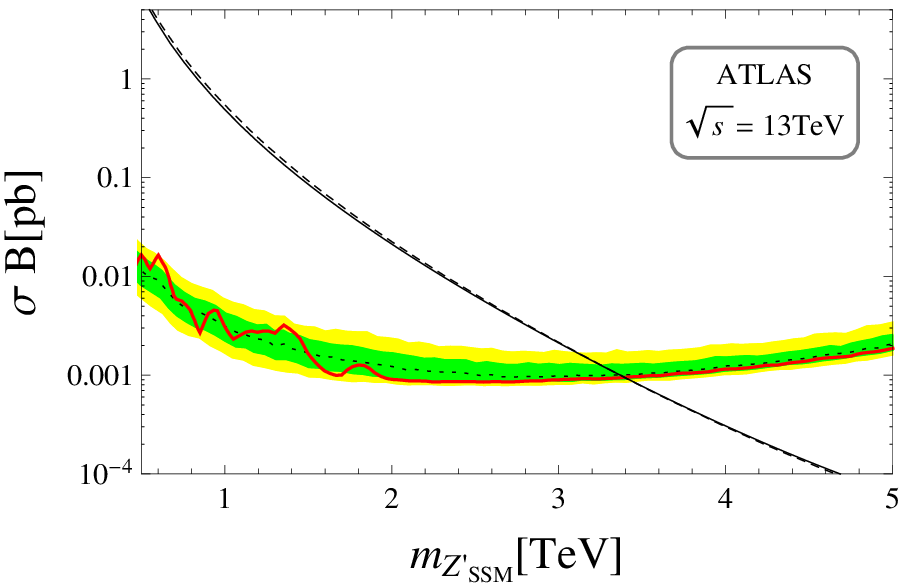}
(b)
\end{center}
\end{minipage}
\caption
{
(a) The cross section as a function of the $Z^\prime_{SSM}$ mass (solid line) 
     with $k=1.18$, along with the LHC Run-1 ATLAS result 
     from the combined dielectron and dimuon channels in Ref.~\cite{ATLAS8_Zp}. 
(b) Same as (a), but with $k=1.19$,  along with the LHC Run-2 ATLAS result
     in Ref.~\cite{ATLAS13_Zp}.  
}
\label{Fig:ATLAS}
\end{figure}
%%%%%%%%%%%%%%%%%%%%%%%%%%%%%%%%%%%%%%%%%%%%%%%%%%%%
%%%%%%%%%%%%%%%%%%%   Sequential SM Z' boson mass by CMS  %%%%%%%%%%%%%%%%%%%%
\begin{figure}[htbp]
%%%%%%%%%%%%%   8 TeV result  %%%%%%%%%%%%%%%%%%%%%%%%%%
\begin{minipage}{0.5\linewidth}
\begin{center}
\includegraphics[width=0.95\linewidth]{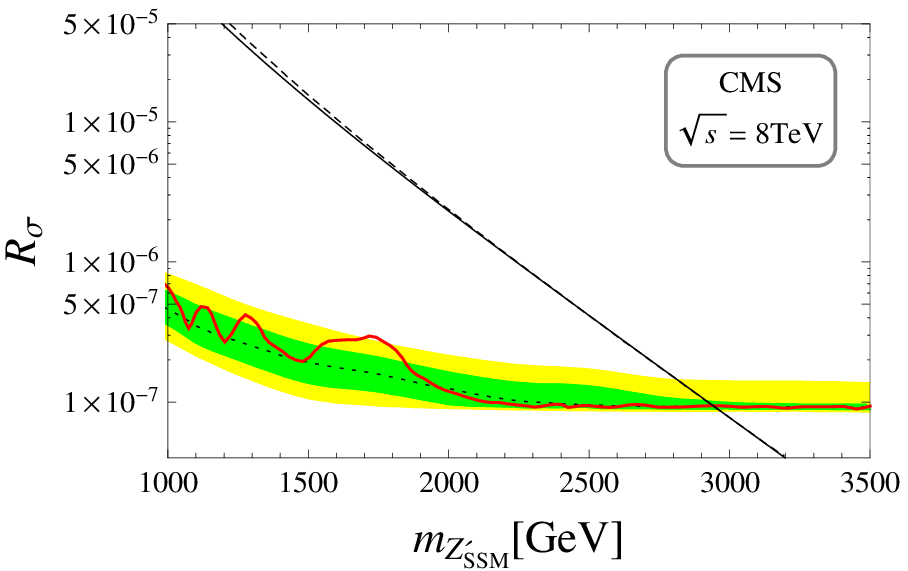}
(a)
\end{center}
\end{minipage}
%%%%%%%%%%%%%   13 TeV result  %%%%%%%%%%%%%%%%%%%%%%%%%%
\begin{minipage}{0.5\linewidth}
\begin{center}
\includegraphics[width=0.95\linewidth]{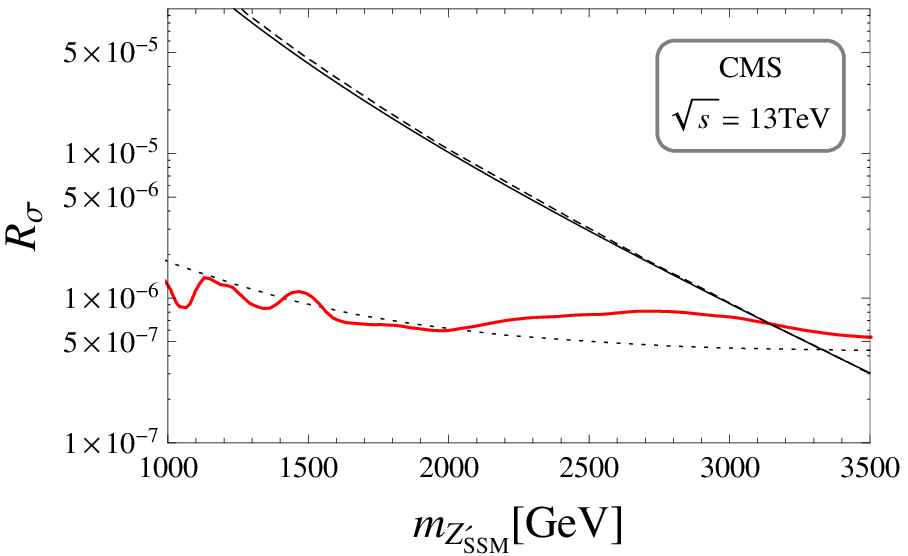}
(b)
\end{center}
\end{minipage}
\caption
{
(a) The cross section ratio as a function of the $Z^\prime_{SSM}$ mass (solid line) 
     with $k=1.01$, along with the LHC Run-1 CMS result from the combined dielectron and dimuon channels 
     in Ref.~\cite{CMS8_Zp}. 
(b) Same as (a), but with $k=1.65$,  along with the LHC Run-2 CMS result
     in Ref.~\cite{CMS13_Zp}. 
}
\label{Fig:CMS}
\end{figure}
%%%%%%%%%%%%%%%%%%%%%%%%%%%%%%%%%%%%%%%%%%%%%%%%%%%%

In interpreting the ATLAS and the CMS results for the U(1)$^\prime$ $Z^\prime$ boson,  
   we follow the strategy in \cite{NOSO}, where the minimal U(1)$_{B-L}$ model has been 
   investigated and an upper bound on the U(1)$_{B-L}$ gauge coupling 
   as a function of the $Z^\prime$ boson mass has been obtained 
   from the ATLAS and the CMS results at the LHC Run-2.
We first analyze the sequential SM $Z^\prime$ model to check the consistency of our analysis 
   with the one by the ATLAS and the CMS Collaborations.  
With the  same couplings as the SM, we calculate the differential cross section of the process
   $pp \to Z^\prime_{SSM}+X \to \ell^+ \ell^- +X$ like Eq.~(\ref{CrossLHC}). 
According to the analysis by the ATLAS Collaboration at the LHC Run-1 (Run-2),  
   we integrate the differential cross section for the range of 128 GeV$\leq M_{\ell \ell} \leq 4500$ GeV \cite{ATLAS8_Zp}
   (128 GeV$\leq M_{\ell \ell} \leq 6000$ GeV \cite{ATLAS13_Zp})  
   and obtain the cross section of the dilepton production process 
   as a function of the $Z^\prime_{SSM}$ boson mass.  
Our results are shown as solid lines in Fig.~\ref{Fig:ATLAS}(a) for the LHC Run-1 
  and \ref{Fig:ATLAS}(b) for the LHC Run-2, respectively, 
  along with the plots presented by the ATLAS Collaborations at the LHC Run-1 \cite{ATLAS8_Zp} 
  and the LHC Run-2 \cite{ATLAS13_Zp}. 
In Figs.~\ref{Fig:ATLAS}(a) and \ref{Fig:ATLAS}(b),
  the experimental upper bounds on the $Z^\prime$ boson production 
   cross section are depicted as the horizontal solid (red) curves. 
The theoretical $Z^\prime$ boson production cross section is shown as the diagonal dashed lines, 
   and  the lower limits of the $Z^\prime_{SSM}$ boson mass obtained by the ATLAS Collaborations      
   are found to be $2.90$ TeV for the LHC Run-1 and $3.4$ TeV  for the LHC Run-2, respectively, 
   which can be read off from the intersection points of the theoretical predictions (diagonal dashed lines) 
   and the experimental cross section bounds (horizontal solid (red) curves).  
In order to take into account the difference of the parton distribution functions
  used in the ATLAS analysis and our analysis and QCD corrections of the process,
  we have scaled our resultant cross sections by a factor $k=1.18$ in Fig.~\ref{Fig:ATLAS}(a)
  and by $k=1.19$ in Fig.~\ref{Fig:ATLAS}(b),
  with which we can obtain the same lower limits of the $Z^\prime_{SSM}$ boson mass
  as $2.90$ TeV and $3.4$ TeV. 
We can see that our results (solid lines) in Fig.~\ref{Fig:ATLAS} 
  with the factors of $k=1.18$ and $k=1.19$, respectively, 
  are very consistent with the theoretical predictions (diagonal dashed lines) presented
  by the ATLAS Collaboration. 
We use these factors in the following analysis for the U(1)$^\prime$ $Z^\prime$ production process.

Now we calculate the cross section of the process $pp \to Z^\prime+X \to \ell^+ \ell^- +X$ 
    for various values of $g_X$, $x_H$ and $v_\phi$, 
    and read off the constraints on these parameters from the cross section bounds 
    given by the ATLAS Collaboration. 
In Figs.~\ref{Fig:scan1a_enlarge}-\ref{Fig:scan2_xH=-2.5},
   our results from the ATLAS bounds at the LHC Run-1 
   and Run-2 are depicted as thin solid lines and thick solid lines, respectively. 
We can see that the LHC Run-2 results have dramatically improved 
   the bounds from those obtained by the LHC Run-1 results.

We apply the same strategy and compare our results for the $Z^\prime_{SSM}$ model 
   with those by the CMS Run-1 analysis~\cite{CMS8_Zp} and the CMS Run-2 one~\cite{CMS13_Zp}.
According to the analysis by the CMS Collaboration at the LHC Run-1 (Run-2),  
   we integrate the differential cross section for the range of 
   $0.6 \; m_{Z^\prime_{SSM}} \leq  M_{\ell \ell} \leq  1.4 \; m_{Z^\prime_{SSM}}$ \cite{CMS8_Zp} 
  ($0.97 \; m_{Z^\prime_{SSM}} \leq  M_{\ell \ell} \leq  1.03 \; m_{Z^\prime_{SSM}}$ \cite{CMS13_Zp})  
  and obtain the cross section. 
In the CMS analysis, the limits are set on the ratio of the $Z^\prime_{SSM}$ boson cross section 
   to the $Z/\gamma^*$ cross section:
\begin{eqnarray}
R_{\sigma} &=&
	\frac{\sigma (pp \to Z^\prime+X \to \ell \ell +X)}
		{\sigma (pp \to Z+X \to \ell \ell +X)},
\end{eqnarray}
   where the $Z/\gamma^*$ production cross sections 
   in the mass window of $60$ GeV$\leq  M_{\ell \ell} \leq 120$ GeV 
   are predicted to be $1117$ pb at the LHC Run-1 \cite{CMS8_Zp}
   and $1928$ pb at the LHC Run-2 \cite{CMS13_Zp}, respectively.   
Our results for the $Z^\prime_{SSM}$ model are shown as the solid lines
   in Figs.~\ref{Fig:CMS}(a) and \ref{Fig:CMS}(b),  
   along with the plots presented in \cite{CMS8_Zp} and \cite{CMS13_Zp}, respectively.   
The analyses in these CMS papers lead to the lower limits of the $Z^\prime_{SSM}$ boson mass
  as $2.96$ TeV for the LHC Run-1 and $3.15$ TeV for the LHC Run-2, 
  which are read off from the intersection points of the theoretical predictions (diagonal dashed lines) 
  and  the experimental cross section bounds (horizontal solid (red) curves).  
In order to obtain the same lower mass limits, 
   we have scaled our resultant cross sections by a factor $k=1.01$ in Fig.~\ref{Fig:CMS}(a)   
  and by $k=1.65$ in Fig.~\ref{Fig:CMS}(b), respectively. 
With these $k$ factors, our results (solid lines) are very consistent
   with the theoretical predictions (diagonal dashed lines) presented 
   in Refs.~\cite{CMS8_Zp} and~\cite{CMS13_Zp}.    
We use these $k$ factors in our analysis to interpret the CMS results 
    for the U(1)$^\prime$ $Z^\prime$ boson case.  
In Figs.~\ref{Fig:scan1a_enlarge}-\ref{Fig:scan2_xH=-2.5},
   our results from the CMS bounds at the LHC Run-1 
   and Run-2 are depicted as thin dashed lines and thick dashed lines, respectively. 
We can see that the CMS results at the LHC Run-2 have dramatically improved 
   the bounds obtained by the LHC Run-1 results.  
We find that  the ATLAS and the CMS bounds we have obtained are consistent with each other. 
For the LHC Run-2 results, the ATLAS bounds are slightly more severe 
  than the CMS bounds for $m_{Z^\prime} \leq 3.5$ TeV and applicable up to $m_{Z^\prime} =5$ TeV, 
  leading to the most severe LHC bound on the model parameters.

%%%%%%%%%%%%%%%%%%%   LEP 2013 updated bound  %%%%%%%%%%%%%%%%%%%%
\begin{figure}[t]
\begin{center}
\includegraphics[width=0.5\linewidth]{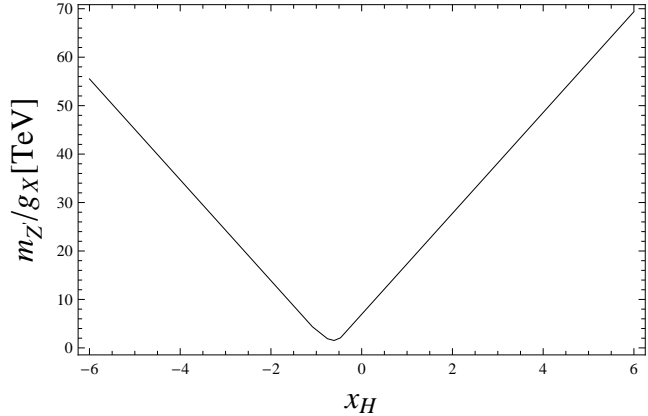}
\caption
{
The lower bound on $m_{Z^\prime}/g_X$ as a function of $x_H$ (with a fixed $x_\Phi$ =2), 
   obtained by the limits from the final LEP 2 data~\cite{LEP2Aupdate} 
   at 95\% confidence level. 
}
\label{Fig:LEP}
\end{center}
\end{figure}
%%%%%%%%%%%%%%%%%%%%%%%%%%%%%%%%%%%%%%%%%%%%%%%%%%%%

The search for effective 4-Fermi interactions mediated by the $Z^\prime$ boson at the LEP
  leads to a lower bound on $m_{Z^\prime}/g_X$~\cite{LEP2A,LEP2B,LEP2Aupdate}.
Employing the limits from the final LEP 2 data~\cite{LEP2Aupdate} at 95\% confidence level, 
   we follow \cite{LEP2B} and derive a lower bound on $m_{Z^\prime}/g_X$ 
  as a function $x_H$. 
Our result is shown in Fig.~\ref{Fig:LEP}. 
In Figs.~\ref{Fig:scan1a_enlarge}-\ref{Fig:scan2_xH=-2.5}, the LEP bounds are depicted as 
  the dashed-dotted lines.

%%%%%%%%%%%%%%%%%%%%%%%%%%%%%%%%%%%%%%%%%%%%%%%
\section{Naturalness bounds from SM Higgs mass corrections}
%%%%%%%%%%%%%%%%%%%%%%%%%%%%%%%%%%%%%%%%%%%%%%%
\label{Sec_naturalness}
Once the classical conformal symmetry is radiatively broken by the Coleman-Weinberg mechanism,
    the masses for the $Z^\prime$ boson and the Majorana neutrinos are generated 
    and they contribute to self-energy corrections of the SM Higgs doublet. 
If the U(1)$^\prime$ gauge symmetry breaking scale is very large, 
  the self-energy corrections may exceed the electroweak scale
  and require us to fine-tune the model parameters in reproducing the correct electroweak scale.
See \cite{Casas} for related discussions. 
We consider two heavy states, the right-handed neutrino and $Z^\prime$ boson, 
  whose masses are generated by the U(1)$^\prime$ gauge symmetry breaking.

Since the original theory is classically conformal and defined as a massless theory,  
   the self-energy corrections to the SM Higgs doublet originate 
   from corrections to the mixing quartic coupling $\lambda_{\rm mix}$.     
Thus, what we calculate to derive the naturalness bounds are quantum corrections to 
   the term $\lambda_{\rm mix} h^2 \phi^2$ in the effective Higgs potential 
\begin{eqnarray} 
  V_{\rm eff} \supset
    \frac{\lambda_{\rm mix}}{4} h^2 \phi^2 
	+ \frac{\beta_{\lambda_{\rm mix}}}{8} h^2 \phi^2 \left( \ln\left[\phi^2\right] + C \right),
\end{eqnarray}
   where the logarithmic divergence and the terms independent of $\phi$ are all encoded in $C$.
Here, the major contributions to quantum corrections are found to be 
\begin{eqnarray}
\beta_{\lambda_{\rm mix}} &\supset& - \frac{48 |y_M|^2 |Y_\nu|^2}{16 \pi^2}
	+ \frac{12 x_H^2 x_\Phi^2 g_X^4}{16 \pi^2} 
	- \frac{4 \left( 19 x_H^2 + 10 x_H x_\Phi  + x_\Phi^2 \right) x_\Phi^2 y_t^2 g_X^4}{\left(16 \pi^2\right)^2},
\end{eqnarray}
  where the first term comes from the one-loop diagram involving the Majorana neutrinos,
  the second one is from the one-loop diagram involving the $Z^\prime$ boson,
  and the third one is from the two-loop diagram \cite{IOO1}
  involving the $Z^\prime$ boson and the top quark.
By adding a counterterm, we renormalize the coupling $\lambda_{\rm mix}$ with the renormalization condition, 
\begin{eqnarray}
   \frac{\partial^4  V_{\rm eff}}{\partial h^2 \partial \phi^2} \Big|_{h=0, \phi=v_\phi} = \lambda_{\rm mix}, 
\end{eqnarray}
   where $\lambda_{\rm mix}$ is the renormalized coupling.
As a result, we obtain 
\begin{eqnarray} 
  V_{\rm eff} \supset
    \frac{\lambda_{\rm mix}}{4} h^2 \phi^2 
	+ \frac{\beta_{\lambda_{\rm mix}}}{8} h^2 \phi^2 \left( \ln\left[\frac{\phi^2}{v_\phi}\right] - 3 \right).
\end{eqnarray}
Substituting $\phi=v_\phi$, we obtain the SM Higgs self-energy correction as 
\begin{eqnarray}    
   \Delta m_h^2 &=& - \frac{3}{4} \beta_{\lambda_{\rm mix}} v_\phi^2 \nonumber \\
	&\sim& \frac{9 m_\nu m_N^3}{4 \pi^2 v_h^2} - \frac{9}{4 \pi} x_H^2 \alpha_{g_X} m_{Z^\prime}^2
		+ \frac{3 m_t^2} {32 \pi^3 v_h^2} \left(19x_H^2 + 20x_H + 4\right) \alpha_{g_X} m_{Z^\prime}^2
 \label{Eq:Delta_mh2}
\end{eqnarray}
  where we have used the seesaw formula, $m_\nu \sim Y_\nu^2 v_h^2/2m_N$ \cite{seesaw},
  and set $x_\Phi=2$. 
For the stability of the electroweak vacuum, we impose $\Delta m_h^2 \lesssim m_h^2$ as the naturalness. 
For example, when the light neutrino mass scale is around $m_\nu \sim 0.1$ eV, 
  we have an upper bound from the first term of Eq.~(\ref{Eq:Delta_mh2})
  for the  Majorana mass as $m_N \lesssim 3 \times 10^6$ GeV. 
This bound is much larger than the scale that we are interested in, $m_N \lesssim 1$ TeV. 
The most important contribution to $\Delta m_h^2$ is the second term of Eq.~(\ref{Eq:Delta_mh2})
  generated through the one-loop diagram with the $Z^\prime$ gauge boson,
  and the third term becomes important in the case of the U(1)$_{B-L}$ model
  because the $x_H=0$ condition makes the second term vanish.

If $\Delta m_h^2$ is much larger than the electroweak scale,  
  we need a fine-tuning of the tree-level Higgs mass ($|\lambda_{\rm mix}| v_\phi^2/2$) 
  to reproduce the correct SM Higgs VEV, $v_h=246$ GeV. 
We simply evaluate a fine-tuning level as 
\begin{eqnarray}
  \delta  = \frac{m_h^2}{2 |\Delta m_h^2|}. 
\end{eqnarray}
Here, $\delta =0.1$, for example, indicates that we need to fine-tune the tree-level Higgs mass squared 
   at the 10\% accuracy level.  
In Figs.~\ref{Fig:scan1a_enlarge}-\ref{Fig:scan2_xH=-2.5}, 
   the naturalness bounds for 10\% and 30\% fine-tuning levels are plotted as the dotted lines.  
Interestingly, the naturalness bounds from the 30\% fine-tuning level are found to be 
  compatible with the ALTAS bounds from the LHC Run-2.

%%%%%%%%%%%%%%%%%%%%%%%%%%%%%%%%%%%%%
\section{Conclusions}
%%%%%%%%%%%%%%%%%%%%%%%%%%%%%%%%%%%%%
\label{Sec_conclusion}

We have considered the classically conformal U(1)$^\prime$ extended SM 
    with three right-handed neutrinos and a U(1)$^\prime$ Higgs singlet field.  
The U(1)$^\prime$ symmetry is radiatively broken by the Coleman-Weinberg mechanism, 
  by which the $Z^\prime$ boson as well as the right-handed (Majorana) neutrinos acquire their masses. 
With the Majorana heavy neutrinos, the seesaw mechanism is automatically implemented.   
Through a mixing quartic term between the U(1)$^\prime$ Higgs and the SM Higgs doublet fields, 
  a negative mass squared for the SM Higgs doublet is generated and, as a result,   
  the electroweak symmetry breaking is triggered associated with the radiative U(1)$^\prime$ 
  gauge symmetry breaking. 
Therefore, all mass generations occur through the dimensional transmutation in our model.

In the context of the classically conformal U(1)$^\prime$ model, 
  we have investigated the possibility to resolve the electroweak vacuum instability.  
Since the gauge symmetry is broken by the Coleman-Weinberg mechanism, 
  all quartic couplings in the Higgs potential, except the SM Higgs one, are very small, 
  and hence their positive contributions to the U(1)$^\prime$ model 
  are not effective in resolving the SM Higgs vacuum instability.  
On the other hand, in the U(1)$^\prime$ model, the SM Higgs doublet has a nonzero U(1)$^\prime$ charge, 
  and this gauge interaction positively contributes to the beta function. 
In addition, the U(1)$^\prime$ gauge interaction negatively contributes to the beta function of  
  the top Yukawa coupling, so the running top Yukawa coupling is decreasing faster 
  than in the SM case, and its negative contribution to the beta function 
  of the SM Higgs quartic coupling becomes milder.  
For three free parameters of the model, $m_{Z^\prime}$, $\alpha_{g_X}$ and $x_H$,
  we have performed a parameter scan by analyzing the renormalization group evolutions
  of the model parameters at the two-loop level,
  and we have identified parameter regions which can solve the electroweak vacuum instability problem 
  and keep all coupling values in the perturbative regime up to the Planck mass.  
We have found that the resultant parameter regions are very severely constrained,   
  and also that the U(1)$_{B-L}$ model and the orthogonal model are excluded
  from having the electroweak vacuum stability with the current world average of the experimental data,
  $m_t =173.34$ GeV~\cite{MtCombine} and $m_h=125.09$ GeV~\cite{MhCombine}.

We have also considered the current collider bounds on the U(1)$^\prime$ $Z^\prime$ boson mass
   from the recent ATLAS and CMS results at the LHC Run-2 with $\sqrt{s}=13$ TeV.  
We have interpreted the $Z^\prime$ boson resonance search results at the LHC Run-1 and Run-2
  to the U(1)$^\prime$ $Z^\prime$ boson case
  and obtained the collider bound on the U(1)$^\prime$ charge of the SM Higgs doublet $x_H$ 
  for a fixed U(1)$^\prime$ gauge coupling,
  the collider bound on $x_H$ for a fixed VEV of the U(1)$^\prime$ Higgs,
  or the upper bound on the U(1)$^\prime$ gauge coupling 
  for a fixed $x_H$ as a function of  the U(1)$^\prime$ $Z^\prime$ boson mass $m_{Z^\prime}$.  
The LEP results in the search for effective 4-Fermi interactions mediated by the U(1)$^\prime$ $Z^\prime$ boson
   can also constrain the model parameter space, but the constraint is found to be weaker than 
   those obtained from the LHC Run-2 results.

Once the U(1)$^\prime$ gauge symmetry is broken, 
   the $Z^\prime$ boson and the right-handed neutrinos become heavy
   and contribute to the SM Higgs self-energy through quantum corrections. 
Therefore, the SM Higgs self-energy can exceed the electroweak scale, 
   if the states are so heavy. 
Since the SM Higgs doublet has nonzero U(1)$^\prime$ charge, 
  the self-energy corrections from the $Z^\prime$ boson occur at the one-loop level. 
This is in sharp contrast to the classically conformal U(1)$_{B-L}$ model~\cite{IOO1},  
  where the Higgs doublet has no U(1)$_{B-L}$ charge, and the self-energy corrections 
   from the $Z^\prime$ boson occur at the two-loop level.  
We have evaluated the Higgs self-energy corrections and found the naturalness bounds 
  to reproduce the right electroweak scale
  for the fine-tuning level better than 10\%.
We have found that the naturalness bounds for the 30\% fine-tuning level are compatible 
  with the ALTAS constraint from the LHC Run-2 results, and requiring a fine-tuning level $> 10$\% 
  leads to the upper bound on the U(1)$^\prime$ $Z^\prime$ boson mass as  $m_{Z^\prime} \lesssim 7$ TeV.

Putting all our results together in Figs.~\ref{Fig:scan1a_enlarge}-\ref{Fig:scan2_xH=-2.5},  
  we have found that the U(1)$^\prime$ $Z^\prime$ boson mass lies in 
  the range of $3.5$ TeV$\lesssim m_{Z^\prime} \lesssim 7$ TeV.  
This region can be explored by the LHC Run-2 in the near future.

%%%%%%%%%%%%%%%%%%%%%%%%%%%%%%%%%%%%%%%%%
\section*{Acknowledgements}
%%%%%%%%%%%%%%%%%%%%%%%%%%%%%%%%%%%%%%%%%
The work of D.T. and S.O. was supported by the Sugawara unit and Hikami unit, respectively, 
   of the Okinawa Institute of Science and Technology Graduate University.
The work of N.O. was supported in part by the United States Department of Energy (DE-SC0013680).

%%%%%%%%%%%%%%%%%%%%%%%%%%%%%%%%%%%%%%%%%%%%%%%%%%%%%
\appendix
%%%%%%%%%%%%%%%%%%%%%%%%%%%%%%%%%%%%%%%%%%%%%%%%%%%%%
\section{U(1)$^\prime$ RGES AT THE TWO-LOOP LEVEL} 
\label{Sec:U(1)'_RGEs}
%%%%%%%%%%%%%%%%%%%%%%%%%%%%%%%%%%%%%%%%%%%%%%%%%%%%%
In this appendix we present the two-loop RGEs for the U(1)$^\prime$ extension 
   of the SM, which are used in our analysis.  
The definitions of the covariant derivative, the Yukawa interactions and the scalar potential are given 
   by Eqs.~(\ref{Eq:covariant_derivative}), (\ref{Eq:L_Yukawa}) and (\ref{Eq:classical_potential}), respectively.
We only include the top quark Yukawa coupling $y_t$
   and the right-handed neutrino Majorana Yukawa coupling $y_M=Y_M^i$ ($i=1,2,3$) 
   since the other Yukawa couplings are negligibly small. 
The U(1)$^\prime$ charges $x_i$ are defined in Table~\ref{Tab:particle_contents}. 
The U(1)$^\prime$ RGEs at the two-loop level have been generated by using SARAH \cite{SARAH}.

%%%%%%%%%%%%%%%%%%%%%%%%%%%%%%%%%%%%%%%%%%%%%%%%%%%%%
\subsection{U(1)$^\prime$ RGEs for the gauge couplings} 
%%%%%%%%%%%%%%%%%%%%%%%%%%%%%%%%%%%%%%%%%%%%%%%%%%%%%
The RGEs for the gauge couplings at the two-loop level are given by  
\begin{eqnarray}
\mu \frac{d g_i}{d\mu} &=& \beta_{g_i}^{(1)}+\beta_{g_i}^{(2)}, 
\end{eqnarray}
  where $\beta_{g_i}^{(1)}$ and $\beta_{g_i}^{(2)}$ are the one-loop and two-loop beta functions
  for the gauge couplings, respectively, and $g_i$ represents $g_3$, $g_2$, $g_1$, $g_{X1}$, $g_{1X}$ and $g_X$.
Here, the one-loop beta functions for the gauge couplings are given by
\begin{eqnarray}
%%%%% U(1)' RGE for g_3 at 1-loop %%%%%
\beta_{g_3}^{(1)} &=& \frac{g_3^3}{16\pi^2} \Big[ -7 \Big], 
\nonumber \\
%%%%% U(1)' RGE for g_2 at 1-loop %%%%%
\beta_{g_2}^{(1)} &=& \frac{g_2^3}{16\pi^2} \left[ -\frac{19}{6} \right],  
\nonumber \\  
%%%%% U(1)' RGE for g_1 at 1-loop %%%%%
\beta_{g_1}^{(1)} &=& \frac{1}{16\pi^2} \left[ g_1 
		\left\{ \frac{41}{6}g_1^2 + \frac{1}{3}\big(82x_H + 16x_\Phi \big) g_1 g_{X1}
				+ \frac{1}{3}\big(82x_H^2 + 32x_H x_\Phi + 9x_\Phi^2 \big) g_{X1}^2 \right\} \right.
	\nonumber \\  
	&+& g_{1X} 
		\left\{ \frac{41}{6}g_1 g_{1X} 
			+ \frac{1}{3}\big(41x_H + 8x_\Phi \big) g_1 g_X 
			+ \frac{1}{3}\big(41x_H + 8x_\Phi \big) g_{1X} g_{X1} \right.
	\nonumber \\  
	&+&  \left. \left. \frac{1}{3}\big(82x_H^2 + 32x_H x_\Phi + 9x_\Phi^2 \big) g_{X1} g_X \right\} \right],
\nonumber \\  
%%%%% U(1)' RGE for g_{X1} at 1-loop %%%%%
\beta_{g_{X1}}^{(1)} &=& \frac{1}{16\pi^2} \left[ g_{X1} 
		\left\{ \frac{41}{6}g_1^2 + \frac{1}{3}\big(82x_H + 16x_\Phi \big) g_1 g_{X1}
				+ \frac{1}{3}\big(82x_H^2 + 32x_H x_\Phi + 9x_\Phi^2 \big) g_{X1}^2 \right\} \right.
	\nonumber \\  
	&+& g_X 
		\left\{ \frac{41}{6}g_1 g_{1X} 
			+ \frac{1}{3}\big(41x_H + 8x_\Phi \big) g_1 g_X 
			+ \frac{1}{3}\big(41x_H + 8x_\Phi \big) g_{1X} g_{X1} \right.
	\nonumber \\  
	&+&  \left. \left. \frac{1}{3}\big(82x_H^2 + 32x_H x_\Phi + 9x_\Phi^2 \big) g_{X1} g_X \right\} \right],
\nonumber \\  
%%%%% U(1)' RGE for g_{1X} at 1-loop %%%%%
\beta_{g_{1X}}^{(1)} &=& \frac{1}{16\pi^2} \left[ g_{1X} 
		\left\{ \frac{41}{6}g_{1X}^2 + \frac{1}{3}\big(82x_H + 16x_\Phi \big) g_{1X} g_X
				+ \frac{1}{3}\big(82x_H^2 + 32x_H x_\Phi + 9x_\Phi^2 \big) g_X^2 \right\} \right.
	\nonumber \\  
	&+& g_1 
		\left\{ \frac{41}{6}g_1 g_{1X} 
			+ \frac{1}{3}\big(41x_H + 8x_\Phi \big) g_1 g_X 
			+ \frac{1}{3}\big(41x_H + 8x_\Phi \big) g_{1X} g_{X1} \right.
	\nonumber \\  
	&+&  \left. \left. \frac{1}{3}\big(82x_H^2 + 32x_H x_\Phi + 9x_\Phi^2 \big) g_{X1} g_X \right\} \right],
\nonumber \\  
%%%%% U(1)' RGE for g_X at 1-loop %%%%%
\beta_{g_X}^{(1)} &=& \frac{1}{16\pi^2} \left[ g_X 
		\left\{ \frac{41}{6}g_{1X}^2 + \frac{1}{3}\big(82x_H + 16x_\Phi \big) g_{1X} g_X
				+ \frac{1}{3}\big(82x_H^2 + 32x_H x_\Phi + 9x_\Phi^2 \big) g_X^2 \right\} \right.
	\nonumber \\  
	&+& g_{X1} 
		\left\{ \frac{41}{6}g_1 g_{1X} 
			+ \frac{1}{3}\big(41x_H + 8x_\Phi \big) g_1 g_X 
			+ \frac{1}{3}\big(41x_H + 8x_\Phi \big) g_{1X} g_{X1} \right.
	\nonumber \\  
	&+&  \left. \left. \frac{1}{3}\big(82x_H^2 + 32x_H x_\Phi + 9x_\Phi^2 \big) g_{X1} g_X \right\} \right],
\end{eqnarray}
and the two-loop beta functions for the gauge couplings are given by
\begin{eqnarray}
%%%%% U(1)' RGE for g_3 at 2-loop %%%%%
\beta_{g_3}^{(2)} &=& \frac{1}{{(16 \pi^2)}^2} \cdot \frac{g_3^3}{6}
	\Big[11 g_1^2+27 g_2^2-156 g_3^2+11 g_{1X}^2+44 g_{1X} g_X x_H+44 g_1 g_{X1}
x_H+44 g_X^2 x_H^2\Big. 
	\nonumber \\ 
	&+&
	\Big.44 g_{X1}^2 x_H^2+4 g_{1X} g_X x_\Phi +4 g_1 g_{X1} x_\Phi +8 g_X^2 x_H x_\Phi +8 g_{X1}^2 x_H
x_\Phi +2 g_X^2 x_\Phi ^2+2 g_{X1}^2 x_\Phi ^2-12 y_t^2\Big],
	\nonumber \\ 
%%%%% U(1)' RGE for g_2 at 2-loop %%%%%
\beta_{g_2}^{(2)} &=& \frac{1}{{(16 \pi^2)}^2} \cdot \frac{g_2^3}{6}
	\Big[ 9 g_1^2+35 g_2^2+72 g_3^2+9 g_{1X}^2+36 g_{1X} g_X x_H+36 g_1 g_{X1} x_H+36
g_X^2 x_H^2+36 g_{X1}^2 x_H^2\Big. 
	\nonumber \\ 
	&+&
	\Big.12 g_{1X} g_X x_\Phi +12 g_1 g_{X1} x_\Phi +24 g_X^2 x_H x_\Phi +24 g_{X1}^2 x_H x_\Phi +6 g_X^2
x_\Phi ^2+6 g_{X1}^2 x_\Phi ^2-9 y_t^2\Big],
	\nonumber \\
%%%%% U(1)' RGE for g_1 at 2-loop %%%%%
\beta_{g_1}^{(2)} &=& \frac{1}{{(16 \pi^2)}^2} \cdot \frac{1}{18}
	\Big[199 g_1^5+81 g_1^3 g_2^2+264 g_1^3 g_3^2+398 g_1^3 g_{1X}^2+81 g_1 g_2^2 g_{1X}^2
		+264 g_1 g_3^2 g_{1X}^2\Big. 
	\nonumber \\ 
	&+&
	199 g_1 g_{1X}^4+1194 g_1^3 g_{1X} g_X x_H+162 g_1 g_2^2 g_{1X} g_X x_H
		+528 g_1 g_3^2 g_{1X} g_X x_H 
	\nonumber \\ 
	&+&
	1194 g_1 g_{1X}^3 g_X x_H+1592 g_1^4 g_{X1} x_H+324 g_1^2 g_2^2 g_{X1} x_H
		+1056 g_1^2 g_3^2 g_{X1} x_H+1990 g_1^2 g_{1X}^2 g_{X1} x_H 
	\nonumber \\ 
	&+&
	162 g_2^2 g_{1X}^2 g_{X1} x_H+528 g_3^2 g_{1X}^2 g_{X1} x_H+398 g_{1X}^4 g_{X1} x_H
		+796 g_1^3 g_X^2 x_H^2+2388 g_1 g_{1X}^2 g_X^2 x_H^2 
	\nonumber \\ 
	&+&
	5572 g_1^2 g_{1X} g_X g_{X1} x_H^2+324 g_2^2 g_{1X} g_X g_{X1} x_H^2
		+1056 g_3^2 g_{1X} g_X g_{X1} x_H^2+2388 g_{1X}^3 g_X g_{X1} x_H^2 
	\nonumber \\ 
	&+&
	4776 g_1^3 g_{X1}^2 x_H^2+324 g_1 g_2^2 g_{X1}^2 x_H^2+1056 g_1 g_3^2 g_{X1}^2 x_H^2
		+3184 g_1 g_{1X}^2 g_{X1}^2 x_H^2+1592 g_1 g_{1X} g_X^3 x_H^3 
	\nonumber \\ 
	&+&
	3184 g_1^2 g_X^2 g_{X1} x_H^3+4776 g_{1X}^2 g_X^2 g_{X1} x_H^3
		+7960 g_1 g_{1X} g_X g_{X1}^2 x_H^3+6368 g_1^2 g_{X1}^3 x_H^3 
	\nonumber \\ 
	&+&
	1592 g_{1X}^2 g_{X1}^3 x_H^3+3184 g_{1X} g_X^3 g_{X1} x_H^4+3184 g_1 g_X^2 g_{X1}^2 x_H^4
		+3184 g_{1X} g_X g_{X1}^3 x_H^4 
	\nonumber \\ 
	&+&
	3184 g_1 g_{X1}^4 x_H^4+246 g_1^3 g_{1X} g_X x_\Phi +54 g_1 g_2^2 g_{1X} g_X x_\Phi 
		+48 g_1 g_3^2 g_{1X} g_X x_\Phi  
	\nonumber \\ 
	&+&
	246 g_1 g_{1X}^3 g_X x_\Phi +328 g_1^4 g_{X1} x_\Phi +108 g_1^2 g_2^2 g_{X1} x_\Phi 
		+96 g_1^2 g_3^2 g_{X1} x_\Phi +410 g_1^2 g_{1X}^2 g_{X1} x_\Phi  
	\nonumber \\ 
	&+&
	54 g_2^2 g_{1X}^2 g_{X1} x_\Phi +48 g_3^2 g_{1X}^2 g_{X1} x_\Phi +82 g_{1X}^4 g_{X1} x_\Phi 
		+328 g_1^3 g_X^2 x_H x_\Phi +984 g_1 g_{1X}^2 g_X^2 x_H x_\Phi  
	\nonumber \\ 
	&+&
	2296 g_1^2 g_{1X} g_X g_{X1} x_H x_\Phi +216 g_2^2 g_{1X} g_X g_{X1} x_H x_\Phi 
		+192 g_3^2 g_{1X} g_X g_{X1} x_H x_\Phi  
	\nonumber \\ 
	&+&
	984 g_{1X}^3 g_X g_{X1} x_H x_\Phi +1968 g_1^3 g_{X1}^2 x_H x_\Phi +216 g_1 g_2^2 g_{X1}^2 x_H x_\Phi 
		+192 g_1 g_3^2 g_{X1}^2 x_H x_\Phi  
	\nonumber \\ 
	&+&
	1312 g_1 g_{1X}^2 g_{X1}^2 x_H x_\Phi +984 g_1 g_{1X} g_X^3 x_H^2 x_\Phi 
		+1968 g_1^2 g_X^2 g_{X1} x_H^2 x_\Phi +2952 g_{1X}^2 g_X^2 g_{X1} x_H^2 x_\Phi  
	\nonumber \\ 
	&+&
	4920 g_1 g_{1X} g_X g_{X1}^2 x_H^2 x_\Phi +3936 g_1^2 g_{X1}^3 x_H^2 x_\Phi 
		+984 g_{1X}^2 g_{X1}^3 x_H^2 x_\Phi +2624 g_{1X} g_X^3 g_{X1} x_H^3 x_\Phi  
	\nonumber \\ 
	&+&
	2624 g_1 g_X^2 g_{X1}^2 x_H^3 x_\Phi +2624 g_{1X} g_X g_{X1}^3 x_H^3 x_\Phi 
		+2624 g_1 g_{X1}^4 x_H^3 x_\Phi +46 g_1^3 g_X^2 x_\Phi ^2 
	\nonumber \\ 
	&+&
	138 g_1 g_{1X}^2 g_X^2 x_\Phi ^2+322 g_1^2 g_{1X} g_X g_{X1} x_\Phi ^2
		+54 g_2^2 g_{1X} g_X g_{X1} x_\Phi ^2+48 g_3^2 g_{1X} g_X g_{X1} x_\Phi ^2 
	\nonumber \\ 
	&+&
	138 g_{1X}^3 g_X g_{X1} x_\Phi ^2+276 g_1^3 g_{X1}^2 x_\Phi ^2+54 g_1 g_2^2 g_{X1}^2 x_\Phi ^2
		+48 g_1 g_3^2 g_{X1}^2 x_\Phi ^2+184 g_1 g_{1X}^2 g_{X1}^2 x_\Phi ^2 
	\nonumber \\ 
	&+&
	276 g_1 g_{1X} g_X^3 x_H x_\Phi ^2+552 g_1^2 g_X^2 g_{X1} x_H x_\Phi ^2
		+828 g_{1X}^2 g_X^2 g_{X1} x_H x_\Phi ^2+1380 g_1 g_{1X} g_X g_{X1}^2 x_H x_\Phi ^2 
	\nonumber \\ 
	&+&
	1104 g_1^2 g_{X1}^3 x_H x_\Phi ^2+276 g_{1X}^2 g_{X1}^3 x_H x_\Phi ^2
		+1104 g_{1X} g_X^3 g_{X1} x_H^2 x_\Phi ^2+1104 g_1 g_X^2 g_{X1}^2 x_H^2 x_\Phi ^2 
	\nonumber \\ 
	&+&
	1104 g_{1X} g_X g_{X1}^3 x_H^2 x_\Phi ^2+1104 g_1 g_{X1}^4 x_H^2 x_\Phi ^2
		+28 g_1 g_{1X} g_X^3 x_\Phi ^3+56 g_1^2 g_X^2 g_{X1} x_\Phi ^3 
	\nonumber \\ 
	&+&
	84 g_{1X}^2 g_X^2 g_{X1} x_\Phi ^3+140 g_1 g_{1X} g_X g_{X1}^2 x_\Phi ^3
		+112 g_1^2 g_{X1}^3 x_\Phi ^3+28 g_{1X}^2 g_{X1}^3 x_\Phi ^3+224 g_{1X} g_X^3 g_{X1} x_H x_\Phi ^3 
	\nonumber \\ 
	&+&
	224 g_1 g_X^2 g_{X1}^2 x_H x_\Phi ^3+224 g_{1X} g_X g_{X1}^3 x_H x_\Phi ^3
		+224 g_1 g_{X1}^4 x_H x_\Phi ^3+100 g_{1X} g_X^3 g_{X1} x_\Phi ^4 
	\nonumber \\ 
	&+&
	100 g_1 g_X^2 g_{X1}^2 x_\Phi ^4+100 g_{1X} g_X g_{X1}^3 x_\Phi ^4+100 g_1 g_{X1}^4 x_\Phi ^4
		-54 g_{1X} g_X g_{X1} x_\Phi ^2 y_M^2-54 g_1 g_{X1}^2 x_\Phi ^2 y_M^2 
	\nonumber \\ 
	&-&
	51 g_1^3 y_t^2-51 g_1 g_{1X}^2 y_t^2-102 g_1 g_{1X} g_X x_H y_t^2-204 g_1^2 g_{X1} x_H y_t^2
		-102 g_{1X}^2 g_{X1} x_H y_t^2 
	\nonumber \\ 
	&-&
	204 g_{1X} g_X g_{X1} x_H^2 y_t^2-204 g_1 g_{X1}^2 x_H^2 y_t^2-15 g_1 g_{1X} g_X x_\Phi  y_t^2
		-30 g_1^2 g_{X1} x_\Phi  y_t^2 
	\nonumber \\ 
	&-&
	\Big.15 g_{1X}^2 g_{X1} x_\Phi  y_t^2-60 g_{1X} g_X g_{X1} x_H x_\Phi  y_t^2
		-60 g_1 g_{X1}^2 x_H x_\Phi  y_t^2-6 g_{1X} g_X g_{X1} x_\Phi ^2 y_t^2
		-6 g_1 g_{X1}^2 x_\Phi ^2 y_t^2\Big],
	\nonumber  
\end{eqnarray}
\begin{eqnarray}
%%%%% U(1)' RGE for g_{X1} at 2-loop %%%%%
\beta_{g_{X1}}^{(2)} &=& \frac{1}{{(16 \pi^2)}^2} \cdot \frac{1}{18}
	\Big[199 g_1^3 g_{1X} g_X+81 g_1 g_2^2 g_{1X} g_X+264 g_1 g_3^2 g_{1X} g_X
		+199 g_1 g_{1X}^3 g_X+199 g_1^4 g_{X1}\Big. 
	\nonumber \\ 
	&+&
	81 g_1^2 g_2^2 g_{X1}+264 g_1^2 g_3^2 g_{X1}+199 g_1^2 g_{1X}^2 g_{X1}+398 g_1^3 g_X^2 x_H
		+162 g_1 g_2^2 g_X^2 x_H 
	\nonumber \\ 
	&+&
	528 g_1 g_3^2 g_X^2 x_H+1194 g_1 g_{1X}^2 g_X^2 x_H+1990 g_1^2 g_{1X} g_X g_{X1} x_H
		+162 g_2^2 g_{1X} g_X g_{X1} x_H 
	\nonumber \\ 
	&+&
	528 g_3^2 g_{1X} g_X g_{X1} x_H+398 g_{1X}^3 g_X g_{X1} x_H+1592 g_1^3 g_{X1}^2 x_H
		+324 g_1 g_2^2 g_{X1}^2 x_H 
	\nonumber \\ 
	&+&
	1056 g_1 g_3^2 g_{X1}^2 x_H+796 g_1 g_{1X}^2 g_{X1}^2 x_H+2388 g_1 g_{1X} g_X^3 x_H^2
		+3184 g_1^2 g_X^2 g_{X1} x_H^2 
	\nonumber \\ 
	&+&
	324 g_2^2 g_X^2 g_{X1} x_H^2+1056 g_3^2 g_X^2 g_{X1} x_H^2+2388 g_{1X}^2 g_X^2 g_{X1} x_H^2
		+5572 g_1 g_{1X} g_X g_{X1}^2 x_H^2 
	\nonumber \\ 
	&+&
	4776 g_1^2 g_{X1}^3 x_H^2+324 g_2^2 g_{X1}^3 x_H^2+1056 g_3^2 g_{X1}^3 x_H^2
		+796 g_{1X}^2 g_{X1}^3 x_H^2+1592 g_1 g_X^4 x_H^3 
	\nonumber \\ 
	&+&
	4776 g_{1X} g_X^3 g_{X1} x_H^3+7960 g_1 g_X^2 g_{X1}^2 x_H^3+4776 g_{1X} g_X g_{X1}^3 x_H^3
		+6368 g_1 g_{X1}^4 x_H^3 
	\nonumber \\ 
	&+&
	3184 g_X^4 g_{X1} x_H^4+6368 g_X^2 g_{X1}^3 x_H^4+3184 g_{X1}^5 x_H^4+82 g_1^3 g_X^2 x_\Phi 
		+54 g_1 g_2^2 g_X^2 x_\Phi +48 g_1 g_3^2 g_X^2 x_\Phi  
	\nonumber \\ 
	&+&
	246 g_1 g_{1X}^2 g_X^2 x_\Phi +410 g_1^2 g_{1X} g_X g_{X1} x_\Phi 
		+54 g_2^2 g_{1X} g_X g_{X1} x_\Phi +48 g_3^2 g_{1X} g_X g_{X1} x_\Phi  
	\nonumber \\ 
	&+&
	82 g_{1X}^3 g_X g_{X1} x_\Phi +328 g_1^3 g_{X1}^2 x_\Phi +108 g_1 g_2^2 g_{X1}^2 x_\Phi 
		+96 g_1 g_3^2 g_{X1}^2 x_\Phi +164 g_1 g_{1X}^2 g_{X1}^2 x_\Phi  
	\nonumber \\ 
	&+&
	984 g_1 g_{1X} g_X^3 x_H x_\Phi +1312 g_1^2 g_X^2 g_{X1} x_H x_\Phi +216 g_2^2 g_X^2 g_{X1} x_H x_\Phi 
		+192 g_3^2 g_X^2 g_{X1} x_H x_\Phi  
	\nonumber \\ 
	&+&
	984 g_{1X}^2 g_X^2 g_{X1} x_H x_\Phi +2296 g_1 g_{1X} g_X g_{X1}^2 x_H x_\Phi 
		+1968 g_1^2 g_{X1}^3 x_H x_\Phi +216 g_2^2 g_{X1}^3 x_H x_\Phi  
	\nonumber \\ 
	&+&
	192 g_3^2 g_{X1}^3 x_H x_\Phi +328 g_{1X}^2 g_{X1}^3 x_H x_\Phi +984 g_1 g_X^4 x_H^2 x_\Phi 
		+2952 g_{1X} g_X^3 g_{X1} x_H^2 x_\Phi  
	\nonumber \\ 
	&+&
	4920 g_1 g_X^2 g_{X1}^2 x_H^2 x_\Phi +2952 g_{1X} g_X g_{X1}^3 x_H^2 x_\Phi 
		+3936 g_1 g_{X1}^4 x_H^2 x_\Phi +2624 g_X^4 g_{X1} x_H^3 x_\Phi  
	\nonumber \\ 
	&+&
	5248 g_X^2 g_{X1}^3 x_H^3 x_\Phi +2624 g_{X1}^5 x_H^3 x_\Phi +138 g_1 g_{1X} g_X^3 x_\Phi ^2
		+184 g_1^2 g_X^2 g_{X1} x_\Phi ^2+54 g_2^2 g_X^2 g_{X1} x_\Phi ^2 
	\nonumber \\ 
	&+&
	48 g_3^2 g_X^2 g_{X1} x_\Phi ^2+138 g_{1X}^2 g_X^2 g_{X1} x_\Phi ^2
		+322 g_1 g_{1X} g_X g_{X1}^2 x_\Phi ^2+276 g_1^2 g_{X1}^3 x_\Phi ^2+54 g_2^2 g_{X1}^3 x_\Phi ^2 
	\nonumber \\ 
	&+&
	48 g_3^2 g_{X1}^3 x_\Phi ^2+46 g_{1X}^2 g_{X1}^3 x_\Phi ^2+276 g_1 g_X^4 x_H x_\Phi ^2
		+828 g_{1X} g_X^3 g_{X1} x_H x_\Phi ^2+1380 g_1 g_X^2 g_{X1}^2 x_H x_\Phi ^2 
	\nonumber \\ 
	&+&
	828 g_{1X} g_X g_{X1}^3 x_H x_\Phi ^2+1104 g_1 g_{X1}^4 x_H x_\Phi ^2
		+1104 g_X^4 g_{X1} x_H^2 x_\Phi ^2+2208 g_X^2 g_{X1}^3 x_H^2 x_\Phi ^2 
	\nonumber \\ 
	&+&
	1104 g_{X1}^5 x_H^2 x_\Phi ^2+28 g_1 g_X^4 x_\Phi ^3+84 g_{1X} g_X^3 g_{X1} x_\Phi ^3
		+140 g_1 g_X^2 g_{X1}^2 x_\Phi ^3+84 g_{1X} g_X g_{X1}^3 x_\Phi ^3 
	\nonumber \\ 
	&+&
	112 g_1 g_{X1}^4 x_\Phi ^3+224 g_X^4 g_{X1} x_H x_\Phi ^3+448 g_X^2 g_{X1}^3 x_H x_\Phi ^3
		+224 g_{X1}^5 x_H x_\Phi ^3+100 g_X^4 g_{X1} x_\Phi ^4 
	\nonumber \\ 
	&+&
	200 g_X^2 g_{X1}^3 x_\Phi ^4+100 g_{X1}^5 x_\Phi ^4-54 g_X^2 g_{X1} x_\Phi ^2 y_M^2
		-54 g_{X1}^3 x_\Phi ^2 y_M^2-51 g_1 g_{1X} g_X y_t^2-51 g_1^2 g_{X1} y_t^2 
	\nonumber \\ 
	&-&
	102 g_1 g_X^2 x_H y_t^2-102 g_{1X} g_X g_{X1} x_H y_t^2-204 g_1 g_{X1}^2 x_H y_t^2
		-204 g_X^2 g_{X1} x_H^2 y_t^2-204 g_{X1}^3 x_H^2 y_t^2 
	\nonumber \\ 
	&-&
	15 g_1 g_X^2 x_\Phi  y_t^2-15 g_{1X} g_X g_{X1} x_\Phi  y_t^2-30 g_1 g_{X1}^2 x_\Phi  y_t^2
		-60 g_X^2 g_{X1} x_H x_\Phi  y_t^2-60 g_{X1}^3 x_H x_\Phi  y_t^2 
	\nonumber \\ 
	&-&
	\Big.6 g_X^2 g_{X1} x_\Phi ^2 y_t^2-6 g_{X1}^3 x_\Phi ^2 y_t^2\Big],
	\nonumber
\end{eqnarray}
\begin{eqnarray}
%%%%% U(1)' RGE for g_{1X} at 2-loop %%%%%
\beta_{g_{1X}}^{(2)} &=& \frac{1}{{(16 \pi^2)}^2} \cdot \frac{1}{18}
	\Big[199 g_1^4 g_{1X}+81 g_1^2 g_2^2 g_{1X}+264 g_1^2 g_3^2 g_{1X}+398 g_1^2 g_{1X}^3+81 g_2^2 g_{1X}^3\Big. 
	\nonumber \\ 
	&+&
	264 g_3^2 g_{1X}^3+199 g_{1X}^5+398 g_1^4 g_X x_H+162 g_1^2 g_2^2 g_X x_H+528 g_1^2 g_3^2 g_X x_H
		+1990 g_1^2 g_{1X}^2 g_X x_H 
	\nonumber \\ 
	&+&
	324 g_2^2 g_{1X}^2 g_X x_H+1056 g_3^2 g_{1X}^2 g_X x_H+1592 g_{1X}^4 g_X x_H+1194 g_1^3 g_{1X} g_{X1} x_H 
	\nonumber \\ 
	&+&
	162 g_1 g_2^2 g_{1X} g_{X1} x_H+528 g_1 g_3^2 g_{1X} g_{X1} x_H+1194 g_1 g_{1X}^3 g_{X1} x_H
		+3184 g_1^2 g_{1X} g_X^2 x_H^2 
	\nonumber \\ 
	&+&
	324 g_2^2 g_{1X} g_X^2 x_H^2+1056 g_3^2 g_{1X} g_X^2 x_H^2+4776 g_{1X}^3 g_X^2 x_H^2
		+2388 g_1^3 g_X g_{X1} x_H^2 
	\nonumber \\ 
	&+&
	324 g_1 g_2^2 g_X g_{X1} x_H^2+1056 g_1 g_3^2 g_X g_{X1} x_H^2+5572 g_1 g_{1X}^2 g_X g_{X1} x_H^2
		+2388 g_1^2g_{1X} g_{X1}^2 x_H^2 
	\nonumber \\ 
	&+&
	796 g_{1X}^3 g_{X1}^2 x_H^2+1592 g_1^2 g_X^3 x_H^3+6368 g_{1X}^2 g_X^3 x_H^3
		+7960 g_1 g_{1X}g_X^2 g_{X1} x_H^3+4776 g_1^2 g_X g_{X1}^2 x_H^3 
	\nonumber \\ 
	&+&
	3184 g_{1X}^2 g_X g_{X1}^2 x_H^3+1592 g_1 g_{1X} g_{X1}^3 x_H^3+3184 g_{1X} g_X^4 x_H^4
		+3184 g_1 g_X^3 g_{X1} x_H^4 
	\nonumber \\ 
	&+&
	3184 g_{1X} g_X^2 g_{X1}^2 x_H^4+3184 g_1 g_X g_{X1}^3 x_H^4+82 g_1^4 g_X x_\Phi 
		+54 g_1^2 g_2^2 g_X x_\Phi +48 g_1^2 g_3^2 g_X x_\Phi  
	\nonumber \\ 
	&+&
	410 g_1^2 g_{1X}^2 g_X x_\Phi +108 g_2^2 g_{1X}^2 g_X x_\Phi +96 g_3^2 g_{1X}^2 g_X x_\Phi 
		+328 g_{1X}^4 g_X x_\Phi +246 g_1^3 g_{1X} g_{X1} x_\Phi  
	\nonumber \\ 
	&+&
	54 g_1 g_2^2 g_{1X} g_{X1} x_\Phi +48 g_1 g_3^2 g_{1X} g_{X1} x_\Phi 
		+246 g_1 g_{1X}^3 g_{X1} x_\Phi +1312 g_1^2 g_{1X} g_X^2 x_H x_\Phi  
	\nonumber \\ 
	&+&
	216 g_2^2 g_{1X} g_X^2 x_H x_\Phi +192 g_3^2 g_{1X} g_X^2 x_H x_\Phi +1968 g_{1X}^3 g_X^2 x_H x_\Phi 
		+984 g_1^3 g_X g_{X1} x_H x_\Phi  
	\nonumber \\ 
	&+&
	216 g_1 g_2^2 g_X g_{X1} x_H x_\Phi +192 g_1 g_3^2 g_X g_{X1} x_H x_\Phi 
		+2296 g_1 g_{1X}^2 g_X g_{X1} x_H x_\Phi +984 g_1^2 g_{1X} g_{X1}^2 x_H x_\Phi  
	\nonumber \\ 
	&+&
	328 g_{1X}^3 g_{X1}^2 x_H x_\Phi +984 g_1^2 g_X^3 x_H^2 x_\Phi +3936 g_{1X}^2 g_X^3 x_H^2 x_\Phi 
		+4920 g_1 g_{1X} g_X^2 g_{X1} x_H^2 x_\Phi  
	\nonumber \\ 
	&+&
	2952 g_1^2 g_X g_{X1}^2 x_H^2 x_\Phi +1968 g_{1X}^2 g_X g_{X1}^2 x_H^2 x_\Phi 
		+984 g_1 g_{1X} g_{X1}^3 x_H^2 x_\Phi +2624 g_{1X} g_X^4 x_H^3 x_\Phi  
	\nonumber \\ 
	&+&
	2624 g_1 g_X^3 g_{X1} x_H^3 x_\Phi +2624 g_{1X} g_X^2 g_{X1}^2 x_H^3 x_\Phi 
		+2624 g_1 g_X g_{X1}^3 x_H^3 x_\Phi +184 g_1^2 g_{1X} g_X^2 x_\Phi ^2 
	\nonumber \\ 
	&+&
	54 g_2^2 g_{1X} g_X^2 x_\Phi ^2+48 g_3^2 g_{1X} g_X^2 x_\Phi ^2+276 g_{1X}^3 g_X^2 x_\Phi ^2
		+138 g_1^3 g_X g_{X1} x_\Phi ^2+54 g_1 g_2^2 g_X g_{X1} x_\Phi ^2 
	\nonumber \\ 
	&+&
	48 g_1 g_3^2 g_X g_{X1} x_\Phi ^2+322 g_1 g_{1X}^2 g_X g_{X1} x_\Phi ^2
		+138 g_1^2 g_{1X} g_{X1}^2 x_\Phi ^2+46 g_{1X}^3 g_{X1}^2 x_\Phi ^2 
	\nonumber \\ 
	&+&
	276 g_1^2 g_X^3 x_H x_\Phi ^2+1104 g_{1X}^2 g_X^3 x_H x_\Phi ^2
		+1380 g_1 g_{1X} g_X^2 g_{X1} x_H x_\Phi ^2 +828 g_1^2 g_X g_{X1}^2 x_H x_\Phi ^2 
	\nonumber \\ 
	&+&
	552 g_{1X}^2 g_X g_{X1}^2 x_H x_\Phi ^2+276 g_1 g_{1X} g_{X1}^3 x_H x_\Phi ^2
		+1104 g_{1X} g_X^4 x_H^2 x_\Phi ^2+1104 g_1 g_X^3 g_{X1} x_H^2 x_\Phi ^2 
	\nonumber \\ 
	&+&
	1104 g_{1X} g_X^2 g_{X1}^2 x_H^2 x_\Phi ^2+1104 g_1 g_X g_{X1}^3 x_H^2 x_\Phi ^2
		+28 g_1^2 g_X^3 x_\Phi ^3+112 g_{1X}^2 g_X^3 x_\Phi ^3 
	\nonumber \\ 
	&+&
	140 g_1 g_{1X} g_X^2 g_{X1} x_\Phi ^3+84 g_1^2 g_X g_{X1}^2 x_\Phi ^3
		+56 g_{1X}^2 g_X g_{X1}^2 x_\Phi ^3+28 g_1 g_{1X} g_{X1}^3 x_\Phi ^3 
	\nonumber \\ 
	&+&
	224 g_{1X} g_X^4 x_H x_\Phi ^3+224 g_1 g_X^3 g_{X1} x_H x_\Phi ^3
		+224 g_{1X} g_X^2 g_{X1}^2 x_H x_\Phi ^3+224 g_1 g_X g_{X1}^3 x_H x_\Phi ^3 
	\nonumber \\ 
	&+&
	100 g_{1X} g_X^4 x_\Phi ^4+100 g_1 g_X^3 g_{X1} x_\Phi ^4+100 g_{1X} g_X^2 g_{X1}^2 x_\Phi ^4
		+100 g_1 g_X g_{X1}^3 x_\Phi ^4-54 g_{1X} g_X^2 x_\Phi ^2 y_M^2 
	\nonumber \\ 
	&-&
	54 g_1 g_X g_{X1} x_\Phi ^2 y_M^2-51 g_1^2 g_{1X} y_t^2-51 g_{1X}^3 y_t^2-102 g_1^2 g_X x_H y_t^2
		-204 g_{1X}^2 g_X x_H y_t^2 
	\nonumber \\ 
	&-&
	102 g_1 g_{1X} g_{X1} x_H y_t^2-204 g_{1X} g_X^2 x_H^2 y_t^2-204 g_1 g_X g_{X1} x_H^2 y_t^2
		-15 g_1^2 g_X x_\Phi  y_t^2 
	\nonumber \\ 
	&-&
	30 g_{1X}^2 g_X x_\Phi  y_t^2-15 g_1 g_{1X} g_{X1} x_\Phi  y_t^2-60 g_{1X} g_X^2 x_H x_\Phi  y_t^2
		-60 g_1 g_X g_{X1} x_H x_\Phi  y_t^2 
	\nonumber \\ 
	&-&
	\Big.6 g_{1X} g_X^2 x_\Phi ^2 y_t^2-6 g_1 g_X g_{X1} x_\Phi ^2 y_t^2\Big],
	\nonumber 
\end{eqnarray}
\begin{eqnarray}
%%%%% U(1)' RGE for g_X at 2-loop %%%%%
\beta_{g_X}^{(2)} &=& \frac{1}{{(16 \pi^2)}^2} \cdot \frac{1}{18}
	\Big[199 g_1^2 g_{1X}^2 g_X+81 g_2^2 g_{1X}^2 g_X+264 g_3^2 g_{1X}^2 g_X
		+199g_{1X}^4 g_X+199 g_1^3 g_{1X} g_{X1}\Big. 
	\nonumber \\ 
	&+&
	81 g_1 g_2^2 g_{1X} g_{X1}+264 g_1 g_3^2 g_{1X} g_{X1}+199 g_1 g_{1X}^3 g_{X1}
		+796g_1^2 g_{1X} g_X^2 x_H 
	\nonumber \\ 
	&+&
	324 g_2^2 g_{1X} g_X^2 x_H+1056 g_3^2 g_{1X} g_X^2 x_H+1592 g_{1X}^3 g_X^2 x_H+398 g_1^3 g_X g_{X1}x_H
	\nonumber \\ 
	&+&
	162 g_1 g_2^2 g_X g_{X1} x_H+528 g_1 g_3^2 g_X g_{X1} x_H+1990 g_1 g_{1X}^2 g_X g_{X1} x_H
		+1194 g_1^2 g_{1X}g_{X1}^2 x_H 
	\nonumber \\ 
	&+&
	162 g_2^2 g_{1X} g_{X1}^2 x_H+528 g_3^2 g_{1X} g_{X1}^2 x_H+398 g_{1X}^3 g_{X1}^2 x_H
		+796 g_1^2 g_X^3 x_H^2+324 g_2^2 g_X^3 x_H^2 
	\nonumber \\ 
	&+&
	1056 g_3^2 g_X^3 x_H^2+4776 g_{1X}^2 g_X^3 x_H^2+5572 g_1 g_{1X} g_X^2 g_{X1} x_H^2
		+2388 g_1^2 g_X g_{X1}^2x_H^2 
	\nonumber \\ 
	&+&
	324 g_2^2 g_X g_{X1}^2 x_H^2+1056 g_3^2 g_X g_{X1}^2 x_H^2+3184 g_{1X}^2 g_X g_{X1}^2 x_H^2
		+2388 g_1 g_{1X}g_{X1}^3 x_H^2 
	\nonumber \\ 
	&+&
	6368 g_{1X} g_X^4 x_H^3+4776 g_1 g_X^3 g_{X1} x_H^3+7960 g_{1X} g_X^2 g_{X1}^2 x_H^3
		+4776 g_1 g_X g_{X1}^3 x_H^3 
	\nonumber \\ 
	&+&
	1592 g_{1X} g_{X1}^4 x_H^3+3184 g_X^5 x_H^4+6368 g_X^3 g_{X1}^2 x_H^4+3184 g_X g_{X1}^4 x_H^4
		+164 g_1^2 g_{1X} g_X^2 x_\Phi  
	\nonumber \\ 
	&+&
	108 g_2^2 g_{1X} g_X^2 x_\Phi +96 g_3^2 g_{1X} g_X^2 x_\Phi +328 g_{1X}^3 g_X^2 x_\Phi 
		+82 g_1^3 g_X g_{X1} x_\Phi +54 g_1 g_2^2 g_X g_{X1} x_\Phi  
	\nonumber \\ 
	&+&
	48 g_1 g_3^2 g_X g_{X1} x_\Phi +410 g_1 g_{1X}^2 g_X g_{X1} x_\Phi +246 g_1^2 g_{1X} g_{X1}^2 x_\Phi 
		+54 g_2^2 g_{1X} g_{X1}^2 x_\Phi  
	\nonumber \\ 
	&+&
	48 g_3^2 g_{1X} g_{X1}^2 x_\Phi +82 g_{1X}^3 g_{X1}^2 x_\Phi +328 g_1^2 g_X^3 x_H x_\Phi 
		+216 g_2^2 g_X^3 x_H x_\Phi +192 g_3^2 g_X^3 x_H x_\Phi  
	\nonumber \\ 
	&+&
	1968 g_{1X}^2 g_X^3 x_H x_\Phi +2296 g_1 g_{1X} g_X^2 g_{X1} x_H x_\Phi 
		+984 g_1^2 g_X g_{X1}^2 x_H x_\Phi +216 g_2^2 g_X g_{X1}^2 x_H x_\Phi  
	\nonumber \\ 
	&+&
	192 g_3^2 g_X g_{X1}^2 x_H x_\Phi +1312 g_{1X}^2 g_X g_{X1}^2 x_H x_\Phi 
		+984 g_1 g_{1X} g_{X1}^3 x_H x_\Phi +3936 g_{1X} g_X^4 x_H^2 x_\Phi  
	\nonumber \\ 
	&+&
	2952 g_1 g_X^3 g_{X1} x_H^2 x_\Phi +4920 g_{1X} g_X^2 g_{X1}^2 x_H^2 x_\Phi 
		+2952 g_1 g_X g_{X1}^3 x_H^2 x_\Phi +984 g_{1X} g_{X1}^4 x_H^2 x_\Phi  
	\nonumber \\ 
	&+&
	2624 g_X^5 x_H^3 x_\Phi +5248 g_X^3 g_{X1}^2 x_H^3 x_\Phi +2624 g_X g_{X1}^4 x_H^3 x_\Phi 
		+46 g_1^2 g_X^3 x_\Phi ^2+54 g_2^2 g_X^3 x_\Phi ^2 
	\nonumber \\ 
	&+&
	48 g_3^2 g_X^3 x_\Phi ^2+276 g_{1X}^2 g_X^3 x_\Phi ^2+322 g_1 g_{1X} g_X^2 g_{X1} x_\Phi ^2
		+138 g_1^2 g_X g_{X1}^2 x_\Phi ^2+54 g_2^2 g_X g_{X1}^2 x_\Phi ^2 
	\nonumber \\ 
	&+&
	48 g_3^2 g_X g_{X1}^2 x_\Phi ^2+184 g_{1X}^2 g_X g_{X1}^2 x_\Phi ^2
		+138 g_1 g_{1X} g_{X1}^3 x_\Phi ^2+1104 g_{1X} g_X^4 x_H x_\Phi ^2 
	\nonumber \\ 
	&+&
	828 g_1 g_X^3 g_{X1} x_H x_\Phi ^2+1380 g_{1X} g_X^2 g_{X1}^2 x_H x_\Phi ^2
		+828 g_1 g_X g_{X1}^3 x_H x_\Phi ^2+276 g_{1X} g_{X1}^4 x_H x_\Phi ^2 
	\nonumber \\ 
	&+&
	1104 g_X^5 x_H^2 x_\Phi ^2+2208 g_X^3 g_{X1}^2 x_H^2 x_\Phi ^2+1104 g_X g_{X1}^4 x_H^2 x_\Phi ^2
		+112 g_{1X} g_X^4 x_\Phi ^3+84 g_1 g_X^3 g_{X1} x_\Phi ^3 
	\nonumber \\ 
	&+&
	140 g_{1X} g_X^2 g_{X1}^2 x_\Phi ^3+84 g_1 g_X g_{X1}^3 x_\Phi ^3
		+28 g_{1X} g_{X1}^4 x_\Phi ^3+224 g_X^5 x_H x_\Phi ^3+448 g_X^3 g_{X1}^2 x_H x_\Phi ^3 
	\nonumber \\ 
	&+&
	224 g_X g_{X1}^4 x_H x_\Phi ^3+100 g_X^5 x_\Phi ^4+200 g_X^3 g_{X1}^2 x_\Phi ^4
		+100 g_X g_{X1}^4 x_\Phi ^4-54 g_X^3 x_\Phi ^2 y_M^2-54 g_X g_{X1}^2 x_\Phi ^2 y_M^2 
	\nonumber \\ 
	&-&
	51 g_{1X}^2 g_X y_t^2-51 g_1 g_{1X} g_{X1} y_t^2-204 g_{1X} g_X^2 x_H y_t^2
		-102 g_1 g_X g_{X1} x_H y_t^2-102 g_{1X} g_{X1}^2 x_H y_t^2 
	\nonumber \\ 
	&-&
	204 g_X^3 x_H^2 y_t^2-204 g_X g_{X1}^2 x_H^2 y_t^2-30 g_{1X} g_X^2 x_\Phi  y_t^2
		-15 g_1 g_X g_{X1} x_\Phi y_t^2-15 g_{1X} g_{X1}^2 x_\Phi  y_t^2 
	\nonumber \\ 
	&-&
	\Big.60 g_X^3 x_H x_\Phi  y_t^2-60 g_X g_{X1}^2 x_H x_\Phi  y_t^2-6 g_X^3 x_\Phi ^2 y_t^2
		-6 g_X g_{X1}^2 x_\Phi ^2 y_t^2\Big].
\end{eqnarray}

%%%%%%%%%%%%%%%%%%%%%%%%%%%%%%%%%%%%%%%%%%%%%%%%%%%%%
\subsection{U(1)$^\prime$ RGEs for the Yukawa couplings} 
%%%%%%%%%%%%%%%%%%%%%%%%%%%%%%%%%%%%%%%%%%%%%%%%%%%%%
The RGEs for the Yukawa couplings at the two-loop level are given by  
\begin{eqnarray}
\mu \frac{d y_i}{d\mu} &=& \beta_{y_i}^{(1)}+\beta_{y_i}^{(2)}, 
\end{eqnarray}
  where $\beta_{y_i}^{(1)}$ and $\beta_{y_i}^{(2)}$ are the one-loop and two-loop beta functions
  for the Yukawa couplings, respectively, and $y_i$ represents $y_t$ and $y_M$.
Here, the one-loop beta functions for the Yukawa couplings are given by
\begin{eqnarray}
%%%%% U(1)' RGE for y_t at 1-loop %%%%%
\beta_{y_t}^{(1)} &=& \frac{y_t}{16\pi^2} 
	\left[ \frac{9}{2}y_t^2 - 8g_3^2 - \frac{9}{4}g_2^2 
		- \frac{1}{6}\big( g_1 + 2x_H g_{X1} + x_\Phi g_{X1}\big) \big( 4 g_1 + 8x_H g_{X1} + x_\Phi g_{X1} \big) 
	\right. 
\nonumber \\ 
  &-& \frac{3}{4} \big(g_1 + 2x_H g_{X1} \big)^{\!2} 
	- \frac{1}{6} \big( g_{1X} + 2x_H g_X + x_\Phi g_X\big) \big( 4g_{1X} +8x_H g_X + x_\Phi g_X \big) 
\nonumber \\ 
  &-& \bigg. \frac{3}{4} \big(g_{1X} + 2x_H g_X \big)^{\!2} \bigg],   \nonumber \\
%%%%% U(1)' RGE for y_M at 1-loop %%%%%
\beta_{y_M}^{(1)} &=& \frac{y_M}{16\pi^2} 
	\left[ 10 y_M^2 - \frac{3}{2} x_\Phi^2 \left( g_{X1}^2 + g_X^2 \right) \right],
\label{Eq:RGE_Yukawa}  
\end{eqnarray} 
and the two-loop beta functions for the Yukawa couplings are given by
\begin{eqnarray}
%%%%% U(1)' RGE for y_t  at 2-loop %%%%%
\beta_{y_t}^{(2)} &=& \frac{1}{{(16 \pi^2)}^2} \cdot \frac{1}{432}
	\Big[-9\left(-72 y_t^5-y_t^3 \left(223 g_1^2+405 g_2^2+768 g_3^2+223 g_{1X}^2
		+892 g_{1X} g_X x_H \right.\right.\Big.
	\nonumber \\ 
	&+&
	892 g_1 g_{X1} x_H+892 g_X^2 x_H^2+892 g_{X1}^2 x_H^2+50 g_{1X} g_X x_\Phi +50 g_1 g_{X1} x_\Phi 
		+100 g_X^2 x_H x_\Phi  
	\nonumber \\ 
	&+&
	\left.\left.100 g_{X1}^2 x_H x_\Phi +16 g_X^2 x_\Phi ^2+16 g_{X1}^2 x_\Phi ^2
		-324 y_t^2-576 \lambda _H\right)\right) 
	\nonumber \\ 
	&+&
	y_t \left(2374 g_1^4-324 g_1^2 g_2^2-2484 g_2^4+912 g_1^2 g_3^2+3888 g_2^2 g_3^2
		-46656 g_3^4+4748 g_1^2 g_{1X}^2\right. 
	\nonumber \\ 
	&-&
	324 g_2^2 g_{1X}^2+912 g_3^2 g_{1X}^2+2374 g_{1X}^4+18992 g_1^2 g_{1X} g_X x_H
		-1296 g_2^2 g_{1X} g_X x_H 
	\nonumber \\ 
	&+&
	3648 g_3^2 g_{1X} g_X x_H+18992 g_{1X}^3 g_X x_H+18992 g_1^3 g_{X1} x_H-1296 g_1 g_2^2 g_{X1} x_H 
	\nonumber \\ 
	&+&
	3648 g_1 g_3^2 g_{X1} x_H+18992 g_1 g_{1X}^2 g_{X1} x_H+18992 g_1^2 g_X^2 x_H^2-1296 g_2^2 g_X^2 x_H^2 
	\nonumber \\ 
	&+&
	3648 g_3^2 g_X^2 x_H^2+56976 g_{1X}^2 g_X^2 x_H^2+75968 g_1 g_{1X} g_X g_{X1} x_H^2
		+56976 g_1^2 g_{X1}^2 x_H^2 
	\nonumber \\ 
	&-&
	1296 g_2^2 g_{X1}^2 x_H^2+3648 g_3^2 g_{X1}^2 x_H^2+18992 g_{1X}^2 g_{X1}^2 x_H^2
		+75968 g_{1X} g_X^3 x_H^3 
	\nonumber \\ 
	&+&
	75968 g_1 g_X^2 g_{X1} x_H^3+75968 g_{1X} g_X g_{X1}^2 x_H^3+75968 g_1 g_{X1}^3 x_H^3+37984 g_X^4 x_H^4 
	\nonumber \\ 
	&+&
	75968 g_X^2 g_{X1}^2 x_H^4+37984 g_{X1}^4 x_H^4+4016 g_1^2 g_{1X} g_X x_\Phi 
		+486 g_2^2 g_{1X} g_X x_\Phi  
	\nonumber \\ 
	&-&
	480 g_3^2 g_{1X} g_X x_\Phi +4016 g_{1X}^3 g_X x_\Phi +4016 g_1^3 g_{X1} x_\Phi 
		+486 g_1 g_2^2 g_{X1} x_\Phi -480 g_1 g_3^2 g_{X1} x_\Phi  
	\nonumber \\ 
	&+&
	4016 g_1 g_{1X}^2 g_{X1} x_\Phi +8032 g_1^2 g_X^2 x_H x_\Phi +972 g_2^2 g_X^2 x_H x_\Phi 
		-960 g_3^2 g_X^2 x_H x_\Phi  
	\nonumber \\ 
	&+&
	24096 g_{1X}^2 g_X^2 x_H x_\Phi +32128 g_1 g_{1X} g_X g_{X1} x_H x_\Phi 
		+24096 g_1^2 g_{X1}^2 x_H x_\Phi +972 g_2^2 g_{X1}^2 x_H x_\Phi  
	\nonumber \\ 
	&-&
	960 g_3^2 g_{X1}^2 x_H x_\Phi +8032 g_{1X}^2 g_{X1}^2 x_H x_\Phi +48192 g_{1X} g_X^3 x_H^2 x_\Phi 
		+48192 g_1 g_X^2 g_{X1} x_H^2 x_\Phi  
	\nonumber \\ 
	&+&
	48192 g_{1X} g_X g_{X1}^2 x_H^2 x_\Phi +48192 g_1 g_{X1}^3 x_H^2 x_\Phi +32128 g_X^4 x_H^3 x_\Phi 
		+64256 g_X^2 g_{X1}^2 x_H^3 x_\Phi  
	\nonumber \\ 
	&+&
	32128 g_{X1}^4 x_H^3 x_\Phi +819 g_1^2 g_X^2 x_\Phi ^2+81 g_2^2 g_X^2 x_\Phi ^2
		-96 g_3^2 g_X^2 x_\Phi ^2+3255 g_{1X}^2 g_X^2 x_\Phi ^2 
	\nonumber \\ 
	&+&
	4872 g_1 g_{1X} g_X g_{X1} x_\Phi ^2+3255 g_1^2 g_{X1}^2 x_\Phi ^2+81 g_2^2 g_{X1}^2 x_\Phi ^2
		-96 g_3^2 g_{X1}^2 x_\Phi ^2+819 g_{1X}^2 g_{X1}^2 x_\Phi ^2 
	\nonumber \\ 
	&+&
	13020 g_{1X} g_X^3 x_H x_\Phi ^2+13020 g_1 g_X^2 g_{X1} x_H x_\Phi ^2
		+13020 g_{1X} g_X g_{X1}^2 x_H x_\Phi ^2+13020 g_1 g_{X1}^3 x_H x_\Phi ^2 
	\nonumber \\ 
	&+&
	13020 g_X^4 x_H^2 x_\Phi ^2+26040 g_X^2 g_{X1}^2 x_H^2 x_\Phi ^2+13020 g_{X1}^4 x_H^2 x_\Phi ^2
		+1330 g_{1X} g_X^3 x_\Phi ^3 
	\nonumber \\ 
	&+&
	1330 g_1 g_X^2 g_{X1} x_\Phi ^3+1330 g_{1X} g_X g_{X1}^2 x_\Phi ^3+1330 g_1 g_{X1}^3 x_\Phi ^3
		+2660 g_X^4 x_H x_\Phi ^3+5320 g_X^2 g_{X1}^2 x_H x_\Phi ^3 
	\nonumber \\ 
	&+&
	2660 g_{X1}^4 x_H x_\Phi ^3+203 g_X^4 x_\Phi ^4+406 g_X^2 g_{X1}^2 x_\Phi ^4+203 g_{X1}^4 x_\Phi ^4
		+1530 g_1^2 y_t^2+2430 g_2^2 y_t^2 
	\nonumber \\ 
	&+&
	8640 g_3^2 y_t^2+1530 g_{1X}^2 y_t^2+6120 g_{1X} g_X x_H y_t^2+6120 g_1 g_{X1} x_H y_t^2
		+6120 g_X^2 x_H^2 y_t^2 
	\nonumber \\ 
	&+&
	6120 g_{X1}^2 x_H^2 y_t^2+900 g_{1X} g_X x_\Phi  y_t^2+900 g_1 g_{X1} x_\Phi  y_t^2
		+1800 g_X^2 x_H x_\Phi y_t^2+1800 g_{X1}^2 x_H x_\Phi  y_t^2 
	\nonumber \\ 
	&+&
	\Big.\left.180 g_X^2 x_\Phi ^2 y_t^2+180 g_{X1}^2 x_\Phi ^2 y_t^2-2916 y_t^4+2592 \lambda _H^2
		+216 \lambda _{\rm mix}^2\right)\Big],
	\nonumber  
\end{eqnarray} 
\begin{eqnarray}
%%%%% U(1)' RGE for y_M at 2-loop %%%%%
\beta_{y_M}^{(2)} &=& \frac{1}{{(16 \pi^2)}^2} \cdot \frac{1}{48}
	\Big[y_M \left(-70 g_{1X}^2 g_X^2 x_\Phi ^2-140 g_1 g_{1X} g_X g_{X1} x_\Phi ^2
		-70 g_1^2 g_{X1}^2 x_\Phi ^2-280 g_{1X} g_X^3 x_H x_\Phi ^2\right.\Big.
	\nonumber \\ 
	&-&
	280 g_1 g_X^2 g_{X1} x_H x_\Phi ^2-280 g_{1X} g_X g_{X1}^2 x_H x_\Phi ^2
		-280 g_1 g_{X1}^3 x_H x_\Phi ^2-280 g_X^4 x_H^2 x_\Phi ^2 
	\nonumber \\ 
	&-&
	560 g_X^2 g_{X1}^2 x_H^2 x_\Phi ^2-280 g_{X1}^4 x_H^2 x_\Phi ^2-64 g_{1X} g_X^3 x_\Phi ^3
		-64 g_1 g_X^2 g_{X1} x_\Phi ^3-64 g_{1X} g_X g_{X1}^2 x_\Phi ^3 
	\nonumber \\ 
	&-&
	64 g_1 g_{X1}^3 x_\Phi ^3-128 g_X^4 x_H x_\Phi ^3-256 g_X^2 g_{X1}^2 x_H x_\Phi ^3
		-128 g_{X1}^4 x_H x_\Phi ^3-381 g_X^4 x_\Phi ^4-762 g_X^2 g_{X1}^2 x_\Phi ^4 
	\nonumber \\ 
	&-&
	\left.381 g_{X1}^4 x_\Phi ^4+360 \left(g_X^2+g_{X1}^2\right) x_\Phi ^2 y_M^2-1728 y_M^4
		+48 \lambda _{\rm mix}^2+192 \lambda_\Phi ^2\right) 
	\nonumber \\ 
	&+&
	\Big.6 \left(224 y_M^5-32 y_M^3 \left(-11 \left(g_X^2+g_{X1}^2\right) x_\Phi ^2+9 y_M^2
		+8 \lambda_\Phi \right)\right)\Big].
\end{eqnarray}

%%%%%%%%%%%%%%%%%%%%%%%%%%%%%%%%%%%%%%%%%%%%%%%%%%%%%
\subsection{U(1)$^\prime$ RGEs for the scalar quartic couplings} 
%%%%%%%%%%%%%%%%%%%%%%%%%%%%%%%%%%%%%%%%%%%%%%%%%%%%%
Finally, the RGEs for the scalar quartic couplings at the two-loop level are given by 
\begin{eqnarray}
\mu \frac{d \lambda_i}{d\mu} &=& \beta_{\lambda_i}^{(1)}+\beta_{\lambda_i}^{(2)}, 
\end{eqnarray}
  where $\beta_{\lambda_i}^{(1)}$ and $\beta_{\lambda_i}^{(2)}$ are the one-loop and two-loop beta functions
  for the scalar quartic couplings, respectively, 
  and $\lambda_i$ represents $\lambda_H$, $\lambda_\Phi$ and $\lambda_{\rm mix}$.
Here, the one-loop beta functions for the  scalar quartic couplings are given by
\begin{eqnarray}
%%%%% U(1)' RGE for \lambda_H at 1-loop %%%%%
\beta_{\lambda_H}^{(1)} &=& \frac{1}{16\pi^2} 
	\bigg[ \lambda_H \left\{ 24\lambda_H + 12y_t^2 - 9g_2^2 - 3\big(g_1+2x_H g_{X1} \big)^{\!2} 
						- 3\big(g_{1X}+2x_H g_X \big)^{\!2} \right\}
	\bigg. \nonumber \\
	&+& \left.
	    \lambda_{\rm mix}^2 - 6y_t^4 	+ \frac{9}{8}g_2^4 
		+ \frac{3}{8}\left\{\big(g_1+2x_H g_{X1} \big)^{\!2} 
				+ \big(g_{1X}+2x_H g_X \big)^{\!2}\right\}^2
	\right. \nonumber \\
	&+&\left.
		 \frac{3}{4}g_2^2\big( g_1 + 2x_H g_{X1} \big)^{\!2} 
		+ \frac{3}{4}g_2^2\big( g_{1X} + 2x_H g_X \big)^{\!2} 
	 \right],  
	 \nonumber \\
%%%%% U(1)' RGE for \lambda_\Phi at 1-loop %%%%%
\beta_{\lambda_\Phi}^{(1)} &=& \frac{1}{16\pi^2} 
	\bigg[ \lambda_\Phi \Big\{ 20\lambda_\Phi + 24 y_M^2 
							- 12\big( x_\Phi g_{X1} \big)^2 - 12\big( x_\Phi g_X \big)^2 \Big\} 
	\bigg. \nonumber \\
	&+&\bigg.
		 2\lambda_{\rm mix}^2 - 48 y_M^4 
		+ 6 \Big\{ \big( x_\Phi g_{X1} \big)^2 + \big( x_\Phi g_X \big)^2 \Big\}^2 
	\bigg], 
	 \nonumber \\  
%%%%% U(1)' RGE for \lambda_{mix} at 1-loop %%%%%
\beta_{\lambda_{\rm mix}}^{(1)} &=& \frac{1}{16\pi^2} 
	\left[ \lambda_{\rm mix} \bigg\{ 12\lambda_H + 8\lambda_\Phi + 4\lambda_{\rm mix} 
							+ 6y_t^2 + 12 y_M^2 \bigg. 
	\right. \nonumber \\
	&-& \left.
		 \frac{9}{2}g_2^2 - \frac{3}{2}\big( g_1+ 2x_H g_{X1} \big)^{\!2} - 6\big(x_\Phi g_{X1} \big)^2 
		- \frac{3}{2}\big(g_{1X} + 2x_H g_X \big)^{\!2} - 6\big( x_\Phi g_X \big)^2 
	 \right\} \nonumber \\
	&+& \left.
		3 \Big\{ \big( g_1 + 2x_H g_{X1} \big) \! \big( x_\Phi g_{X1} \big) 
						+ \big(g_{1X} + 2x_H g_X \big) \! \big( x_\Phi g_X \big) \Big\}^{2} 
	\right],
\label{Eq:RGE_lambda}
\end{eqnarray}
and the two-loop beta functions for the  scalar quartic couplings are given by
\begin{eqnarray}
%%%%% U(1)' RGE for \lambda_H at 2-loop %%%%%
\beta_{\lambda_H}^{(2)} &=& \frac{1}{{(16 \pi^2)}^2} 
	\bigg[-\frac{379 g_1^6}{48}-\frac{559}{48} g_1^4 g_2^2-\frac{289}{48} g_1^2 g_2^4
		+\frac{305 g_2^6}{16}-\frac{469}{48} g_1^4 g_{1X}^2-\frac{75}{8} g_1^2 g_2^2 g_{1X}^2\bigg. 
	\nonumber \\ 
	&-&
	\frac{289}{48} g_2^4 g_{1X}^2-\frac{469}{48} g_1^2 g_{1X}^4-\frac{559}{48} g_2^2 g_{1X}^4
		-\frac{379 g_{1X}^6}{48}-\frac{469}{12} g_1^4 g_{1X} g_X x_H 
	\nonumber \\ 
	&-&
	\frac{75}{2} g_1^2 g_2^2 g_{1X} g_X x_H-\frac{289}{12} g_2^4 g_{1X} g_X x_H
		-\frac{469}{6} g_1^2 g_{1X}^3 g_X x_H-\frac{559}{6} g_2^2 g_{1X}^3 g_X x_H 
	\nonumber \\ 
	&-&
	\frac{379}{4} g_{1X}^5 g_X x_H-\frac{379}{4} g_1^5 g_{X1} x_H-\frac{559}{6} g_1^3 g_2^2 g_{X1} x_H
		-\frac{289}{12} g_1 g_2^4 g_{X1} x_H-\frac{469}{6} g_1^3 g_{1X}^2 g_{X1} x_H 
	\nonumber \\ 
	&-&
	\frac{75}{2} g_1 g_2^2 g_{1X}^2 g_{X1} x_H-\frac{469}{12} g_1 g_{1X}^4 g_{X1} x_H
		-\frac{469}{12} g_1^4 g_X^2 x_H^2-\frac{75}{2} g_1^2 g_2^2 g_X^2 x_H^2-\frac{289}{12} g_2^4 g_X^2 x_H^2 
	\nonumber \\ 
	&-&
	\frac{469}{2} g_1^2 g_{1X}^2 g_X^2 x_H^2-\frac{559}{2} g_2^2 g_{1X}^2 g_X^2 x_H^2
		-\frac{1895}{4} g_{1X}^4 g_X^2 x_H^2-\frac{938}{3} g_1^3 g_{1X} g_X g_{X1} x_H^2 
	\nonumber \\ 
	&-&
	150 g_1 g_2^2 g_{1X} g_X g_{X1} x_H^2-\frac{938}{3} g_1 g_{1X}^3 g_X g_{X1} x_H^2
		-\frac{1895}{4} g_1^4 g_{X1}^2 x_H^2-\frac{559}{2} g_1^2 g_2^2 g_{X1}^2 x_H^2 
	\nonumber \\ 
	&-&
	\frac{289}{12} g_2^4 g_{X1}^2 x_H^2-\frac{469}{2} g_1^2 g_{1X}^2 g_{X1}^2 x_H^2
		-\frac{75}{2} g_2^2 g_{1X}^2 g_{X1}^2 x_H^2-\frac{469}{12} g_{1X}^4 g_{X1}^2 x_H^2
		-\frac{938}{3} g_1^2 g_{1X} g_X^3 x_H^3 
	\nonumber \\ 
	&-&
	\frac{1118}{3} g_2^2 g_{1X} g_X^3 x_H^3-\frac{3790}{3} g_{1X}^3 g_X^3 x_H^3
		-\frac{938}{3} g_1^3 g_X^2 g_{X1} x_H^3-150 g_1 g_2^2 g_X^2 g_{X1} x_H^3 
	\nonumber \\ 
	&-&
	938 g_1 g_{1X}^2 g_X^2 g_{X1} x_H^3-938 g_1^2 g_{1X} g_X g_{X1}^2 x_H^3
		-150 g_2^2 g_{1X} g_X g_{X1}^2 x_H^3-\frac{938}{3} g_{1X}^3 g_X g_{X1}^2 x_H^3 
	\nonumber \\ 
	&-&
	\frac{3790}{3} g_1^3 g_{X1}^3 x_H^3-\frac{1118}{3} g_1 g_2^2 g_{X1}^3 x_H^3
		-\frac{938}{3} g_1 g_{1X}^2 g_{X1}^3 x_H^3-\frac{469}{3} g_1^2 g_X^4 x_H^4-\frac{559}{3} g_2^2 g_X^4 x_H^4 
	\nonumber \\ 
	&-&
	1895 g_{1X}^2 g_X^4 x_H^4-\frac{3752}{3} g_1 g_{1X} g_X^3 g_{X1} x_H^4
		-938 g_1^2 g_X^2 g_{X1}^2 x_H^4-150 g_2^2 g_X^2 g_{X1}^2 x_H^4 
	\nonumber \\ 
	&-&
	938 g_{1X}^2 g_X^2 g_{X1}^2 x_H^4-\frac{3752}{3} g_1 g_{1X} g_X g_{X1}^3 x_H^4
		-1895 g_1^2 g_{X1}^4 x_H^4-\frac{559}{3} g_2^2 g_{X1}^4 x_H^4 
	\nonumber \\ 
	&-&
	\frac{469}{3} g_{1X}^2 g_{X1}^4 x_H^4-1516 g_{1X} g_X^5 x_H^5-\frac{1876}{3} g_1 g_X^4 g_{X1} x_H^5
		-\frac{3752}{3} g_{1X} g_X^3 g_{X1}^2 x_H^5 
	\nonumber \\ 
	&-&
	\frac{3752}{3} g_1 g_X^2 g_{X1}^3 x_H^5-\frac{1876}{3} g_{1X} g_X g_{X1}^4 x_H^5
		-1516 g_1 g_{X1}^5 x_H^5-\frac{1516}{3} g_X^6 x_H^6-\frac{1876}{3} g_X^4 g_{X1}^2 x_H^6 
	\nonumber \\ 
	&-&
	\frac{1876}{3} g_X^2 g_{X1}^4 x_H^6-\frac{1516}{3} g_{X1}^6 x_H^6
		-\frac{16}{3} g_1^2 g_{1X}^3 g_X x_\Phi -\frac{16}{3} g_2^2 g_{1X}^3 g_X x_\Phi 
		-\frac{16}{3} g_{1X}^5 g_X x_\Phi  
	\nonumber \\ 
	&-&
	\frac{16}{3} g_1^5 g_{X1} x_\Phi -\frac{16}{3} g_1^3 g_2^2 g_{X1} x_\Phi 
		-\frac{16}{3} g_1^3 g_{1X}^2 g_{X1} x_\Phi -32 g_1^2 g_{1X}^2 g_X^2 x_H x_\Phi 
		-32 g_2^2 g_{1X}^2 g_X^2 x_H x_\Phi  
	\nonumber \\ 
	&-&
	\frac{160}{3} g_{1X}^4 g_X^2 x_H x_\Phi -\frac{64}{3} g_1^3 g_{1X} g_X g_{X1} x_H x_\Phi 
		-\frac{64}{3} g_1 g_{1X}^3 g_X g_{X1} x_H x_\Phi -\frac{160}{3} g_1^4 g_{X1}^2 x_H x_\Phi  
	\nonumber \\ 
	&-&
	32 g_1^2 g_2^2 g_{X1}^2 x_H x_\Phi -32 g_1^2 g_{1X}^2 g_{X1}^2 x_H x_\Phi 
		-64 g_1^2 g_{1X} g_X^3 x_H^2 x_\Phi -64 g_2^2 g_{1X} g_X^3 x_H^2 x_\Phi  
	\nonumber \\ 
	&-&
	\frac{640}{3} g_{1X}^3 g_X^3 x_H^2 x_\Phi -\frac{64}{3} g_1^3 g_X^2 g_{X1} x_H^2 x_\Phi 
		-128 g_1 g_{1X}^2 g_X^2 g_{X1} x_H^2 x_\Phi -128 g_1^2 g_{1X} g_X g_{X1}^2 x_H^2 x_\Phi  
	\nonumber \\ 
	&-&
	\frac{64}{3} g_{1X}^3 g_X g_{X1}^2 x_H^2 x_\Phi -\frac{640}{3} g_1^3 g_{X1}^3 x_H^2 x_\Phi 
		-64 g_1 g_2^2 g_{X1}^3 x_H^2 x_\Phi -64 g_1 g_{1X}^2 g_{X1}^3 x_H^2 x_\Phi  
	\nonumber \\ 
	&-&
	\frac{128}{3} g_1^2 g_X^4 x_H^3 x_\Phi -\frac{128}{3} g_2^2 g_X^4 x_H^3 x_\Phi 
		-\frac{1280}{3} g_{1X}^2 g_X^4 x_H^3 x_\Phi -256 g_1 g_{1X} g_X^3 g_{X1} x_H^3 x_\Phi  
	\nonumber  
\end{eqnarray}
\begin{eqnarray}
	&-&
	128 g_1^2 g_X^2 g_{X1}^2 x_H^3 x_\Phi -128 g_{1X}^2 g_X^2 g_{X1}^2 x_H^3 x_\Phi 
		-256 g_1 g_{1X} g_X g_{X1}^3 x_H^3 x_\Phi -\frac{1280}{3} g_1^2 g_{X1}^4 x_H^3 x_\Phi  
	\nonumber \\ 
	&-&
	\frac{128}{3} g_2^2 g_{X1}^4 x_H^3 x_\Phi -\frac{128}{3} g_{1X}^2 g_{X1}^4 x_H^3 x_\Phi 
		-\frac{1280}{3} g_{1X} g_X^5 x_H^4 x_\Phi -\frac{512}{3} g_1 g_X^4 g_{X1} x_H^4 x_\Phi  
	\nonumber \\ 
	&-&
	256 g_{1X} g_X^3 g_{X1}^2 x_H^4 x_\Phi -256 g_1 g_X^2 g_{X1}^3 x_H^4 x_\Phi 
		-\frac{512}{3} g_{1X} g_X g_{X1}^4 x_H^4 x_\Phi -\frac{1280}{3} g_1 g_{X1}^5 x_H^4 x_\Phi  
	\nonumber \\ 
	&-&
	\frac{512}{3} g_X^6 x_H^5 x_\Phi -\frac{512}{3} g_X^4 g_{X1}^2 x_H^5 x_\Phi 
		-\frac{512}{3} g_X^2 g_{X1}^4 x_H^5 x_\Phi -\frac{512}{3} g_{X1}^6 x_H^5 x_\Phi 
		-\frac{13}{4} g_1^2 g_{1X}^2 g_X^2 x_\Phi ^2 
	\nonumber \\ 
	&-&
	\frac{13}{4} g_2^2 g_{1X}^2 g_X^2 x_\Phi ^2-\frac{13}{4} g_{1X}^4 g_X^2 x_\Phi ^2
		-\frac{13}{4} g_1^4 g_{X1}^2 x_\Phi ^2-\frac{13}{4} g_1^2 g_2^2 g_{X1}^2 x_\Phi ^2
		-\frac{13}{4} g_1^2 g_{1X}^2 g_{X1}^2 x_\Phi ^2 
	\nonumber \\
	&-&
	13 g_1^2 g_{1X} g_X^3 x_H x_\Phi ^2-13 g_2^2 g_{1X} g_X^3 x_H x_\Phi ^2
		-26 g_{1X}^3 g_X^3 x_H x_\Phi ^2-13 g_1 g_{1X}^2 g_X^2 g_{X1} x_H x_\Phi ^2 
	\nonumber \\ 
	&-&
	13 g_1^2 g_{1X} g_X g_{X1}^2 x_H x_\Phi ^2-26 g_1^3 g_{X1}^3 x_H x_\Phi ^2
		-13 g_1 g_2^2 g_{X1}^3 x_H x_\Phi ^2-13 g_1 g_{1X}^2 g_{X1}^3 x_H x_\Phi ^2 
	\nonumber \\ 
	&-&
	13 g_1^2 g_X^4 x_H^2 x_\Phi ^2-13 g_2^2 g_X^4 x_H^2 x_\Phi ^2-78 g_{1X}^2 g_X^4 x_H^2 x_\Phi ^2
		-52 g_1 g_{1X} g_X^3 g_{X1} x_H^2 x_\Phi ^2-13 g_1^2 g_X^2 g_{X1}^2 x_H^2 x_\Phi ^2 
	\nonumber \\ 
	&-&
	13 g_{1X}^2 g_X^2 g_{X1}^2 x_H^2 x_\Phi ^2-52 g_1 g_{1X} g_X g_{X1}^3 x_H^2 x_\Phi ^2
		-78 g_1^2 g_{X1}^4 x_H^2 x_\Phi ^2-13 g_2^2 g_{X1}^4 x_H^2 x_\Phi ^2 
	\nonumber \\ 
	&-&
	13 g_{1X}^2 g_{X1}^4 x_H^2 x_\Phi ^2-104 g_{1X} g_X^5 x_H^3 x_\Phi ^2
		-52 g_1 g_X^4 g_{X1} x_H^3 x_\Phi ^2-52 g_{1X} g_X^3 g_{X1}^2 x_H^3 x_\Phi ^2 
	\nonumber \\ 
	&-&
	52 g_1 g_X^2 g_{X1}^3 x_H^3 x_\Phi ^2-52 g_{1X} g_X g_{X1}^4 x_H^3 x_\Phi ^2
		-104 g_1 g_{X1}^5 x_H^3 x_\Phi ^2-52 g_X^6 x_H^4 x_\Phi ^2-52 g_X^4 g_{X1}^2 x_H^4 x_\Phi ^2 
	\nonumber \\ 
	&-&
	52 g_X^2 g_{X1}^4 x_H^4 x_\Phi ^2-52 g_{X1}^6 x_H^4 x_\Phi ^2-\frac{19}{4} g_1^4 y_t^2
		+\frac{21}{2} g_1^2 g_2^2 y_t^2-\frac{9}{4} g_2^4 y_t^2-\frac{19}{2} g_1^2 g_{1X}^2 y_t^2 
	\nonumber \\ 
	&+&
	\frac{21}{2} g_2^2 g_{1X}^2 y_t^2-\frac{19}{4} g_{1X}^4 y_t^2-38 g_1^2 g_{1X} g_X x_H y_t^2
		+42 g_2^2 g_{1X} g_X x_H y_t^2-38 g_{1X}^3 g_X x_H y_t^2 
	\nonumber \\ 
	&-&
	38 g_1^3 g_{X1} x_H y_t^2+42 g_1 g_2^2 g_{X1} x_H y_t^2-38 g_1 g_{1X}^2 g_{X1} x_H y_t^2
		-38 g_1^2 g_X^2 x_H^2 y_t^2+42 g_2^2 g_X^2 x_H^2 y_t^2 
	\nonumber \\ 
	&-&
	114 g_{1X}^2 g_X^2 x_H^2 y_t^2-152 g_1 g_{1X} g_X g_{X1} x_H^2 y_t^2-114 g_1^2 g_{X1}^2 x_H^2 y_t^2
		+42 g_2^2 g_{X1}^2 x_H^2 y_t^2 
	\nonumber \\ 
	&-&
	38 g_{1X}^2 g_{X1}^2 x_H^2 y_t^2-152 g_{1X} g_X^3 x_H^3 y_t^2-152 g_1 g_X^2 g_{X1} x_H^3 y_t^2
		-152 g_{1X} g_X g_{X1}^2 x_H^3 y_t^2 
	\nonumber \\ 
	&-&
	152 g_1 g_{X1}^3 x_H^3 y_t^2-76 g_X^4 x_H^4 y_t^2-152 g_X^2 g_{X1}^2 x_H^4 y_t^2
		-76 g_{X1}^4 x_H^4 y_t^2-5 g_1^2 g_{1X} g_X x_\Phi  y_t^2 
	\nonumber \\ 
	&+&
	3 g_2^2 g_{1X} g_X x_\Phi  y_t^2-5 g_{1X}^3 g_X x_\Phi  y_t^2-5 g_1^3 g_{X1} x_\Phi  y_t^2
		+3 g_1 g_2^2 g_{X1} x_\Phi  y_t^2-5 g_1 g_{1X}^2 g_{X1} x_\Phi  y_t^2 
	\nonumber \\ 
	&-&
	10 g_1^2 g_X^2 x_H x_\Phi  y_t^2+6 g_2^2 g_X^2 x_H x_\Phi  y_t^2-30 g_{1X}^2 g_X^2 x_H x_\Phi  y_t^2
		-40 g_1 g_{1X} g_X g_{X1} x_H x_\Phi  y_t^2 
	\nonumber \\ 
	&-&
	30 g_1^2 g_{X1}^2 x_H x_\Phi  y_t^2+6 g_2^2 g_{X1}^2 x_H x_\Phi  y_t^2
		-10 g_{1X}^2 g_{X1}^2 x_H x_\Phi y_t^2-60 g_{1X} g_X^3 x_H^2 x_\Phi  y_t^2 
	\nonumber \\ 
	&-&
	60 g_1 g_X^2 g_{X1} x_H^2 x_\Phi  y_t^2-60 g_{1X} g_X g_{X1}^2 x_H^2 x_\Phi  y_t^2
		-60 g_1 g_{X1}^3 x_H^2 x_\Phi  y_t^2-40 g_X^4 x_H^3 x_\Phi  y_t^2 
	\nonumber \\ 
	&-&
	80 g_X^2 g_{X1}^2 x_H^3 x_\Phi  y_t^2-40 g_{X1}^4 x_H^3 x_\Phi  y_t^2-g_{1X}^2 g_X^2 x_\Phi ^2 y_t^2
		-2 g_1 g_{1X} g_X g_{X1} x_\Phi ^2 y_t^2-g_1^2 g_{X1}^2 x_\Phi ^2 y_t^2 
	\nonumber \\ 
	&-&
	4 g_{1X} g_X^3 x_H x_\Phi ^2 y_t^2-4 g_1 g_X^2 g_{X1} x_H x_\Phi ^2 y_t^2
		-4 g_{1X} g_X g_{X1}^2 x_H x_\Phi ^2 y_t^2-4 g_1 g_{X1}^3 x_H x_\Phi ^2 y_t^2 
	\nonumber \\ 
	&-&
	4 g_X^4 x_H^2 x_\Phi ^2 y_t^2-8 g_X^2 g_{X1}^2 x_H^2 x_\Phi ^2 y_t^2-4 g_{X1}^4 x_H^2 x_\Phi ^2 y_t^2
		-\frac{8}{3} g_1^2 y_t^4-32 g_3^2 y_t^4-\frac{8}{3} g_{1X}^2 y_t^4 
	\nonumber \\ 
	&-&
	\frac{32}{3} g_{1X} g_X x_H y_t^4-\frac{32}{3} g_1 g_{X1} x_H y_t^4-\frac{32}{3} g_X^2 x_H^2 y_t^4
		-\frac{32}{3} g_{X1}^2 x_H^2 y_t^4-\frac{10}{3} g_{1X} g_X x_\Phi  y_t^4 
	\nonumber \\ 
	&-&
	\frac{10}{3} g_1 g_{X1} x_\Phi  y_t^4-\frac{20}{3} g_X^2 x_H x_\Phi  y_t^4
		-\frac{20}{3} g_{X1}^2 x_H x_\Phi  y_t^4-\frac{2}{3} g_X^2 x_\Phi ^2 y_t^4
		-\frac{2}{3} g_{X1}^2 x_\Phi ^2 y_t^4+30 y_t^6 
	\nonumber \\ 
	&+&
	\frac{629}{24} g_1^4 \lambda _H+\frac{39}{4} g_1^2 g_2^2 \lambda _H-\frac{73}{8} g_2^4 \lambda _H
		+\frac{69}{4} g_1^2 g_{1X}^2 \lambda _H+\frac{39}{4} g_2^2 g_{1X}^2 \lambda _H
		+\frac{629}{24} g_{1X}^4 \lambda _H 
	\nonumber \\ 
	&+&
	69 g_1^2 g_{1X} g_X x_H \lambda _H+39 g_2^2 g_{1X} g_X x_H \lambda _H
		+\frac{629}{3} g_{1X}^3 g_X x_H \lambda _H+\frac{629}{3} g_1^3 g_{X1} x_H \lambda _H 
	\nonumber \\ 
	&+&
	39 g_1 g_2^2 g_{X1} x_H \lambda _H+69 g_1 g_{1X}^2 g_{X1} x_H \lambda _H
		+69 g_1^2 g_X^2 x_H^2 \lambda _H+39 g_2^2 g_X^2 x_H^2 \lambda _H
		+629 g_{1X}^2 g_X^2 x_H^2 \lambda _H 
	\nonumber  
\end{eqnarray}
\begin{eqnarray}
	&+&
	276 g_1 g_{1X} g_X g_{X1} x_H^2 \lambda _H+629 g_1^2 g_{X1}^2 x_H^2 \lambda _H
		+39 g_2^2 g_{X1}^2 x_H^2 \lambda _H+69 g_{1X}^2 g_{X1}^2 x_H^2 \lambda _H 
	\nonumber \\ 
	&+&
	\frac{2516}{3} g_{1X} g_X^3 x_H^3 \lambda _H+276 g_1 g_X^2 g_{X1} x_H^3 \lambda _H
		+276 g_{1X} g_X g_{X1}^2 x_H^3 \lambda _H+\frac{2516}{3} g_1 g_{X1}^3 x_H^3 \lambda _H 
	\nonumber \\ 
	&+&
	\frac{1258}{3} g_X^4 x_H^4 \lambda _H+276 g_X^2 g_{X1}^2 x_H^4 \lambda _H
		+\frac{1258}{3} g_{X1}^4 x_H^4 \lambda _H+\frac{40}{3} g_{1X}^3 g_X x_\Phi  \lambda _H
		+\frac{40}{3} g_1^3 g_{X1} x_\Phi  \lambda _H 
	\nonumber \\ 
	&+&
	80 g_{1X}^2 g_X^2 x_H x_\Phi  \lambda _H+80 g_1^2 g_{X1}^2 x_H x_\Phi  \lambda _H
		+160 g_{1X} g_X^3 x_H^2 x_\Phi  \lambda _H+160 g_1 g_{X1}^3 x_H^2 x_\Phi  \lambda _H 
	\nonumber \\ 
	&+&
	\frac{320}{3} g_X^4 x_H^3 x_\Phi  \lambda _H+\frac{320}{3} g_{X1}^4 x_H^3 x_\Phi  \lambda _H
		+\frac{17}{2} g_{1X}^2 g_X^2 x_\Phi ^2 \lambda _H+\frac{17}{2} g_1^2 g_{X1}^2 x_\Phi ^2 \lambda _H
		+34 g_{1X} g_X^3 x_H x_\Phi ^2 \lambda _H 
	\nonumber \\ 
	&+&
	34 g_1 g_{X1}^3 x_H x_\Phi ^2 \lambda _H+34 g_X^4 x_H^2 x_\Phi ^2 \lambda _H
		+34 g_{X1}^4 x_H^2 x_\Phi ^2 \lambda _H+\frac{85}{6} g_1^2 y_t^2 \lambda _H
		+\frac{45}{2} g_2^2 y_t^2 \lambda _H 
	\nonumber \\ 
	&+&
	80 g_3^2 y_t^2 \lambda _H+\frac{85}{6} g_{1X}^2 y_t^2 \lambda _H
		+\frac{170}{3} g_{1X} g_X x_H y_t^2 \lambda _H+\frac{170}{3} g_1 g_{X1} x_H y_t^2 \lambda _H
		+\frac{170}{3} g_X^2 x_H^2 y_t^2 \lambda _H 
	\nonumber \\ 
	&+&
	\frac{170}{3} g_{X1}^2 x_H^2 y_t^2 \lambda _H+\frac{25}{3} g_{1X} g_X x_\Phi  y_t^2 \lambda _H
		+\frac{25}{3} g_1 g_{X1} x_\Phi  y_t^2 \lambda _H+\frac{50}{3} g_X^2 x_H x_\Phi  y_t^2 \lambda _H 
	\nonumber \\ 
	&+&
	\frac{50}{3} g_{X1}^2 x_H x_\Phi  y_t^2 \lambda _H+\frac{5}{3} g_X^2 x_\Phi ^2 y_t^2 \lambda _H
		+\frac{5}{3} g_{X1}^2 x_\Phi ^2 y_t^2 \lambda _H-3 y_t^4 \lambda _H+36 g_1^2 \lambda _H^2
		+108 g_2^2 \lambda _H^2 
	\nonumber \\ 
	&+&
	36 g_{1X}^2 \lambda _H^2+144 g_{1X} g_X x_H \lambda _H^2+144 g_1 g_{X1} x_H \lambda _H^2
		+144 g_X^2 x_H^2 \lambda _H^2+144 g_{X1}^2 x_H^2 \lambda _H^2-144 y_t^2 \lambda _H^2 
	\nonumber \\ 
	&-&
	312 \lambda _H^3+5 g_{1X}^2 g_X^2 x_\Phi ^2 \lambda _{\rm mix}
		+10 g_1 g_{1X} g_X g_{X1} x_\Phi ^2 \lambda _{\rm mix}+5 g_1^2 g_{X1}^2 x_\Phi ^2 \lambda _{\rm mix}
		+20 g_{1X} g_X^3 x_H x_\Phi ^2 \lambda _{\rm mix} 
	\nonumber \\ 
	&+&
	20 g_1 g_X^2 g_{X1} x_H x_\Phi ^2 \lambda _{\rm mix}
		+20 g_{1X} g_X g_{X1}^2 x_H x_\Phi ^2 \lambda _{\rm mix}
		+20 g_1 g_{X1}^3 x_H x_\Phi ^2 \lambda _{\rm mix}+20 g_X^4 x_H^2 x_\Phi ^2 \lambda _{\rm mix} 
	\nonumber \\ 
	&+&
	40 g_X^2 g_{X1}^2 x_H^2 x_\Phi ^2 \lambda _{\rm mix}+20 g_{X1}^4 x_H^2 x_\Phi ^2 \lambda _{\rm mix}
		+8 g_X^2 x_\Phi ^2 \lambda _{\rm mix}^2+8 g_{X1}^2 x_\Phi ^2 \lambda _{\rm mix}^2
		-12 y_M^2 \lambda _{\rm mix}^2
	\nonumber \\ 
	&-&
	\bigg.10 \lambda _H \lambda _{\rm mix}^2-4 \lambda _{\rm mix}^3 \bigg],
	\nonumber  
\end{eqnarray}
\begin{eqnarray}
%%%%% U(1)' RGE for \lambda_\Phi at 2-loop %%%%%
\beta_{\lambda_\Phi}^{(2)} &=& \frac{1}{{(16 \pi^2)}^2} \cdot \frac{1}{3}
	\bigg[-334 g_{1X}^2 g_X^4 x_\Phi ^4-334 g_1^2 g_X^2 g_{X1}^2 x_\Phi ^4
		-334 g_{1X}^2 g_X^2 g_{X1}^2 x_\Phi ^4-334 g_1^2 g_{X1}^4 x_\Phi ^4\bigg. 
	\nonumber \\ 
	&-&
	1336 g_{1X} g_X^5 x_H x_\Phi ^4-1336 g_{1X} g_X^3 g_{X1}^2 x_H x_\Phi ^4
		-1336 g_1 g_X^2 g_{X1}^3 x_H x_\Phi ^4-1336 g_1 g_{X1}^5 x_H x_\Phi ^4 
	\nonumber \\ 
	&-&
	1336 g_X^6 x_H^2 x_\Phi ^4-1336 g_X^4 g_{X1}^2 x_H^2 x_\Phi ^4-1336 g_X^2 g_{X1}^4 x_H^2 x_\Phi ^4
		-1336 g_{X1}^6 x_H^2 x_\Phi ^4-256 g_{1X} g_X^5 x_\Phi ^5 
	\nonumber \\ 
	&-&
	256 g_{1X} g_X^3 g_{X1}^2 x_\Phi ^5-256 g_1 g_X^2 g_{X1}^3 x_\Phi ^5-256 g_1 g_{X1}^5 x_\Phi ^5
		-512 g_X^6 x_H x_\Phi ^5-512 g_X^4 g_{X1}^2 x_H x_\Phi ^5 
	\nonumber \\ 
	&-&
	512 g_X^2 g_{X1}^4 x_H x_\Phi ^5-512 g_{X1}^6 x_H x_\Phi ^5-336 g_X^6 x_\Phi ^6
		-696 g_X^4 g_{X1}^2 x_\Phi ^6-696 g_X^2 g_{X1}^4 x_\Phi ^6-336 g_{X1}^6 x_\Phi ^6 
	\nonumber \\ 
	&+&
	144 g_X^4 x_\Phi ^4 y_M^2+288 g_X^2 g_{X1}^2 x_\Phi ^4 y_M^2+144 g_{X1}^4 x_\Phi ^4 y_M^2
		+144 g_X^2 x_\Phi ^2 y_M^4+144 g_{X1}^2 x_\Phi ^2 y_M^4+2304 y_M^6 
	\nonumber \\ 
	&+&
	30 g_{1X}^2 g_X^2 x_\Phi ^2 \lambda _{\rm mix}+60 g_1 g_{1X} g_X g_{X1} x_\Phi ^2 \lambda _{\rm mix}
		+30 g_1^2 g_{X1}^2 x_\Phi ^2 \lambda _{\rm mix}+120 g_{1X} g_X^3 x_H x_\Phi ^2 \lambda _{\rm mix} 
	\nonumber \\ 
	&+&
	120 g_1 g_X^2 g_{X1} x_H x_\Phi ^2 \lambda _{\rm mix}
		+120 g_{1X} g_X g_{X1}^2 x_H x_\Phi ^2 \lambda _{\rm mix}
		+120 g_1 g_{X1}^3 x_H x_\Phi ^2 \lambda _{\rm mix}+120 g_X^4 x_H^2 x_\Phi ^2 \lambda _{\rm mix} 
	\nonumber \\ 
	&+&
	240 g_X^2 g_{X1}^2 x_H^2 x_\Phi ^2 \lambda _{\rm mix}+120 g_{X1}^4 x_H^2 x_\Phi ^2 \lambda _{\rm mix}
		+12 g_1^2 \lambda _{\rm mix}^2+36 g_2^2 \lambda _{\rm mix}^2+12 g_{1X}^2 \lambda _{\rm mix}^2 
	\nonumber \\ 
	&+&
	48 g_{1X} g_X x_H \lambda _{\rm mix}^2+48 g_1 g_{X1} x_H \lambda _{\rm mix}^2
		+48 g_X^2 x_H^2 \lambda _{\rm mix}^2+48 g_{X1}^2 x_H^2 \lambda _{\rm mix}^2
		-36 y_t^2 \lambda _{\rm mix}^2-24 \lambda _{\rm mix}^3 
	\nonumber \\ 
	&+&
	211 g_{1X}^2 g_X^2 x_\Phi ^2 \lambda_\Phi +211 g_1^2 g_{X1}^2 x_\Phi ^2 \lambda_\Phi 
		+844 g_{1X} g_X^3 x_H x_\Phi ^2 \lambda_\Phi +844 g_1 g_{X1}^3 x_H x_\Phi ^2 \lambda_\Phi  
	\nonumber \\ 
	&+&
	844 g_X^4 x_H^2 x_\Phi ^2 \lambda_\Phi +844 g_{X1}^4 x_H^2 x_\Phi ^2 \lambda_\Phi 
		+160 g_{1X} g_X^3 x_\Phi ^3 \lambda_\Phi +160 g_1 g_{X1}^3 x_\Phi ^3 \lambda_\Phi 
		+320 g_X^4 x_H x_\Phi ^3 \lambda_\Phi  
	\nonumber \\ 
	&+&
	320 g_{X1}^4 x_H x_\Phi ^3 \lambda_\Phi +396 g_X^4 x_\Phi ^4 \lambda_\Phi 
		+588 g_X^2 g_{X1}^2 x_\Phi ^4 \lambda _\Phi +396 g_{X1}^4 x_\Phi ^4 \lambda_\Phi 
		+90 g_X^2 x_\Phi ^2 y_M^2 \lambda_\Phi  
	\nonumber \\ 
	&+&
	90 g_{X1}^2 x_\Phi ^2 y_M^2 \lambda_\Phi +144 y_M^4 \lambda_\Phi 
		-60 \lambda _{\rm mix}^2 \lambda_\Phi +336 g_X^2 x_\Phi ^2 \lambda_\Phi ^2
		+336 g_{X1}^2 x_\Phi ^2 \lambda_\Phi ^2
	\nonumber \\ 
	&-&
	\bigg.720 y_M^2 \lambda_\Phi ^2-720 \lambda_\Phi ^3\bigg],
	\nonumber  
\end{eqnarray}
\begin{eqnarray}
%%%%% U(1)' RGE for \lambda_{mix} at 2-loop %%%%%
\beta_{\lambda_{\rm mix}}^{(2)} &=& \frac{1}{{(16 \pi^2)}^2}
	\bigg[-\frac{15}{4} g_1^2 g_{1X}^2 g_X^2 x_\Phi ^2-\frac{45}{4} g_2^2 g_{1X}^2 g_X^2 x_\Phi ^2
		-\frac{713}{12} g_{1X}^4 g_X^2 x_\Phi ^2-\frac{379}{6} g_1^3 g_{1X} g_X g_{X1} x_\Phi ^2\bigg. 
	\nonumber \\ 
	&-&
	\frac{45}{2} g_1 g_2^2 g_{1X} g_X g_{X1} x_\Phi ^2-\frac{379}{6} g_1 g_{1X}^3 g_X g_{X1} x_\Phi ^2
		-\frac{713}{12} g_1^4 g_{X1}^2 x_\Phi ^2-\frac{45}{4} g_1^2 g_2^2 g_{X1}^2 x_\Phi ^2 
	\nonumber \\ 
	&-&
	\frac{15}{4} g_1^2 g_{1X}^2 g_{X1}^2 x_\Phi ^2-15 g_1^2 g_{1X} g_X^3 x_H x_\Phi ^2
		-45 g_2^2 g_{1X} g_X^3 x_H x_\Phi ^2-\frac{1426}{3} g_{1X}^3 g_X^3 x_H x_\Phi ^2 
	\nonumber \\ 
	&-&
	\frac{379}{3} g_1^3 g_X^2 g_{X1} x_H x_\Phi ^2-45 g_1 g_2^2 g_X^2 g_{X1} x_H x_\Phi ^2
		-394 g_1 g_{1X}^2 g_X^2 g_{X1} x_H x_\Phi ^2-394 g_1^2 g_{1X} g_X g_{X1}^2 x_H x_\Phi ^2 
	\nonumber \\ 
	&-&
	45 g_2^2 g_{1X} g_X g_{X1}^2 x_H x_\Phi ^2-\frac{379}{3} g_{1X}^3 g_X g_{X1}^2 x_H x_\Phi ^2
		-\frac{1426}{3} g_1^3 g_{X1}^3 x_H x_\Phi ^2-45 g_1 g_2^2 g_{X1}^3 x_H x_\Phi ^2 
	\nonumber \\ 
	&-&
	15 g_1 g_{1X}^2 g_{X1}^3 x_H x_\Phi ^2-15 g_1^2 g_X^4 x_H^2 x_\Phi ^2-45 g_2^2 g_X^4 x_H^2 x_\Phi ^2
		-1426 g_{1X}^2 g_X^4 x_H^2 x_\Phi ^2 
	\nonumber \\ 
	&-&
	818 g_1 g_{1X} g_X^3 g_{X1} x_H^2 x_\Phi ^2-773 g_1^2 g_X^2 g_{X1}^2 x_H^2 x_\Phi ^2
		-90 g_2^2 g_X^2 g_{X1}^2 x_H^2 x_\Phi ^2-773 g_{1X}^2 g_X^2 g_{X1}^2 x_H^2 x_\Phi ^2 
	\nonumber \\ 
	&-&
	818 g_1 g_{1X} g_X g_{X1}^3 x_H^2 x_\Phi ^2-1426 g_1^2 g_{X1}^4 x_H^2 x_\Phi ^2
		-45 g_2^2 g_{X1}^4 x_H^2 x_\Phi ^2-15 g_{1X}^2 g_{X1}^4 x_H^2 x_\Phi ^2 
	\nonumber \\ 
	&-&
	\frac{5704}{3} g_{1X} g_X^5 x_H^3 x_\Phi ^2-\frac{1696}{3} g_1 g_X^4 g_{X1} x_H^3 x_\Phi ^2
		-1576 g_{1X}g_X^3 g_{X1}^2 x_H^3 x_\Phi ^2-1576 g_1 g_X^2 g_{X1}^3 x_H^3 x_\Phi ^2 
	\nonumber \\ 
	&-&
	\frac{1696}{3} g_{1X} g_X g_{X1}^4 x_H^3 x_\Phi ^2-\frac{5704}{3} g_1 g_{X1}^5 x_H^3 x_\Phi ^2
		-\frac{2852}{3} g_X^6 x_H^4 x_\Phi ^2-\frac{3212}{3} g_X^4 g_{X1}^2 x_H^4 x_\Phi ^2 
	\nonumber \\ 
	&-&
	\frac{3212}{3} g_X^2 g_{X1}^4 x_H^4 x_\Phi ^2-\frac{2852}{3} g_{X1}^6 x_H^4 x_\Phi ^2
		-\frac{128}{3} g_{1X}^3 g_X^3 x_\Phi ^3-\frac{128}{3} g_1 g_{1X}^2 g_X^2 g_{X1} x_\Phi ^3 
	\nonumber \\ 
	&-&
	\frac{128}{3} g_1^2 g_{1X} g_X g_{X1}^2 x_\Phi ^3-\frac{128}{3} g_1^3 g_{X1}^3 x_\Phi ^3
		-256 g_{1X}^2 g_X^4 x_H x_\Phi ^3-\frac{512}{3} g_1 g_{1X} g_X^3 g_{X1} x_H x_\Phi ^3 
	\nonumber \\ 
	&-&
	\frac{256}{3} g_1^2 g_X^2 g_{X1}^2 x_H x_\Phi ^3-\frac{256}{3} g_{1X}^2 g_X^2 g_{X1}^2 x_H x_\Phi ^3
		-\frac{512}{3} g_1 g_{1X} g_X g_{X1}^3 x_H x_\Phi ^3-256 g_1^2 g_{X1}^4 x_H x_\Phi ^3 
	\nonumber \\ 
	&-&
	512 g_{1X} g_X^5 x_H^2 x_\Phi ^3-\frac{512}{3} g_1 g_X^4 g_{X1} x_H^2 x_\Phi ^3
		-\frac{1024}{3} g_{1X} g_X^3 g_{X1}^2 x_H^2 x_\Phi ^3-\frac{1024}{3} g_1 g_X^2 g_{X1}^3 x_H^2 x_\Phi ^3 
	\nonumber \\ 
	&-&
	\frac{512}{3} g_{1X} g_X g_{X1}^4 x_H^2 x_\Phi ^3-512 g_1 g_{X1}^5 x_H^2 x_\Phi ^3
		-\frac{1024}{3} g_X^6 x_H^3 x_\Phi ^3-\frac{1024}{3} g_X^4 g_{X1}^2 x_H^3 x_\Phi ^3 
	\nonumber \\ 
	&-&
	\frac{1024}{3} g_X^2 g_{X1}^4 x_H^3 x_\Phi ^3-\frac{1024}{3} g_{X1}^6 x_H^3 x_\Phi ^3
		-41 g_{1X}^2 g_X^4 x_\Phi ^4-56 g_1 g_{1X} g_X^3 g_{X1} x_\Phi ^4-15 g_1^2 g_X^2 g_{X1}^2 x_\Phi ^4 
	\nonumber \\ 
	&-&
	15 g_{1X}^2 g_X^2 g_{X1}^2 x_\Phi ^4-56 g_1 g_{1X} g_X g_{X1}^3 x_\Phi ^4
		-41 g_1^2 g_{X1}^4 x_\Phi ^4-164 g_{1X} g_X^5 x_H x_\Phi ^4-112 g_1 g_X^4 g_{X1} x_H x_\Phi ^4 
	\nonumber \\ 
	&-&
	172 g_{1X} g_X^3 g_{X1}^2 x_H x_\Phi ^4-172 g_1 g_X^2 g_{X1}^3 x_H x_\Phi ^4
		-112 g_{1X} g_X g_{X1}^4 x_H x_\Phi ^4-164 g_1 g_{X1}^5 x_H x_\Phi ^4 
	\nonumber \\ 
	&-&
	164 g_X^6 x_H^2 x_\Phi ^4-284 g_X^4 g_{X1}^2 x_H^2 x_\Phi ^4-284 g_X^2 g_{X1}^4 x_H^2 x_\Phi ^4
		-164 g_{X1}^6 x_H^2 x_\Phi ^4+12 g_{1X}^2 g_X^2 x_\Phi ^2 y_M^2 
	\nonumber \\ 
	&+&
	24 g_1 g_{1X} g_X g_{X1} x_\Phi ^2 y_M^2+12 g_1^2 g_{X1}^2 x_\Phi ^2 y_M^2
		+48 g_{1X} g_X^3 x_H x_\Phi ^2 y_M^2+48 g_1 g_X^2 g_{X1} x_H x_\Phi ^2 y_M^2 
	\nonumber \\ 
	&+&
	48 g_{1X} g_X g_{X1}^2 x_H x_\Phi ^2 y_M^2+48 g_1 g_{X1}^3 x_H x_\Phi ^2 y_M^2
		+48 g_X^4 x_H^2 x_\Phi ^2 y_M^2+96 g_X^2 g_{X1}^2 x_H^2 x_\Phi ^2 y_M^2 
	\nonumber \\ 
	&+&
	48 g_{X1}^4 x_H^2 x_\Phi ^2 y_M^2-19 g_{1X}^2 g_X^2 x_\Phi ^2 y_t^2
		-38 g_1 g_{1X} g_X g_{X1} x_\Phi ^2 y_t^2-19 g_1^2 g_{X1}^2 x_\Phi ^2 y_t^2 
	\nonumber \\ 
	&-&
	76 g_{1X} g_X^3 x_H x_\Phi ^2 y_t^2-76 g_1 g_X^2 g_{X1} x_H x_\Phi ^2 y_t^2
		-76 g_{1X} g_X g_{X1}^2 x_H x_\Phi ^2 y_t^2-76 g_1 g_{X1}^3 x_H x_\Phi ^2 y_t^2 
	\nonumber \\ 
	&-&
	76 g_X^4 x_H^2 x_\Phi ^2 y_t^2-152 g_X^2 g_{X1}^2 x_H^2 x_\Phi ^2 y_t^2
		-76 g_{X1}^4 x_H^2 x_\Phi ^2 y_t^2-20 g_{1X} g_X^3 x_\Phi ^3 y_t^2-20 g_1 g_X^2 g_{X1} x_\Phi ^3 y_t^2 
	\nonumber \\ 
	&-&
	20 g_{1X} g_X g_{X1}^2 x_\Phi ^3 y_t^2-20 g_1 g_{X1}^3 x_\Phi ^3 y_t^2-40 g_X^4 x_H x_\Phi ^3 y_t^2
		-80 g_X^2 g_{X1}^2 x_H x_\Phi ^3 y_t^2-40 g_{X1}^4 x_H x_\Phi ^3 y_t^2 
	\nonumber \\ 
	&-&
	4 g_X^4 x_\Phi ^4 y_t^2-8 g_X^2 g_{X1}^2 x_\Phi ^4 y_t^2-4 g_{X1}^4 x_\Phi ^4 y_t^2
		+30 g_{1X}^2 g_X^2 x_\Phi ^2 \lambda _H+60 g_1 g_{1X} g_X g_{X1} x_\Phi ^2 \lambda _H 
	\nonumber \\ 
	&+&
	30 g_1^2 g_{X1}^2 x_\Phi ^2 \lambda _H+120 g_{1X} g_X^3 x_H x_\Phi ^2 \lambda _H
		+120 g_1 g_X^2 g_{X1} x_H x_\Phi ^2 \lambda _H+120 g_{1X} g_X g_{X1}^2 x_H x_\Phi ^2 \lambda _H 
	\nonumber \\ 
	&+&
	120 g_1 g_{X1}^3 x_H x_\Phi ^2 \lambda _H+120 g_X^4 x_H^2 x_\Phi ^2 \lambda _H
		+240 g_X^2 g_{X1}^2 x_H^2 x_\Phi ^2 \lambda _H+120 g_{X1}^4 x_H^2 x_\Phi ^2 \lambda _H
		+\frac{557}{48} g_1^4 \lambda _{\rm mix} 
	\nonumber  
\end{eqnarray}
\begin{eqnarray}
	&+&
	\frac{15}{8} g_1^2 g_2^2 \lambda _{\rm mix}-\frac{145}{16} g_2^4 \lambda _{\rm mix}
		+\frac{45}{8} g_1^2 g_{1X}^2 \lambda _{\rm mix}+\frac{15}{8} g_2^2 g_{1X}^2 \lambda _{\rm mix}
		+\frac{557}{48} g_{1X}^4 \lambda _{\rm mix} 
	\nonumber \\ 
	&+&
	\frac{45}{2} g_1^2 g_{1X} g_X x_H \lambda _{\rm mix}
		+\frac{15}{2} g_2^2 g_{1X} g_X x_H \lambda _{\rm mix}
		+\frac{557}{6} g_{1X}^3 g_X x_H \lambda _{\rm mix}+\frac{557}{6} g_1^3 g_{X1} x_H \lambda _{\rm mix} 
	\nonumber \\ 
	&+&
	\frac{15}{2} g_1 g_2^2 g_{X1} x_H \lambda _{\rm mix}
		+\frac{45}{2} g_1 g_{1X}^2 g_{X1} x_H \lambda _{\rm mix}
		+\frac{45}{2} g_1^2 g_X^2 x_H^2 \lambda _{\rm mix}+\frac{15}{2} g_2^2 g_X^2 x_H^2 \lambda _{\rm mix} 
	\nonumber \\ 
	&+&
	\frac{557}{2} g_{1X}^2 g_X^2 x_H^2 \lambda _{\rm mix}
		+90 g_1 g_{1X} g_X g_{X1} x_H^2 \lambda _{\rm mix}
		+\frac{557}{2} g_1^2 g_{X1}^2 x_H^2 \lambda _{\rm mix}
		+\frac{15}{2} g_2^2 g_{X1}^2 x_H^2 \lambda _{\rm mix} 
	\nonumber \\ 
	&+&
	\frac{45}{2} g_{1X}^2 g_{X1}^2 x_H^2 \lambda _{\rm mix}+\frac{1114}{3} g_{1X} g_X^3 x_H^3 \lambda _{\rm mix}
		+90 g_1 g_X^2 g_{X1} x_H^3 \lambda _{\rm mix}+90 g_{1X} g_X g_{X1}^2 x_H^3 \lambda _{\rm mix} 
	\nonumber \\ 
	&+&
	\frac{1114}{3} g_1 g_{X1}^3 x_H^3 \lambda _{\rm mix}+\frac{557}{3} g_X^4 x_H^4 \lambda _{\rm mix}
		+90 g_X^2 g_{X1}^2 x_H^4 \lambda _{\rm mix}+\frac{557}{3} g_{X1}^4 x_H^4 \lambda _{\rm mix}
		+\frac{20}{3} g_{1X}^3 g_X x_\Phi  \lambda _{\rm mix} 
	\nonumber \\ 
	&+&
	\frac{20}{3} g_1^3 g_{X1} x_\Phi  \lambda _{\rm mix}+40 g_{1X}^2 g_X^2 x_H x_\Phi  \lambda _{\rm mix}
		+40 g_1^2 g_{X1}^2 x_H x_\Phi  \lambda _{\rm mix}+80 g_{1X} g_X^3 x_H^2 x_\Phi  \lambda _{\rm mix} 
	\nonumber \\ 
	&+&
	80 g_1 g_{X1}^3 x_H^2 x_\Phi  \lambda _{\rm mix}
		+\frac{160}{3} g_X^4 x_H^3 x_\Phi  \lambda _{\rm mix}
		+\frac{160}{3} g_{X1}^4 x_H^3 x_\Phi  \lambda _{\rm mix}
		+\frac{497}{12} g_{1X}^2 g_X^2 x_\Phi ^2 \lambda _{\rm mix} 
	\nonumber \\ 
	&+&
	4 g_1 g_{1X} g_X g_{X1} x_\Phi ^2 \lambda _{\rm mix}
		+\frac{497}{12} g_1^2 g_{X1}^2 x_\Phi ^2 \lambda _{\rm mix}
		+\frac{497}{3} g_{1X} g_X^3 x_H x_\Phi ^2 \lambda _{\rm mix}
		+8 g_1 g_X^2 g_{X1} x_H x_\Phi ^2 \lambda _{\rm mix} 
	\nonumber \\ 
	&+&
	8 g_{1X} g_X g_{X1}^2 x_H x_\Phi ^2 \lambda _{\rm mix}
		+\frac{497}{3} g_1 g_{X1}^3 x_H x_\Phi ^2 \lambda _{\rm mix}
		+\frac{497}{3} g_X^4 x_H^2 x_\Phi ^2 \lambda _{\rm mix}
		+16 g_X^2 g_{X1}^2 x_H^2 x_\Phi ^2 \lambda _{\rm mix} 
	\nonumber \\ 
	&+&
	\frac{497}{3} g_{X1}^4 x_H^2 x_\Phi ^2 \lambda _{\rm mix}
		+\frac{80}{3} g_{1X} g_X^3 x_\Phi ^3 \lambda _{\rm mix}
		+\frac{80}{3} g_1 g_{X1}^3 x_\Phi ^3 \lambda _{\rm mix}
		+\frac{160}{3} g_X^4 x_H x_\Phi ^3 \lambda _{\rm mix} 
	\nonumber \\ 
	&+&
	\frac{160}{3} g_{X1}^4 x_H x_\Phi ^3 \lambda _{\rm mix}+42 g_X^4 x_\Phi ^4 \lambda _{\rm mix}
		+50 g_X^2 g_{X1}^2 x_\Phi ^4 \lambda _{\rm mix}+42 g_{X1}^4 x_\Phi ^4 \lambda _{\rm mix}
		+15 g_X^2 x_\Phi ^2 y_M^2 \lambda _{\rm mix} 
	\nonumber \\ 
	&+&
	15 g_{X1}^2 x_\Phi ^2 y_M^2 \lambda _{\rm mix}-72 y_M^4 \lambda _{\rm mix}
		+\frac{85}{12} g_1^2 y_t^2 \lambda _{\rm mix}+\frac{45}{4} g_2^2 y_t^2 \lambda _{\rm mix}
		+40 g_3^2 y_t^2 \lambda _{\rm mix} 
	\nonumber \\ 
	&+&
	\frac{85}{12} g_{1X}^2 y_t^2 \lambda _{\rm mix}+\frac{85}{3} g_{1X} g_X x_H y_t^2 \lambda _{\rm mix}
		+\frac{85}{3} g_1 g_{X1} x_H y_t^2 \lambda _{\rm mix}+\frac{85}{3} g_X^2 x_H^2 y_t^2 \lambda _{\rm mix} 
	\nonumber \\ 
	&+&
	\frac{85}{3} g_{X1}^2 x_H^2 y_t^2 \lambda _{\rm mix}
		+\frac{25}{6} g_{1X} g_X x_\Phi  y_t^2 \lambda _{\rm mix}
		+\frac{25}{6} g_1 g_{X1} x_\Phi  y_t^2 \lambda _{\rm mix}
		+\frac{25}{3} g_X^2 x_H x_\Phi  y_t^2 \lambda _{\rm mix} 
	\nonumber \\ 
	&+&
	\frac{25}{3} g_{X1}^2 x_H x_\Phi  y_t^2 \lambda _{\rm mix}
		+\frac{5}{6} g_X^2 x_\Phi ^2 y_t^2 \lambda _{\rm mix}
		+\frac{5}{6} g_{X1}^2 x_\Phi ^2 y_t^2 \lambda _{\rm mix}
		-\frac{27}{2} y_t^4 \lambda _{\rm mix}+24 g_1^2 \lambda _H \lambda _{\rm mix} 
	\nonumber \\ 
	&+&
	72 g_2^2 \lambda _H \lambda _{\rm mix}+24 g_{1X}^2 \lambda _H \lambda _{\rm mix}
		+96 g_{1X} g_X x_H \lambda _H \lambda _{\rm mix}+96 g_1 g_{X1} x_H \lambda _H \lambda _{\rm mix}
		+96 g_X^2 x_H^2 \lambda _H \lambda _{\rm mix} 
	\nonumber \\ 
	&+&
	96 g_{X1}^2 x_H^2 \lambda _H \lambda _{\rm mix}-72 y_t^2 \lambda _H \lambda _{\rm mix}
		-60 \lambda _H^2 \lambda _{\rm mix}+g_1^2 \lambda _{\rm mix}^2+3 g_2^2 \lambda _{\rm mix}^2
		+g_{1X}^2 \lambda _{\rm mix}^2+4 g_{1X} g_X x_H \lambda _{\rm mix}^2 
	\nonumber \\ 
	&+&
	4 g_1 g_{X1} x_H \lambda _{\rm mix}^2+4 g_X^2 x_H^2 \lambda _{\rm mix}^2
		+4 g_{X1}^2 x_H^2 \lambda _{\rm mix}^2+4 g_X^2 x_\Phi ^2 \lambda _{\rm mix}^2
		+4 g_{X1}^2 x_\Phi ^2 \lambda _{\rm mix}^2-24 y_M^2 \lambda _{\rm mix}^2 
	\nonumber \\ 
	&-&
	12 y_t^2 \lambda _{\rm mix}^2-72 \lambda _H \lambda _{\rm mix}^2-11 \lambda _{\rm mix}^3
		+20 g_{1X}^2 g_X^2 x_\Phi ^2 \lambda_\Phi +40 g_1 g_{1X} g_X g_{X1} x_\Phi ^2 \lambda_\Phi 
		+20 g_1^2 g_{X1}^2 x_\Phi ^2 \lambda _\Phi  
	\nonumber \\ 
	&+&
	80 g_{1X} g_X^3 x_H x_\Phi ^2 \lambda_\Phi +80 g_1 g_X^2 g_{X1} x_H x_\Phi ^2 \lambda_\Phi 
		+80 g_{1X} g_X g_{X1}^2 x_H x_\Phi ^2 \lambda_\Phi +80 g_1 g_{X1}^3 x_H x_\Phi ^2 \lambda_\Phi  
	\nonumber \\ 
	&+&
	80 g_X^4 x_H^2 x_\Phi ^2 \lambda_\Phi +160 g_X^2 g_{X1}^2 x_H^2 x_\Phi ^2 \lambda_\Phi 
		+80 g_{X1}^4 x_H^2 x_\Phi ^2 \lambda_\Phi +64 g_X^2 x_\Phi ^2 \lambda _{\rm mix} \lambda_\Phi 
		+64 g_{X1}^2 x_\Phi ^2 \lambda _{\rm mix} \lambda _\Phi  
	\nonumber \\ 
	&-&
	\bigg.96 y_M^2 \lambda _{\rm mix} \lambda_\Phi -48 \lambda _{\rm mix}^2 \lambda_\Phi 
		-40 \lambda _{\rm mix} \lambda_\Phi ^2\bigg].
\end{eqnarray}

%%%%%%%%%%%%%%%%%%%%%%%%%%%%%%%%%%%%%%%%
\section{SM RGES AT THE TWO-LOOP LEVEL}
\label{Sec:SM_RGEs}
%%%%%%%%%%%%%%%%%%%%%%%%%%%%%%%%%%%%%%%%
The RGEs for coupling constants of the SM up to two-loop level \cite{RGErun} are given by 
\begin{eqnarray}
%%%%% 2-loop SM RGE for g_3 %%%%%
\mu \frac{d g_3}{d\mu} &=& \frac{g_3^3}{(4\pi)^2} \Big[ -7 \Big]
	+ \frac{g_3^3}{(4\pi)^4} \left[ - 26g_3^2   + \frac{9}{2}g_2^2 + \frac{11}{6}g_1^2 - 2y_t^2 \right],
 \nonumber \\	
%\label{Eq:SM_RGE_g_3}\\
%%%%% 2-loop SM RGE for g_2 %%%%%
\mu \frac{d g_2}{d\mu} &=& \frac{g_2^3}{(4\pi)^2} \left[ -\frac{19}{6} \right]
	+ \frac{g_2^3}{(4\pi)^4} \left[12g_3^2 + \frac{35}{6}g_2^2 + \frac{3}{2}g_1^2 - \frac{3}{2}y_t^2 \right],
 \nonumber \\
%%%%% 2-loop SM RGE for g_1 %%%%%
\mu \frac{d g_1}{d\mu} &=& \frac{g_1^3}{(4\pi)^2} \left[ \frac{41}{6} \right] 
	+ \frac{g_1^3}{(4\pi)^4} 
		\left[ \frac{44}{3}g_3^2 + \frac{9}{2}g_2^2 + \frac{199}{18}g_1^2 - \frac{17}{6}y_t^2 \right],
  \nonumber\\
%%%%% 2-loop SM RGE for y_t %%%%%
\mu \frac{d y_t}{d\mu} &=& \frac{y_t}{(4\pi)^2} 
		\left[ \frac{9}{2}y_t^2 - 8g_3^2 - \frac{9}{4}g_2^2 - \frac{17}{12}g_1^2 \right] \nonumber\\
	&+& \frac{y_t}{(4\pi)^4} \left[ 
			y_t^2 \! \left( - 12y_t^2 - 12\lambda_H 
						+ 36g_3^2 + \frac{225}{16}g_2^2 + \frac{131}{16}g_1^2 \right) 
		\right. \nonumber\\
	&+& \left.
			6\lambda_H^2 - 108 g_3^4 - \frac{23}{4}g_2^4 + \frac{1187}{216}g_1^4 
			+ 9g_3^2 g_2^2 + \frac{19}{9}g_3^2 g_1^2- \frac{3}{4}g_2^2 g_1^2 
	 	\right], 
	\nonumber\\
%%%%% 2-loop SM RGE for \lambda_H %%%%%
\mu \frac{d \lambda_H}{d\mu} &=& \frac{1}{(4\pi)^2} 
		\left[ \lambda_H \Big( 24\lambda_H + 12y_t^2 - 9g_2^2 - 3g_1^2 \Big)
			- 6y_t^4 + \frac{9}{8}g_2^4 + \frac{3}{8}g_1^4 + \frac{3}{4}g_2^2 g_1^2 
		\right] \nonumber\\
	&+& \frac{1}{(4\pi)^4} \left[ 
			\lambda_H^2 \Big( - 312\lambda_H - 144y_t^2 + 108g_2^2 + 36g_1^2 \Big)  
			\right. \nonumber\\
	&+& \lambda_H y_t^2 \! \left( - 3y_t^2 + 80g_3^2 + \frac{45}{2}g_2^2 + \frac{85}{6}g_1^2 \right) 
			+ \lambda_H \! \left( - \frac{73}{8}g_2^4 + \frac{629}{24}g_1^4 + \frac{39}{4}g_1^2 g_2^2 \right) 
			\nonumber\\
	&+& y_t^4 \! \left( 30y_t^2 - 32g_3^2 - \frac{8}{3}g_1^2 \right) 
			+ y_t^2 \! \left( - \frac{9}{4}g_2^4 - \frac{19}{4}g_1^4 + \frac{21}{2}g_2^2 g_1^2 \right)
			\nonumber\\
	&+& \left.
		 \frac{305}{16}g_2^6 - \frac{379}{48}g_1^6 
			- \frac{289}{48}g_2^4 g_1^2 - \frac{559}{48}g_2^2 g_1^4 
		\right] .
\label{Eq:SM_RGE_lambda_H}
\end{eqnarray}
In our analysis, we numerically solve these SM RGEs with the following boundary conditions at $\mu=m_t$ 
  \cite{RGErun} \footnote{
We employed the boundary conditions in arXiv version 4 of Ref. \cite{RGErun}.
} 
\begin{eqnarray}
%%%%% Boundary conditions for the SM RGEs at M_t (Version 4) %%%%%
g_3(m_t) &=& 1.1666 + 0.00314 \left( \frac{\alpha_3(m_Z) - 0.1184}{0.0007} \right) 
			-  0.00046 \left( \frac{m_t}{\rm GeV} - 173.34 \right), \nonumber \\
g_2(m_t) &=& 0.64779 + 0.00004 \left( \frac{m_t}{\rm GeV} - 173.34 \right) 
			+ 0.00011 \left( \frac{m_W - 80.384{\rm GeV}}{0.014{\rm GeV}} \right), \nonumber \\
g_1(m_t) &=& 0.35830 + 0.00011 \left( \frac{m_t}{\rm GeV} - 173.34 \right) 
			- 0.00020\left(  \frac{m_W - 80.384{\rm GeV}}{0.014{\rm GeV}} \right),\nonumber \\
y_t(m_t) &=& 0.93690 + 0.00556 \left( \frac{m_t}{\rm GeV} - 173.34 \right) 
			- 0.00042 \left( \frac{\alpha_3(m_Z) - 0.1184}{0.0007} \right), \nonumber \\
\lambda_H(m_t) &=& 0.12604 + 0.00206 \left( \frac{m_h}{\rm GeV} - 125.15 \right) 
			- 0.00004 \left( \frac{m_t}{\rm GeV} - 173.34\right), 
\label{Eq:SM_BC_lambda_H}
\end{eqnarray}
using the inputs $\alpha_3(m_Z) = 0.1184$, $m_t=173.34$ GeV, $m_h=125.09$ GeV, 
and $m_W=80.384$ GeV. 

%%%%%%%%%%%%%%%%%%%%%%%%%%%%%%%%%%%%

\end{document}